\documentclass[journal,10pt]{IEEEtran}
\usepackage{cite}
\usepackage[font=small]{caption}
\usepackage{ifpdf}
\usepackage{epstopdf}
\usepackage{authblk}
\usepackage{commath}
\usepackage{cuted} 
\usepackage{stix}
\usepackage{pgfplots}
\usetikzlibrary{arrows,decorations.markings}
\usepackage{bigints}
\newif\ifCLASSOPTIONonecolumn       \CLASSOPTIONonecolumnfalse
\newif\ifCLASSOPTIONtwocolumn       \CLASSOPTIONtwocolumntrue
\ifCLASSINFOpdf
\usepackage{multicol}
\usepackage{subcaption}
\usepackage{amsmath}
\usepackage{dsfont}
\usepackage{bbold}
\usepackage{amssymb}
\usepackage{algorithmic}
\usepackage{array}
\usepackage{stfloats}
\fnbelowfloat
\usepackage{bigints}
\usepackage{subcaption}
\usepackage{amsmath}
\usepackage{dsfont}
\usepackage{bbold}
\usepackage{amssymb}
\usepackage{algorithmic}
\usepackage{array}
\usepackage{stix}
\usepackage{stfloats}
\fnbelowfloat
\usepackage{bigints}
\usepackage{booktabs} 
\usepackage{multirow}
\usepackage{siunitx} 
\usepackage{longtable}

\newcolumntype{P}[1]{>{\centering\arraybackslash}p{#1}}

\begin{document}
%
\title{Unleashing the Potential of Networked Tethered Flying Platforms for B5G/6G:
Prospects, Challenges, and Applications}
%
%
%
\author[1]{Baha Eddine Youcef~Belmekki}
\author[2]{Mohamed-Slim~Alouini}
\affil[1]{IRIT Laboratory, School
of ENSEEIHT, Institut National Polytechnique de Toulouse, France,}
\affil[ ]{ e-mail:bahaeddine.belmekki@enseeiht.fr}
\affil[2]{Computer, Electrical and Mathematical
Sciences and Engineering (CEMSE) Division, King Abdullah University of Science and Technology (KAUST), Kingdom of Saudi Arabia,}
\affil[ ]{email: slim.alouini@kaust.edu.sa}
\setcounter{Maxaffil}{0}
\renewcommand\Affilfont{\small}


\markboth{}%
{Shell \MakeLowercase{\textit{et al.}}: Bare Demo of IEEEtran.cls for IEEE Journals}

{}
\maketitle

\IEEEpeerreviewmaketitle

\begin{abstract}

Researchers are currently speculating about what the role of beyond fifth-generation (B5G) or sixth generation (6G) wireless systems will be. Several use cases and applications are proposed, ranging from enhanced mobile broadband communications and massive ultra-reliable low-latency Communication to holographic telepresence and tactile communications. One crucial and pivotal prospect of 6G is worldwide connectivity. Today, nearly 4 billion people do not have access to an internet connection. 6G not only intends to hyper-connect the already connected, but to bridge the digital divide by connecting the unconnected. For that purpose, 6G architectures will integrate airborne communications as they are considered key enablers for the next generation of wireless systems. Airborne communications can be done via free-flying platforms and tethered flying platforms. Free-flying platforms, such as unmanned aerial vehicles (UAVs) and high-altitude platforms (HAPs) have received great interest during the last decade. However, they lack endurance and backhaul capacity.
Networked Tethered Flying Platforms (NTFPs), on the other hand, are enjoying the benefits of free-flying platforms while overcoming their limitations by providing a continuous supply of power and data via the tether. They are also cost-efficient and have good green credentials. Tether platforms have a wide range of applications beyond communications, such as energy harvesting, entertainment, science, research, public safety, disaster relief, government, and defense.
In this survey paper, we intend to provide an extensive and comprehensive overview of NTFPs.
First, we provide a general overview of this solution for all readers interested irrespective of their background by reviewing all the existing types of NTFPs, including their components, their characteristics, their applications, advantages, challenges, and regulations. We also show several case studies in various fields and applications. 
Then, we investigate NTFPs from a wireless communications perspective.
We briefly provide a basic geometry analysis of NTFPs with respect to Earth. Then, we present the works dealing with the performance of NTFPs. 
We also show how NTFPs will be key enablers in 6G for various types of communications, such as flying car communications, maritime communications, and vehicular communications. 
We present an economic analysis of NTFPs with a capital expenditure (CAPEX) and operating expenditure (OPEX) analysis to emphasize their cost-efficiency.
Finally, we present channel models for NTFPs considering different altitudes, that is, low-altitude platforms (LAPs) and HAPs, and for different links, that is, radio frequency (RF) and free-space optics (FSO). Although these channel models are presented in this survey for NTFPs, they are applicable for free-flying platforms.

\end{abstract}

\IEEEpeerreviewmaketitle

\section{Introduction}
\label{section_intro}

Since the dawn of time, humans always had the need to communication and connect with one another.
In the 1980s, the first generation (1G) of wireless communications allowing voice communication was launched. Forty years later, three generation of wireless communication networks have been launched, namely
the second, third, and fourth generation (2G, 3G, and 4G) of wireless networks.
At the time of writing this paper, the fifth generation (5G) of wireless communications is being commercialized in some country, such as Switzerland, South Korea, United States, and United
Kingdom. Evolving standards throughout this saga of communications have resulted in increased data rates and decreased latency from each generation to next.


Now, researchers have begun speculating about what beyond fifth-generation (B5G) or the sixth generation (6G) will be, or more exactly, what 6G should be \cite{dang2020should,2david20186g,3raghavan2019evolution,4yastrebova2018future,5saad2019vision,6strinati20196g,7tariq2020speculative}. The reasons why researchers have already started thinking about 6G are the expected massive growth of mobile traffic in the next decade and the new type of disruptive services that are envisioned.
Some of the applications and services that 6G can offer, such as, enhancing the conventional mobile communications, increasing the accuracy of indoor positioning, providing holographic telepresence, tactile communications, extended reality, worldwide connectivity, and integrated networking have been presented elsewhere \cite{dang2020should}.


Although these applications and use cases are propelling us into the future, more than half of the people on Earth will not have access to this applications. That is what makes worldwide connectivity a major concern, especially since 3.9 billion people remained unconnected at the end of 2018 \cite{sharma2018itu}.
Indeed, a large proportion of people around the word do not have access to an internet connection, especially in rural, sparse, and poor areas. Services that are lacking in these areas, but can be further facilitated by internet connection, include eHealth, eEducation, farming, eCommernce, and eGovernment. 6G has the potential not only to hyper-connect the already connected, but to bridge the digital divide by connecting the unconnected \cite{yaacoub2020key,6GSummita}. 

In order to facilitate worldwide connectivity, 6G will rely on the following trifecta: terrestrial communications, satellite communications, and airborne communications. Terrestrial communications and tower masts are expensive for telecommunication companies in poor or rural areas. Plus, it takes time to construct terrestrial communications. 
Also, terrestrial communications are only suitable for two-dimensional scenarios in which the height of the users is relatively negligible. For instance, terrestrial communications would be ill-suited to connect flying cars \cite{saeed2020wireless}.   

On the other hand, satellite communications, especially Low Earth orbit (LEO) satellites, have ubiquitous coverage and have lower channel loss compared to Geosynchronous Equatorial Orbit (GEO) satellites. However, the cost of launching a large-scale constellations of LEO satellites is extremely high. 
Additionally, satellites take time to be deployed, and their communications are subject to latency due to the distance between the satellite and the users, which cannot be overlooked for critical communications especially for autonomous vehicles and flying cars. At the time of writing this paper, there are two companies working on solutions to connect regular phones with satellites: Lynk Global \cite{lynk} and AST \& Science \cite{ast}. 
However, development is in the early stages, and these companies plan to offer global satellite communications via regular phone in a decade or two.

\begin{figure}[]
\centering
\includegraphics[scale=0.85]{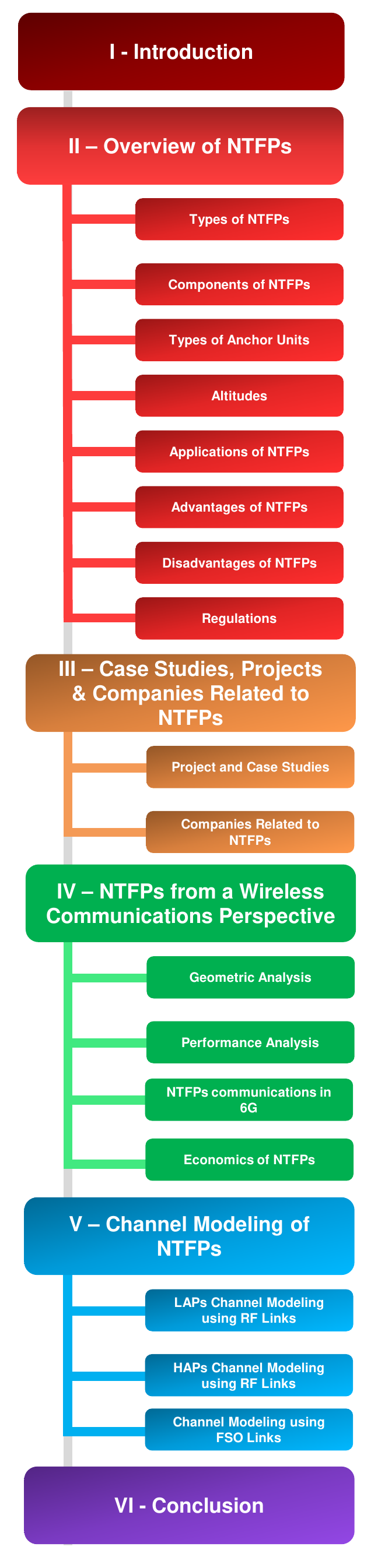}
\caption{Outline and organization of this survey paper.}
\label{Fig.Diag}
\end{figure}

The last type of aforementioned communications is airborne communications which are considered as key enablers in the 6G architecture \cite{han2021abstracted,akhtar2020shift}. Recently this type of communications has recently drawn considerable attention from researchers due to the intrinsic flying properties, which allow borad coverage. These airborne communications include different types of flying platforms. Some fly at a lower altitude, such as unmanned aerial vehicles (UAVs) \cite{saad2020wireless}. Others fly at a higher altitude, such as high-altitude platforms (HAPs) \cite{kurt2020vision}. These flying platforms overcome some of the limitations of satellite and terrestrial communications, such as high cost, delay, and slow deployment. However, these flying platforms also have some limitations in persistence, endurance, and backhaul connection. To deal with these limitations, another type of flying platform is currently used by government, military, and telecommunications companies: Networked Tethered Flying Platforms (NTFPs).


NTFPs, as their name suggests, are flying platforms tethered to a ground unit. The tether continuously supplies the flying platforms with data and power. They are cost-efficient and inexpensive to operate compared to free-flying platforms. Also, from a wireless communications perspective, they are cheaper than other communication infrastructure, such as tower masts and satellites. Another aspects that makes NTFPs an attractive solution compared to free-flying platforms are their endurance and persistence which are crucial to telecommunication and surveillance missions. They are also quick and relatively easy to deploy. But, the most relevant properties of NTFPs are their backhaul capacity and constant power supply. Applications of NTFPs beyond communications include energy harvesting, entertainment, science, research, public safety, disaster relief, government, and defense. 

Although these platforms have great potential as key enablers for 6G, a comprehensive and unifying documentation on this subject is lacking. We intend, through this paper, to provide a comprehensive overview about this solution for readers interested in this solution irrespective of their backgrounds. We also provide a comprehensive analysis on NTFPs from a wireless communications perspectives.
To the best of our knowledge, there are no papers surveying NTFPs in the literature, although there are some papers dealing with tethered aerostats from a design and manufacturing perspective, such as \cite{mahmood2020tethered}, no prior works have considered a comprehensive overview of NTFPs, and we hope that our survey fills this gap.

    \begin{figure}[t!]
    \centering
    \begin{subfigure}[b]{0.2\textwidth}
        \centering
        \includegraphics[scale=0.45]{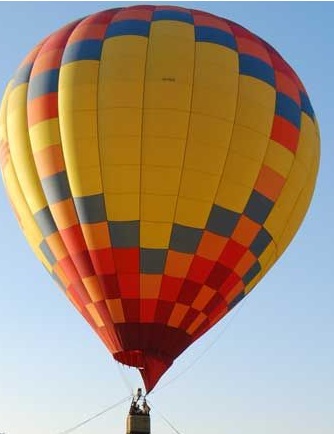}
        \caption{}
       \label{Fig.balloon.a}  
    \end{subfigure}%
    ~ 
    \begin{subfigure}[b]{0.2\textwidth}
        \centering
        \includegraphics[height=5cm,width=3.8cm]{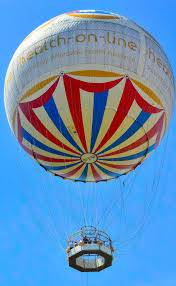}
        \caption{}
       \label{Fig.balloon.b}  
    \end{subfigure}
    \caption{Different shapes of balloons.}
   \label{Fig.balloon}  
\end{figure}

The organization of this paper is depicted in Fig.\ref{Fig.Diag}, and structured as follows:
\begin{itemize}
    \item \textbf{Section \ref{section_2}}: we provide a general and broad overview of NTFPs for readers interested in these type of platforms irrespective of their background and their field. We review the existing types of NTFPs, their main components, what are their different anchor points, and their altitudes.
Also, we highlight their various and divers applications from different fields as well as their main advantages and disadvantages. Finally, we present the current regulations regarding NTFPs.

    \item \textbf{Section \ref{section_3}}: we show the numerous and divers use cases of NTFPs from real-life scenarios. We aim through this section to highlight the various applications from different field. Also, we list the major companies that manufacture that manufacture and sell NTFPs.
    
    \item  \textbf{Section \ref{section_4}}: we address NTFPs from a wireless communications perspective. 
First, we carry out a geometrical modeling between a given NTFP and the Earth.
Second, we present the works that analyze the performance NTFPs in wireless communications. Then, we show how NTFPs are the key enablers to interconnect heterogeneous networks and communications envisioned in 6G. For the sake of completeness, we added a section that deals with the economic aspect of NTFPs in a wireless communications context, with a capital expenditure (CAPEX) and  operating expenditure (OPEX) analysis to emphasize their their cost-efficiency.

    \item  \textbf{Section \ref{section_5}}:  We provided a comprehensive channel modeling for NTFPs. Although this section is titled channel modeling for NTFPs, the model presented are valid for all type of platforms whether they are tethered or untethered (free-flying). 
We split the models according to the altitude of the platforms, that is, low-altitude platforms (LAPs) and HAPs, and the type of link used, that is, radio frequency (RF) and free space optics (FSO).

    \item  \textbf{Section \ref{section_conclusion}}: we conclude the survey paper with a brief summary.
\end{itemize}


\begin{figure*}
        \centering
        \begin{subfigure}[b]{0.475\textwidth}
            \centering
            \includegraphics[height=5cm,width=7cm]{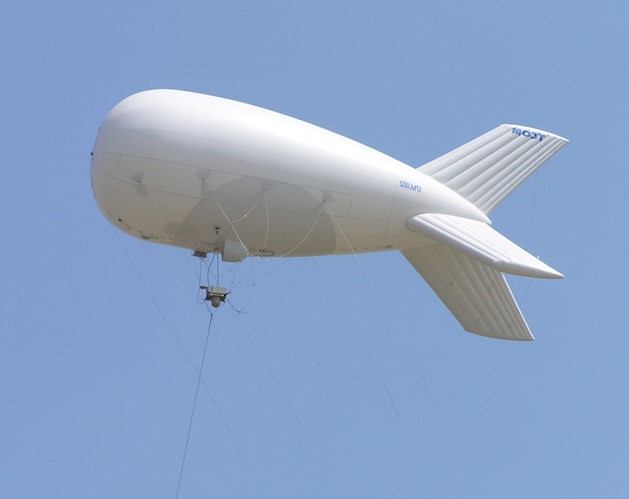}
            \caption[]%
            {Tactical TCOM blimp (17M).}      
            \label{Fig.blimp.17}
        \end{subfigure}
        \hfill
        \begin{subfigure}[b]{0.475\textwidth}  
            \centering 
            \includegraphics[height=5cm,width=7cm]{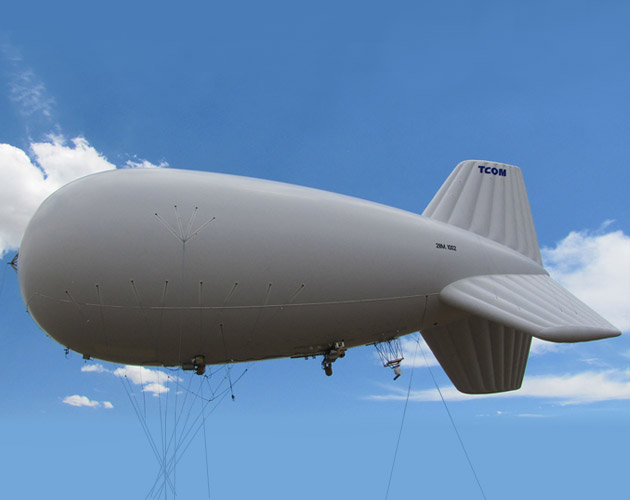}
            \caption[]%
            {Operational TCOM blimp (28M).}    
            \label{Fig.blimp.28}
        \end{subfigure}
        \vskip\baselineskip
        \begin{subfigure}[b]{0.475\textwidth}   
            \centering 
            \includegraphics[height=5cm,width=7cm]{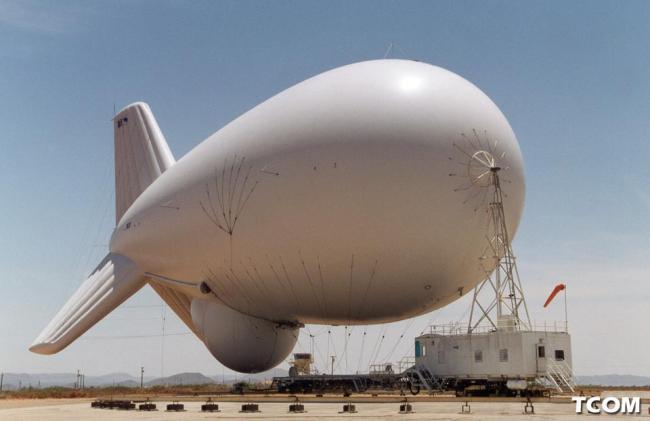}
            \caption[]%
            {Strategic TCOM blimp (71M).}    
            \label{Fig.blimp.71}
        \end{subfigure}
        \quad
        \begin{subfigure}[b]{0.475\textwidth}   
            \centering 
            \includegraphics[height=5cm,width=7cm]{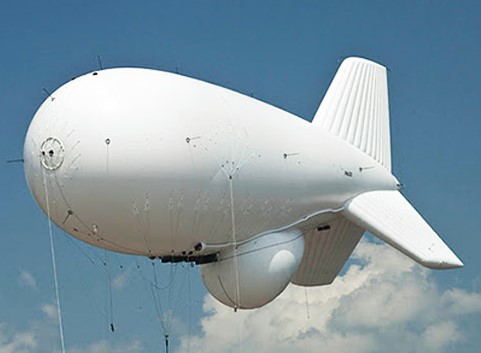}
            \caption[]%
            {Strategic TCOM blimp (74M).}    
            \label{Fig.blimp.74}
        \end{subfigure}
        \caption {Different types of TCOM blimps \cite{TCOM}.} 
        \label{Fig.blimp}
    \end{figure*}

\section{Overview of NTFPs}
\label{section_2}

\subsection{Types of NTFPs}
Airborne platforms are vehicles that can fly in the air by opposing the force of gravity either using a static lift or a dynamic lift. For instance, balloons and blimps use static lift and they belong to the lighter-than-air (LTA) category, whereas airplanes and UAVs use dynamic lift and they belong to the heavier-than-air (HTA) category. There is also a hybrid category of platforms that use both static lift and dynamic lift. In this survey, we are only interested in NTFPs; hence, free-flying platforms are out the scope of this paper.

\subsubsection{LTA NTFPs}
The LTA platforms that use static lift or aerostatic lift are called aerostats. They are filled with a low-density LTA gas such as helium. The difference between the density of the air outside the envelope of the aerostat and the density of the LTA gas produces buoyancy according to Archimedes’ principle. The most popular LTA NTFPs are balloons and blimps.
\paragraph{Balloons} Spherical balloons have been the most commonly used for NTFPs, or more precisely, tethered aerostats (Fig.\ref{Fig.balloon}). They are easy to design and manufacture at a lower cost than the other tethered aerostats. They are also easy to deploy. The maximum altitude a tethered balloon can reach is around 600 m to 700 m, and the maximum payload they can carry is around 50 kg. However, they are not designed to sustain high speed wind since their shape is not designed to cope with it. Tethered balloons can sustain wind speeds around 20 km/h to 40 km/h.

\begin{figure}[]
\centering
\includegraphics[scale=0.2]{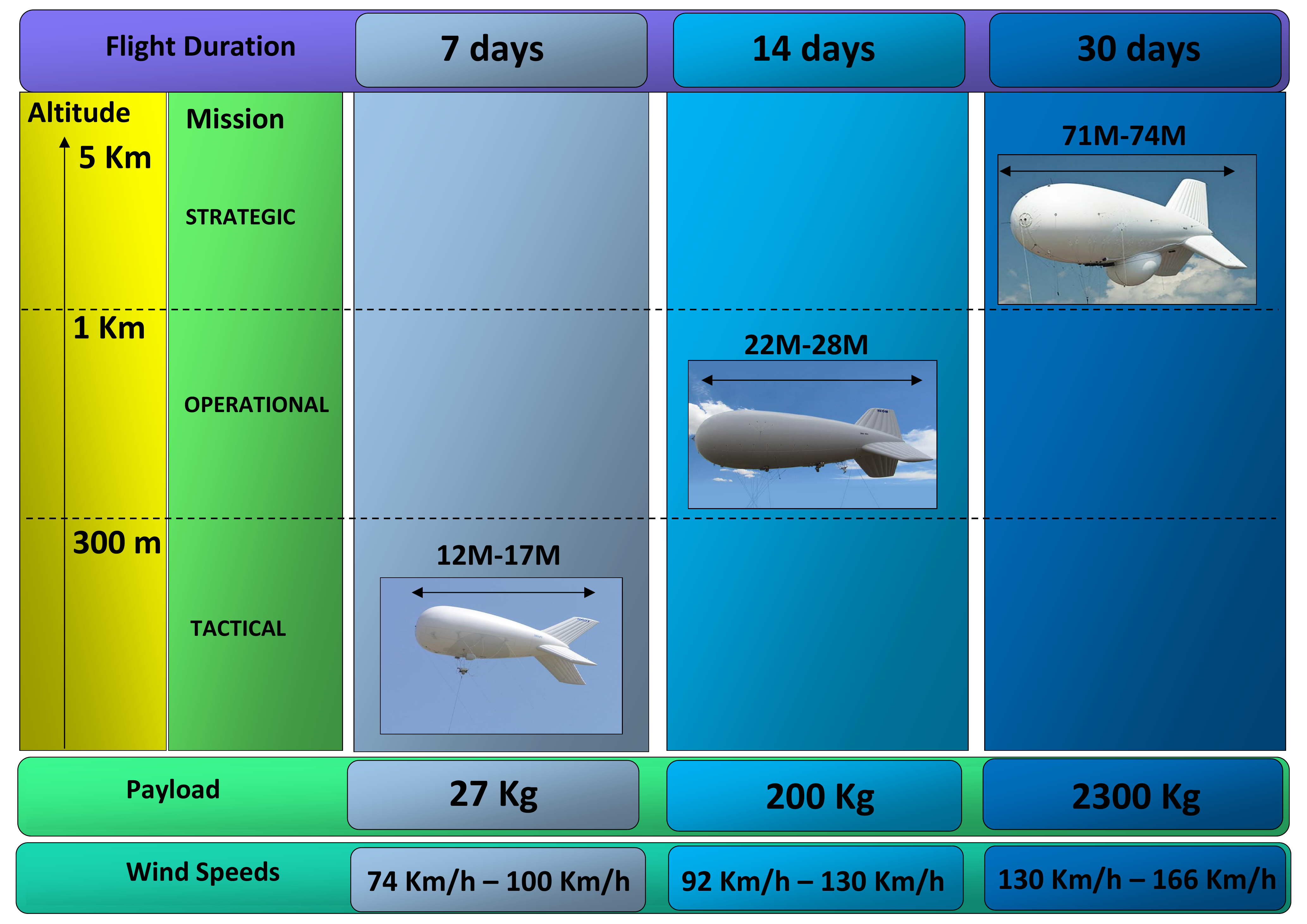}
\caption{TCOM blimps and their type of missions.}
\label{Fig.blimp.type}
\end{figure}

\paragraph{Blimps} Blimps, also known as streamlined aerostats, are high performance platforms that can sustain high speed wind, carry heavy payloads, and stay aloft at high altitudes (Fig.\ref{Fig.blimp}). 
There are several categories of blimps, and they differ in size, altitude, and maximum payload \cite{TCOM}. For instance, TCOM categorizes their blimps into three classes, tactical class, operational class, and strategic class, as depicted in Fig.\ref{Fig.blimp.type}:
\begin{itemize}
    \item Tactical Class: Tactical class blimps are compact and can be deployed rapidly. Their envelope size ranges from 12 m and 17 m. They are suitable for surveillance missions with tactical needs. They have been used in Iraq and Afghanistan by the U.S. army, and for border surveillance between the U.S. and Mexico. They can carry 27 kg of payload and reach an altitude of 300 m. Also, they can stay aloft for seven days and sustain wind speeds up to 100 km/h (see Fig.\ref{Fig.blimp.17}).
    \item Operational Class: Operational class blimps have a medium-sized envelope that ranges between 22 m and 28 m. They combine portability and flexibility for quick deployment and retrieval. They can carry a payload up to 200 kg and can reach an altitude of 1 km. This class of blimps is suitable for surveillance and monitoring operations where land-based surveillance is infeasible. Also, they are suitable for maritime surveillance and border surveillance. Operational class blimps can stay aloft for two weeks and can sustain wind speeds up to 130 km/h (see Fig.\ref{Fig.blimp.28}).
    \item Strategic Class: Strategic class blimps are arguably the largest NTFPs on the market. Their envelope size ranges from 71 m to 74 m and they can carry a large payload of 2300 kg. They are ideal for long surveillance and monitoring missions since they can stay in the air for 30 days and sustain a maximum wind speed of  166 km/h. They can also reach an altitude of 4,6 km allowing them to cover a large area. They have been used to detect low-flying aircraft or cruise missiles (see Fig.\ref{Fig.blimp.71} and Fig.\ref{Fig.blimp.74}).
\end{itemize}

\begin{figure}[]
\centering
\includegraphics[scale=0.25]{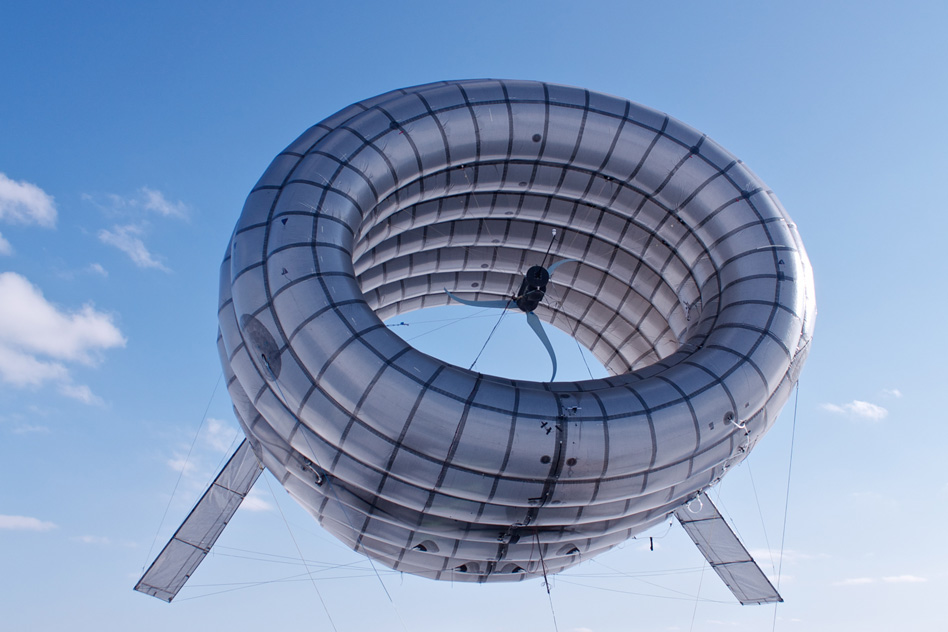}
\caption{Altaeros BATs \cite{Altaeroswebsite}.}
\label{Fig.BAT}
\end{figure}

\paragraph{Buoyant Airborne Turbines (BATs)}
BATs are wind turbines manufactured by Altaeros \cite{Altaeroswebsite}.
They can reach an attitude of 600 m, where wind speeds are higher than on the ground. They can harvest twice as much energy compared to land-based tower turbines \cite{kalabic2013reference,kehs2017online,samson2015adaptive,saleem2018aerodynamic}.
Their envelope is filled with helium, and they have a tether that holds them in place while they transmit the harvested power to a ground station (see Fig.\ref{Fig.BAT}).

\subsubsection{HTA NTFPs}
\begin{figure}[t!]
\centering
\includegraphics[scale=0.2]{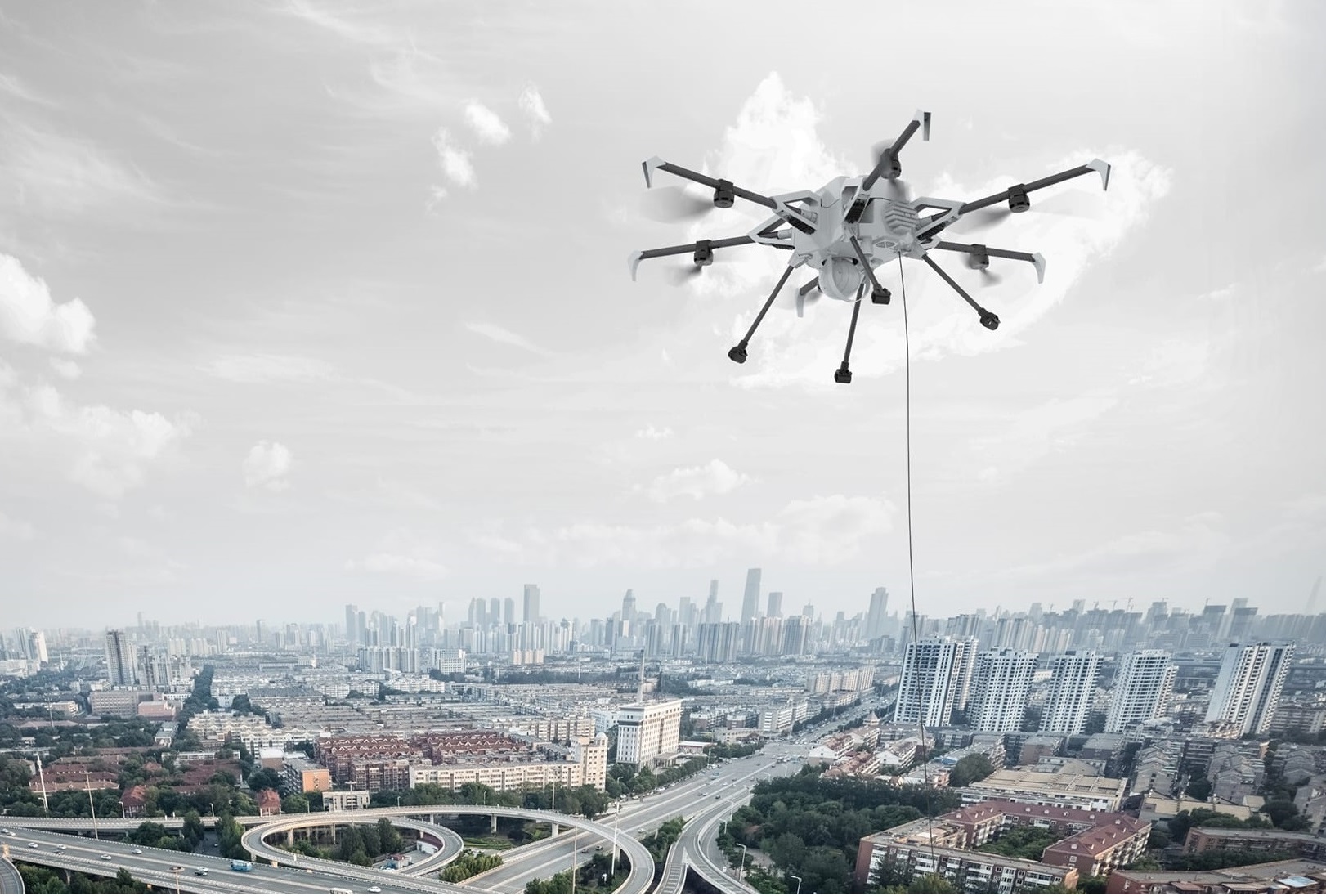}
\caption{tUAVs by Elistair \cite{Elistairwebsite}.}
\label{Fig.UAV}
\end{figure}

The HTA platforms that use dynamic lift or aerodynamic lift are called aerodynes. Their lift is produced by the relative motion between the HTA platform and the air, which pushes the platform upwards by Bernoulli’s principle. The most popular HTA NTFPs are tethered UAVs (tUAVs) and airborne turbine kites.
\paragraph{tUAVs}
tUAVs are UAVs with a physical link called a tether that supplies them with power and data. They can reach an altitude of 200 m and carry a payload up to 15 kg (Fig.\ref{Fig.UAV}). 
They usually have a battery pack in case the tether is damaged or if there is a cut in power.
Since tUAVs have a constant power supply, they can, in theory, stay in the air for an indefinite period of time. However, the main limiting factor is the motor of the UAV, which starts to overheat after two to four days aloft.
 
\begin{figure*}[t!]
    \centering
    \begin{subfigure}[b]{0.5\textwidth}
        \centering
        \includegraphics[scale=0.31]{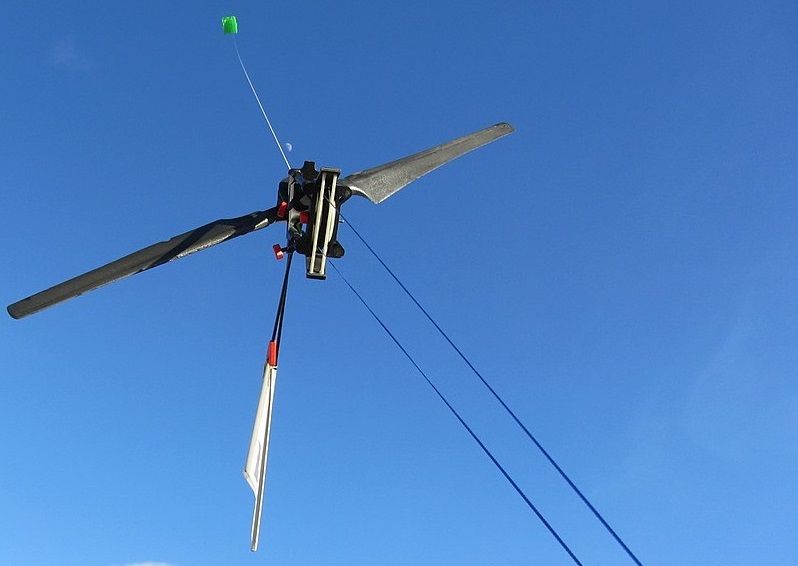} 
        \caption{Kitewinder \cite{Kitewinderwebsite}.}
       \label{Fig.turbine.a}  
    \end{subfigure}%
    ~ 
    \begin{subfigure}[b]{0.5\textwidth}
        \centering
        \includegraphics[scale=0.25]{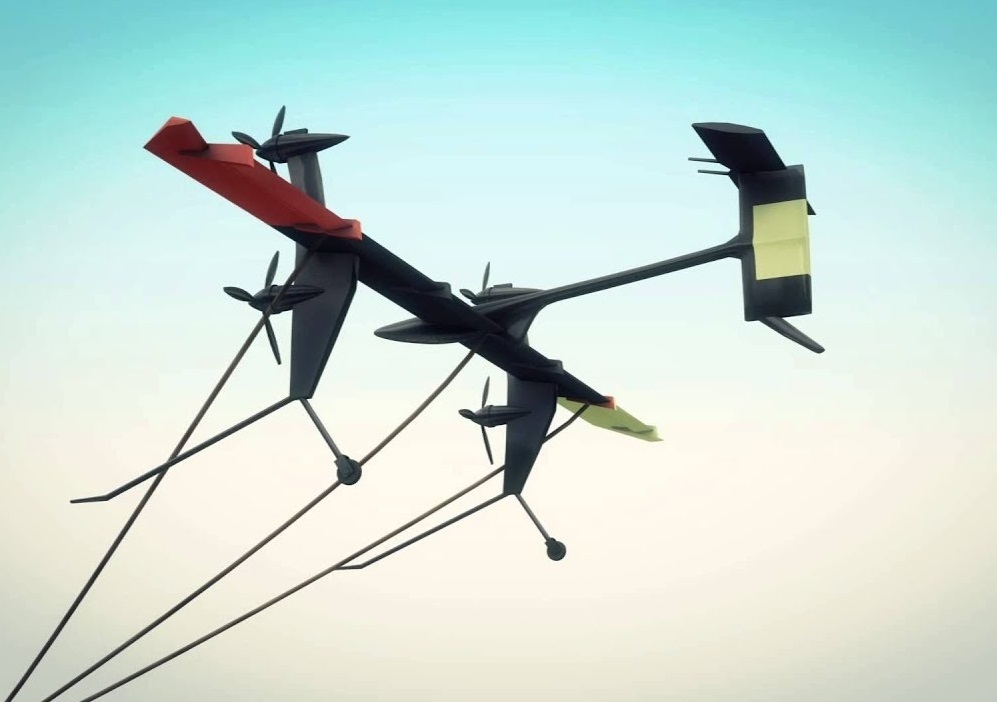}
        \caption{Makani power air turbine \cite{Makaniwebsite}.}
       \label{Fig.turbine.b}  
    \end{subfigure}
    \caption{Airborne wind turbines.}
   \label{Fig.turbine}  
\end{figure*}

\paragraph{Airborne Turbine Kite}
Airborne turbine kites are wind turbines used to harvest wind power in the air, since wind speed is higher at higher altitudes (see Fig.\ref{Fig.turbine}). Thus, they can harvest more energy than a tower, and they are also cheaper to construct. Their electrical generator can be land-based (on the ground) or airborne (in the air). The tether transmits the harvested energy to the ground. They can be maintained aloft at lower or higher altitudes up to 4600 m \cite{williams2008optimal,lansdorp2006laddermill,loyd1980crosswind}.

\subsubsection{Hybrid NTFPs}
As mentioned before, hybrid platforms use both static lift and dynamic lift. The static lift is produced by buoyancy (Archimedes’ principle), and the dynamic lift is produced by the relative motion between the aircraft and the air pushing the aircraft upwards (Bernoulli’s principle). The hybrid NTFP used most often is the Helikite. Also, hybrid airships can be tethered to the ground.

\begin{figure}[]
\centering
\includegraphics[scale=0.9]{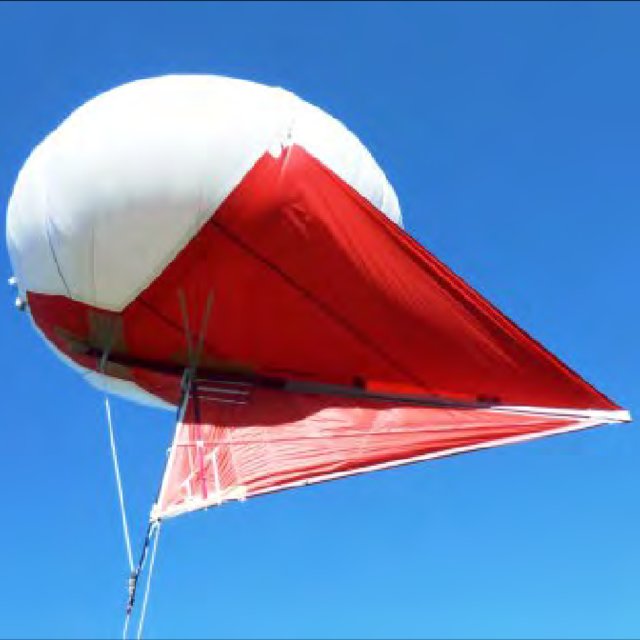}  
\caption{Desert Star Helikites by Allsopp \cite{Allsoppwebsite}.}
\label{Fig.helikite}
\end{figure}

\paragraph{Helikites}
Helikites are hybrid aerostats that benefit from both static lift and dynamic lift. The term Helikite is a portmanteau of helium and kite. Helikites were designed and patented by the company Sandy Allsopp in 1993 (see Fig.\ref{Fig.helikite}). A Helikite is composed of an oblate-spheroid balloon filled with helium to provide static lift, and a kite structure to provide dynamic lift. Combining these two lifts reduces the amount of helium required compared to other similarly sized aerostats, and Helikite can fly at much higher altitudes than other aerostats of the same size. Helikites also offer several advantages compared to other NTFPs: 1) Their compact design allows them to be deployed by fewer personnel. 2) They are not brought down by high speed winds since winds force them to go upward. 3) They are smaller, thus, they have few problems with helium leakage and use less helium because they benefit from dynamic lift. 
Helikites can stay in the air for two weeks at an attitude of 1.5 km carrying a payload of 23 kg. Allsopp claims that their Desert Star Helikites can carry a payload of 220 kg at an altitude of 3.4 Km, but qualify that claim as the estimated performance \cite{Allsoppwebsite}.

\begin{figure}[]
\centering
\includegraphics[scale=0.2]{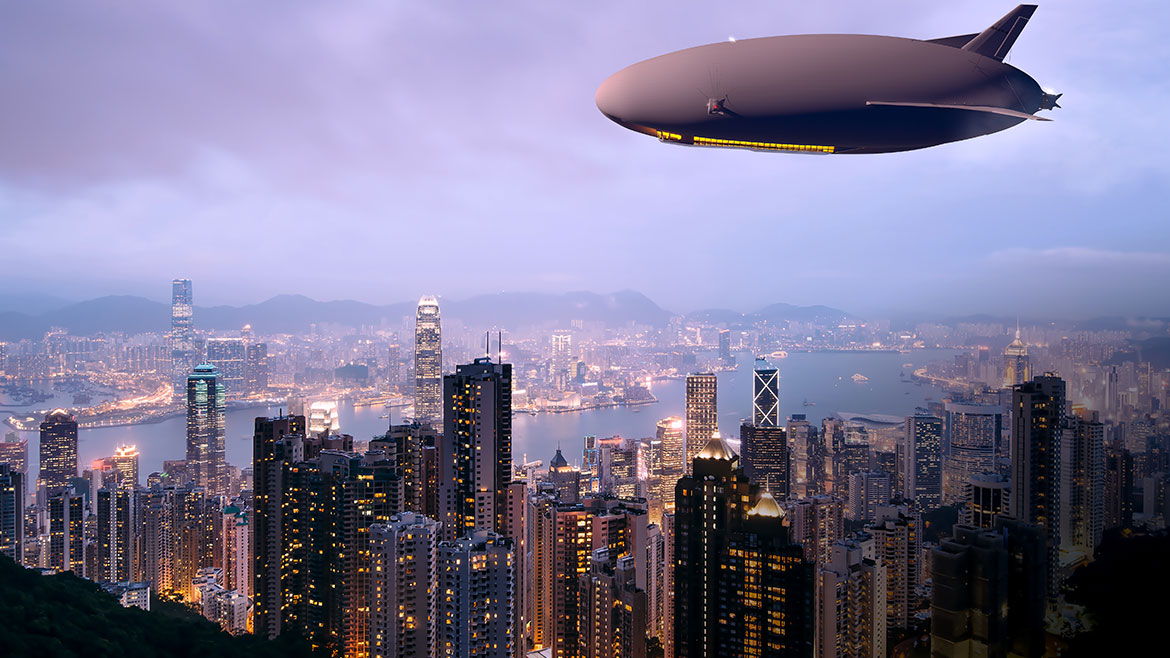}
\caption{Airlander hybrid airship by Hybrid Air Vehicles \cite{HAVwebsite}.}
\label{Fig.airborne}
\end{figure}

\paragraph{Hybrid airships} 
Hybrid airships are hybrid aircraft, which means 60\% of their lift comes from buoyant lift (aerostatic lift), and the remaining 40\% comes from aerodynamic lift (see Fig.\ref{Fig.airborne}). Hybrid airships do not need airports since they can take off and land anywhere with a large open and flat field. They can reach an altitude of 6000 m and carry a payload of 60,000 kg \cite{HAVwebsite,Lockheedwebsite}. Although the main usage of hybrid airships is to transport passengers and deliver heavy cargo, they can still be used for other purposes. They can also be tethered to the ground, then removed and deployed elsewhere if needed. They are expected to in service by 2024 \cite{zhang2010flight,meng2019aerodynamic}.

 \begin{figure*}[th!]
\centering
\includegraphics[scale=0.35]{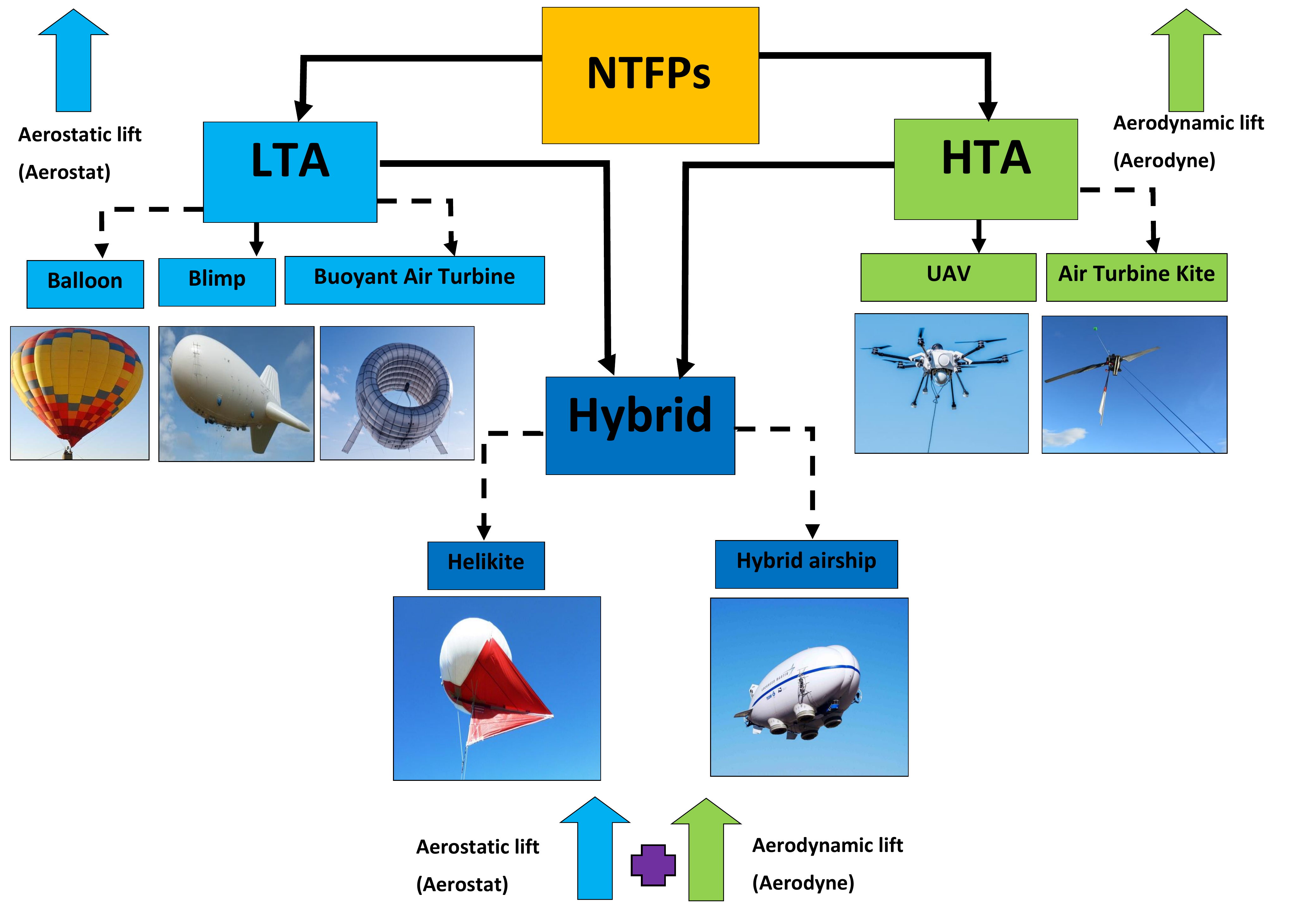}
\caption{Different categories and types of NTFPs.}
\label{}
\end{figure*}

\subsection{Components of NTFPs}
\begin{figure*}[]
\centering
\includegraphics[scale=0.65]{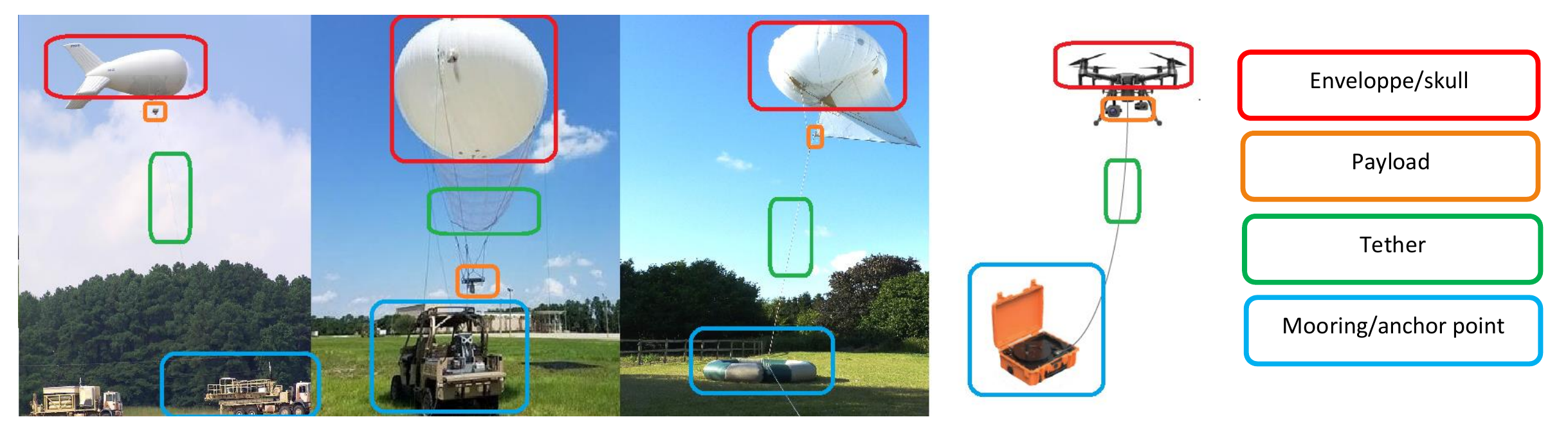}
\caption{Different NTFPs and their components (from left to right : TCOM blimp \cite{TCOM}, Drone Aviation Corp. balloon \cite{Drone_Aviation}, Helikite \cite{Allsoppwebsite}, and Elistair tethered drone \cite{Elistairwebsite}.}
\label{Fig.ssystem1}
\end{figure*}

\begin{figure*}[]
\centering
\includegraphics[scale=0.4]{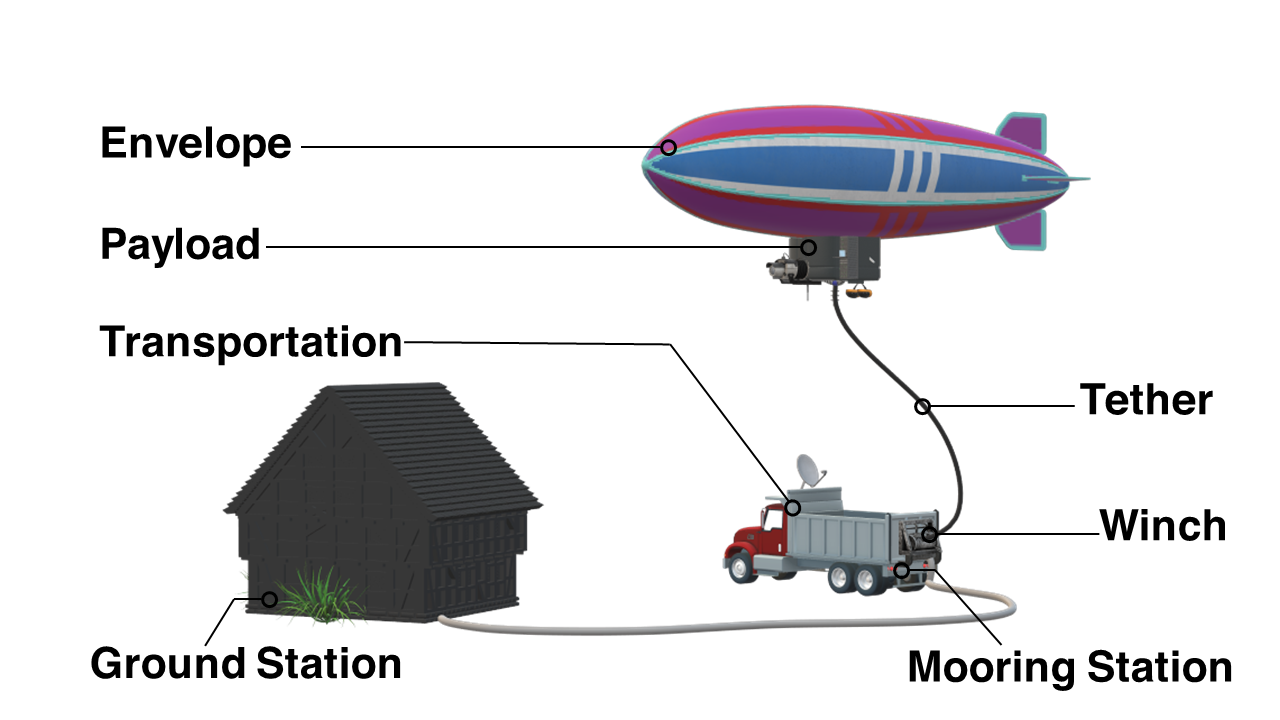}
\caption{Components of NTFPs.}
\label{Fig.ssystem2}
\end{figure*}

We detail in this section, the different components of NTFPs, which are also summarized in Fig.\ref{Fig.ssystem1} and Fig.\ref{Fig.ssystem2}.

\subsubsection{Envelope/Shell}
\begin{figure*}
        \centering
        \begin{subfigure}[b]{0.475\textwidth}
            \centering
            \includegraphics[height=5cm,width=7cm]{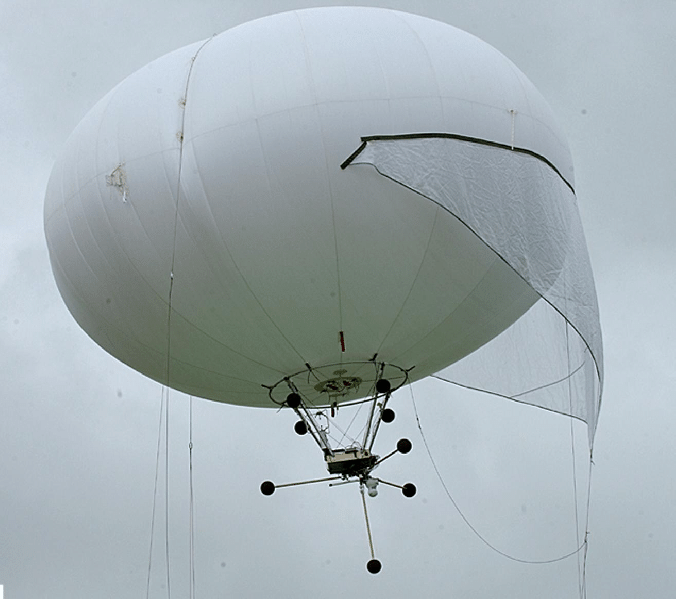} 
            \caption[]%
            {Balloon envelope \cite{Drone_Aviation}.}    
            \label{Fig.envelopp.1}
        \end{subfigure}
        \hfill
        \begin{subfigure}[b]{0.475\textwidth}  
            \centering 
            \includegraphics[height=5cm,width=7cm]{17M.jpg} 
            \caption[]%
            {Blimp envelope \cite{TCOM}.}    
            \label{Fig.envelopp.2}
        \end{subfigure}
        \vskip\baselineskip
        \begin{subfigure}[b]{0.475\textwidth}   
            \centering 
            \includegraphics[height=5cm,width=7cm]{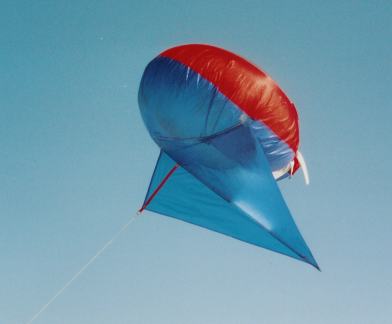} 
            \caption[]%
            {Helikite envelope \cite{Allsoppwebsite}.}    
            \label{Fig.envelopp.3}
        \end{subfigure}
        \quad
        \begin{subfigure}[b]{0.475\textwidth}   
            \centering 
            \includegraphics[height=5cm,width=7cm]{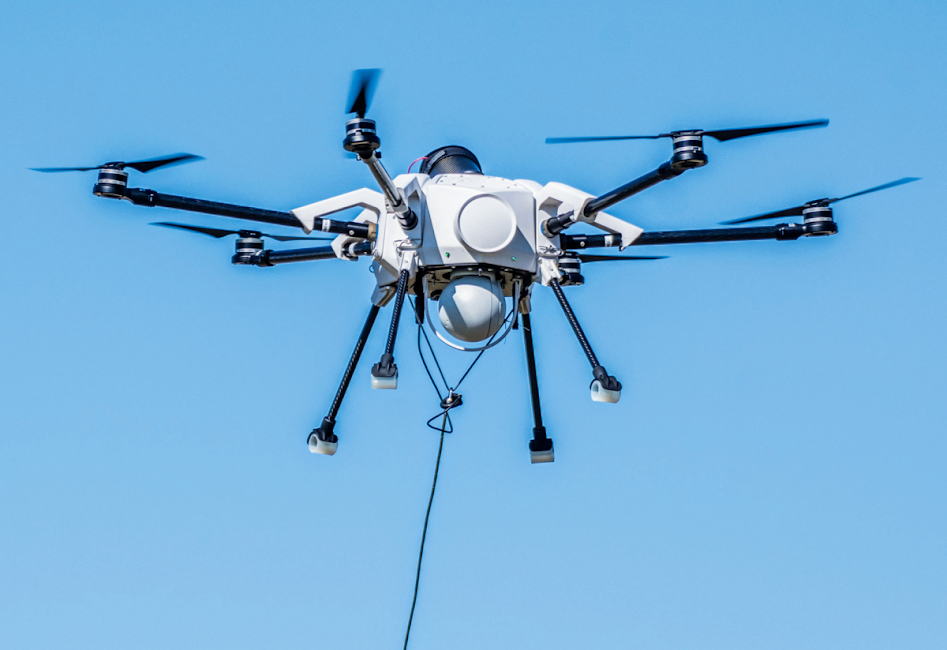}
            \caption[]%
            {UAV shell \cite{Elistairwebsite}.}    
            \label{Fig.envelopp.4}
        \end{subfigure}
  \caption{Illustration of various NTFP envelopes.}
  \label{Fig.envelopp} 
    \end{figure*}

The envelope of NTFPs contains gas, which allows the platforms to soar and stay aloft.
Some envelopes have spherical forms (balloons), as shown in Fig.\ref{Fig.envelopp.1}, other have a fish-shaped or streamlined form (blimps), in Fig.\ref{Fig.envelopp.2}. The lifts of those envelopes rely solely on buoyant gas.
Other envelopes have kites attached to them, which provide an aerodynamic lift to improve their performance in the presence of strong winds, such as Helikites, shown in Fig.\ref{Fig.envelopp.3}. Envelopes are usually made from a synthetic material, such as polyester, polyurethane, or polyvinyl.
Envelopes may contain materials such as laminates to prevent degradation from ultraviolet light exposure or  materials to prevent the envelope from abrasions.

tUAVs have shells or frames instead of envelopes, as shown in Fig.\ref{Fig.envelopp.4}. The shell defines the shape and look of the UAV and also contains the necessary components, such as the motor, blades, the protection cover, landing gears, etc.
The motors provide force and lift to the tUAVs. Generally, UAVs have between four and eight motors.  
An UAV with four motors is called a quadcopter, a UAV with six motors is called a hexacopter, and a UAV with eight motors is called an octacopter. The number of motors used depends on the type of mission. Also, landing gear are used for UAVs that require larger  ground clearance.

\subsubsection{Lifting Gas}
The envelope of the aerostat is filled with a lifting gas, also called a LTA gas, which has a lower density than the air (atmospheric gas); hence it creates buoyancy according to Bernoulli’s principle. Hydrogen and helium are the most common and lightest gases used for aerostats. However, each gas has pros and cons, as shown in Table \ref{Table.gas}.

The main advantages of hydrogen are that it is the lightest existing gas, and it can be easily produced. However, its main disadvantage is its high flammability. Some countries have prohibited its use, especially after the Hindenburg incidents \cite{web1,web2}. Helium, on the other hand, is the second lightest gas, and unlike hydrogen, is a nonflammable gas. However, helium is expensive, scarce, and a non-renewable resource. Although helium is the gas most commonly for NTFPs, its aforementioned disadvantages are leading
researchers and manufacturers to reconsider hydrogen usage and cope with its related safety issues. Also, some vendors are designing aerostat that use both hydrogen and helium.

\begin{table}[h!]
\begin{center}
    \caption{Comparisons between hydrogen and helium. }
    \label{Table.gas}
\begin{tabular}{ |p{1.6cm}||p{2.6cm}|p{2.9cm}| }
 \hline
Gas & Hydrogen & Helium   \\
 \hline
Advantages & 
- Lightest existing gas   \newline
 - Easily produced in \newline large quantities 
&  
- Second lightest gas  \newline
- Non-combustible
\\ \hline
Disadvantages  &  
- Highly flammable
  & 
- Expensive \newline
- Scarce \newline
- Non-renewable resource
\\  
 \hline
\end{tabular}
\end{center}
\end{table}

\subsubsection{Payload}
\begin{figure*}
        \centering
        \begin{subfigure}[b]{0.475\textwidth}
            \centering
            \includegraphics[height=5cm,width=7cm]{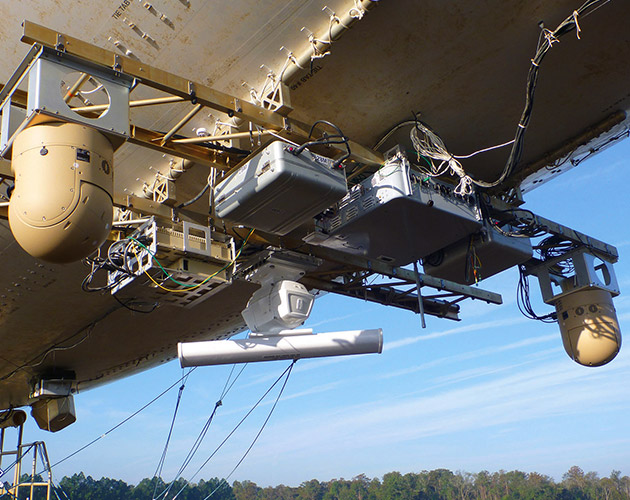}
            \caption[]%
            {Payload of 22M TCOM blimp \cite{TCOM}.}    
            \label{Fig.payload.1}
        \end{subfigure}
        \hfill
        \begin{subfigure}[b]{0.475\textwidth}  
            \centering 
            \includegraphics[height=5cm,width=7cm]{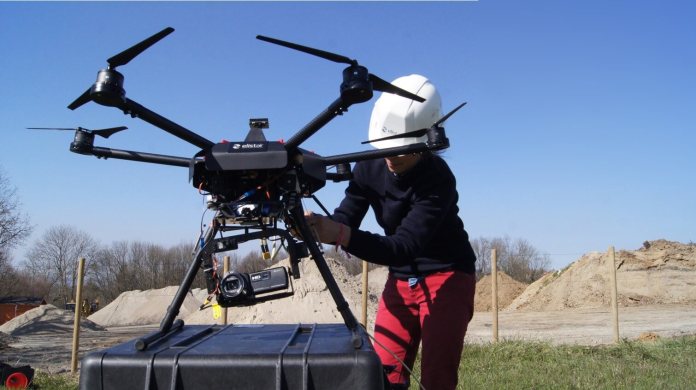}
            \caption[]%
            {Payload of Elistair tethered drone \cite{Elistairwebsite}.}    
            \label{Fig.payload.2}
        \end{subfigure}
        \vskip\baselineskip
        \begin{subfigure}[b]{0.475\textwidth}   
            \centering 
            \includegraphics[height=5cm,width=7cm]{5-balloons.png} 
            \caption[]%
            {Payload of tethered balloon \cite{reiff2009acoustic}.}    
            \label{Fig.payload.3}
        \end{subfigure}
        \quad
        \begin{subfigure}[b]{0.475\textwidth}   
            \centering 
            \includegraphics[height=5cm,width=7cm]{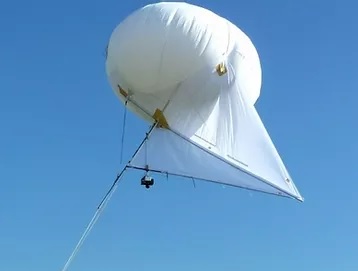}  
            \caption[]%
            {Payload of Allsopp Helikite \cite{Allsoppwebsite}.}    
            \label{Fig.payload.4}
        \end{subfigure}
        \caption {Payload of different NTFPs. } 
        \label{Fig.payload}
    \end{figure*}

The payload is the weight that the NTFP can carry while in the air (Fig.\ref{Fig.payload}). The payload differs from one NTFP to the next as shown in Table \ref{tab2}. To be more specific, we define the total capacity payload as the total weight the platform can lift, excluding its own weight and the weight of its tether, at the desired altitude. The total capacity payload is divided into
\begin{itemize}
    \item The supporting system payload, which includes all the equipment necessary to operate the platform, such as the power system, communication repeaters, backup batteries, lights, etc, and
    \item The operational payload, which includes equipment related to the mission, such as high definition (HD) camera, telescope, electronics, panchromatic imaging camera, electro-optical/infra-red sensors, acoustic detectors. The type of equipment varies from one mission to another.
\end{itemize}

\subsubsection{Tether}
\begin{figure}
 \centering
 \begin{tabular}{@{}cc}
   \includegraphics[height=4cm,width=8cm]{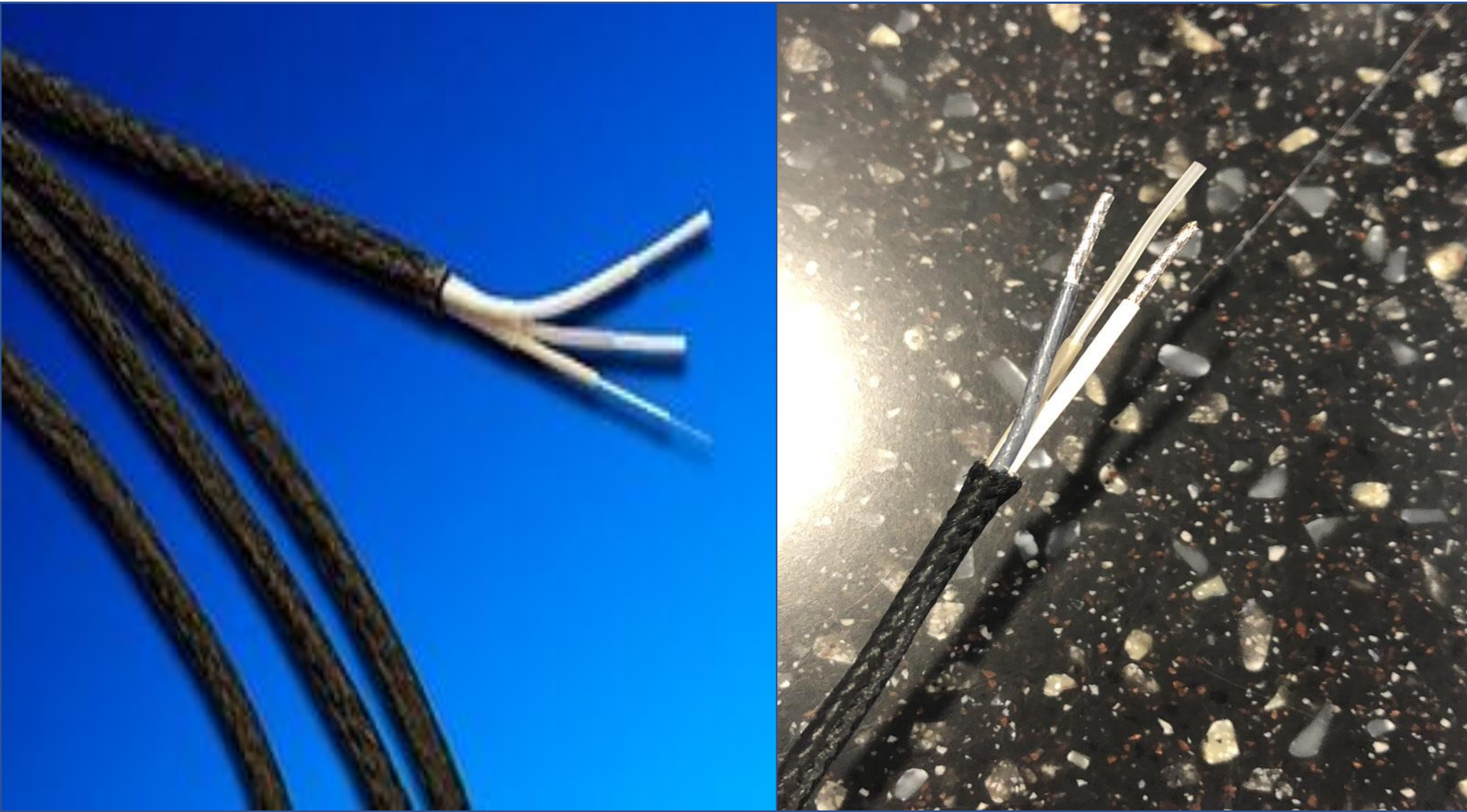} &
   \\
   (a) UAV tether cable \cite{Gore}.& \\
   \includegraphics[height=4cm,width=8cm]{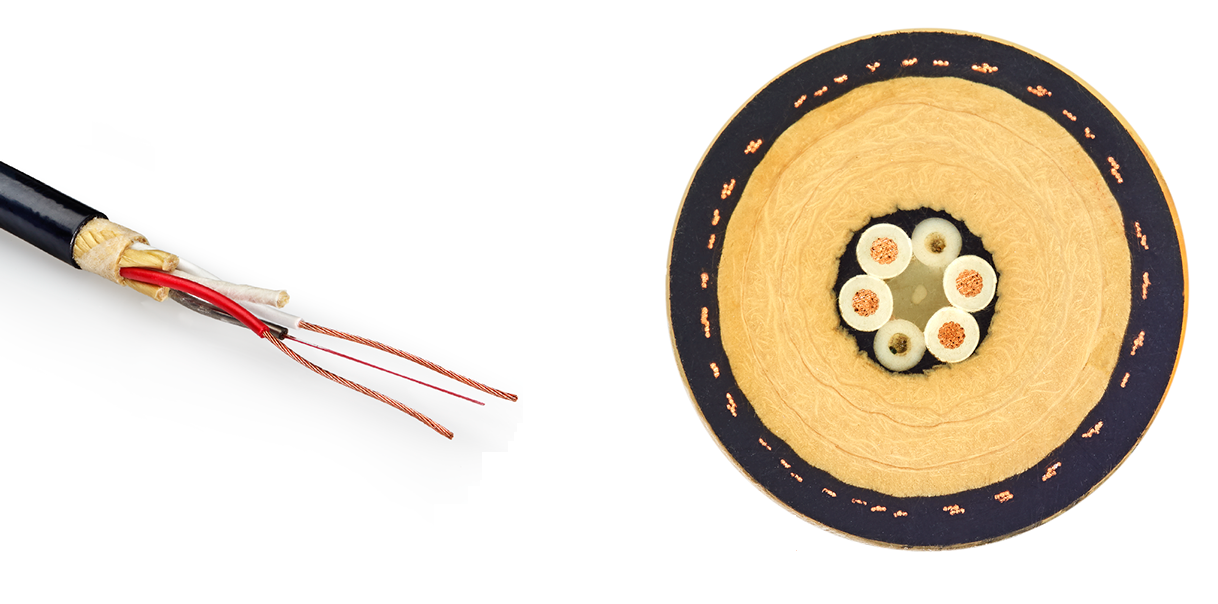}&
   \\
   (b) Aerostats tether cables \cite{Deregtcables}.&  \\
 \end{tabular}
 \caption{Aerostats and UAVs tether cables.}
 \label{Fig.tether}
\end{figure}

The tether is rolled up around a winch at one extremity, and is attached to the envelope/shell at the other. Fig.\ref{Fig.tether} shows pictures of a tether cable. Tethers are usually made of synthetic fibers and, depending on the type and size of the NTFP, the tether’s length, diameter, resistance and weight will differ. For a large platform, several tethers may be used. A tether has the following purposes: 
\begin{itemize}
    \item Maintain and stabilize the platform in the air to the ground;
    \item Provide power to the platform through a power line; and
    \item Provide data to the platform through optical fibers.
\end{itemize}

Also, the tether must be weather-proof to withstand extreme temperatures, humidity, rain, snow, lightning, and other weather conditions. 

\subsubsection{Mooring Stations and Anchor Units}

\paragraph{Mooring stations}

\begin{figure*}
        \centering
        \begin{subfigure}[b]{0.475\textwidth}
            \centering
            \includegraphics[height=5cm,width=7cm]{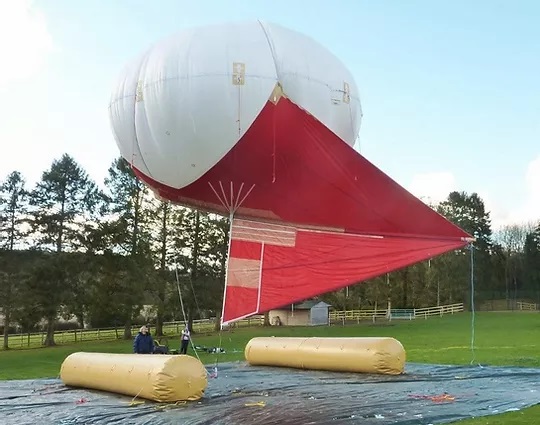}
            \caption[]%
            {Helikite mooring station (Helibase type 1) \cite{Allsoppwebsite1}.}    
            \label{Fig.moor.1}
        \end{subfigure}
        \hfill
        \begin{subfigure}[b]{0.475\textwidth}  
            \centering 
            \includegraphics[height=5cm,width=7cm]{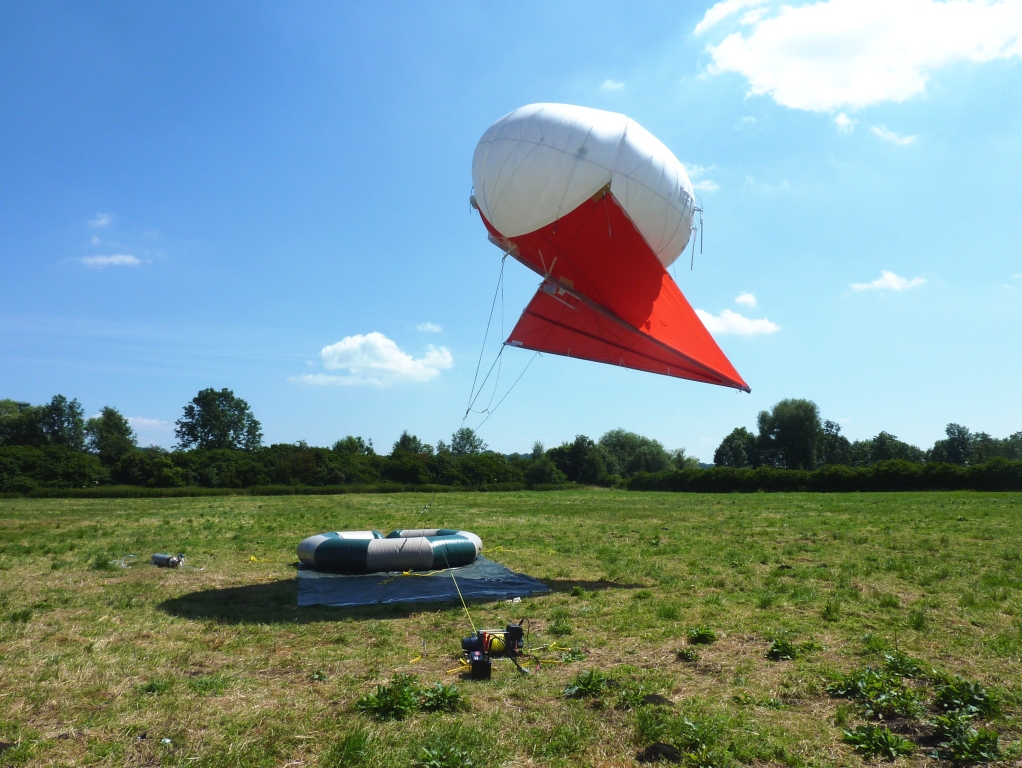}
            \caption[]%
            {Helikite mooring station (Helibase type 2) \cite{AllsoppH}.}    
            \label{Fig.moor.2}
        \end{subfigure}
        \vskip\baselineskip
        \begin{subfigure}[b]{0.475\textwidth}   
            \centering 
            \includegraphics[height=5cm,width=7cm]{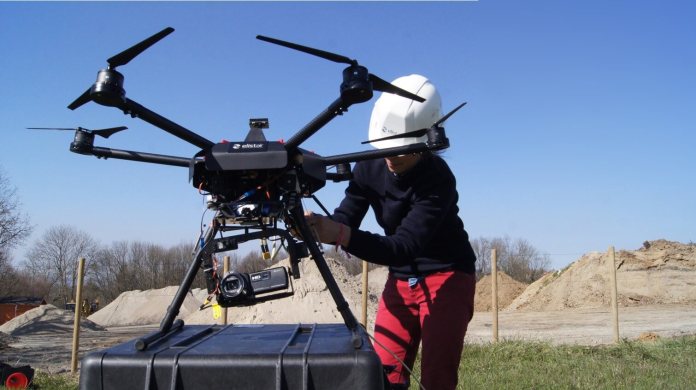}
            \caption[]%
            {tUAV mooring station \cite{Elistairwebsite}.}    
            \label{Fig.moor.3}
        \end{subfigure}
        \quad
        \begin{subfigure}[b]{0.475\textwidth}   
            \centering 
            \includegraphics[height=5cm,width=7cm]{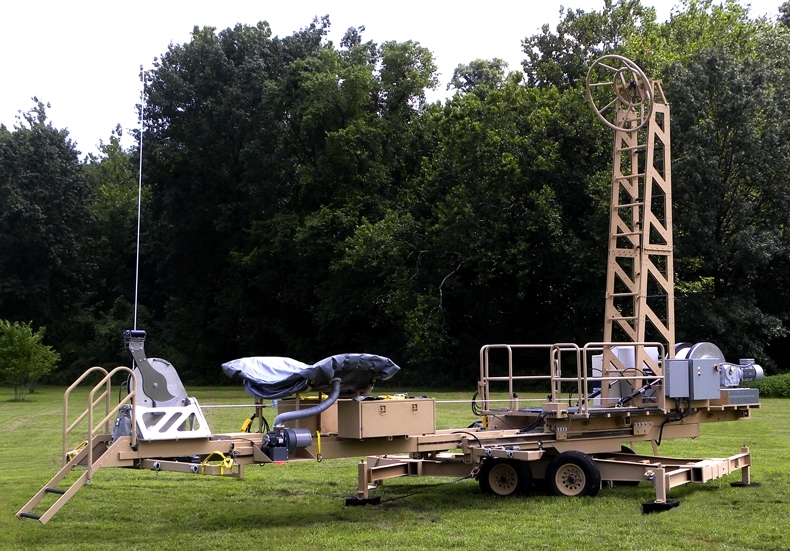}
            \caption[]%
            {Blimp mooring station \cite{Craftsmen}.}    
            \label{Fig.moor.4}
        \end{subfigure}
        \caption {Mooring stations for different types of NTFPs. } 
        \label{Fig.moor}
    \end{figure*}

\begin{figure*}[t!]
    \centering
    \begin{subfigure}[b]{0.4\textwidth}
        \centering
        \includegraphics[scale=0.3]{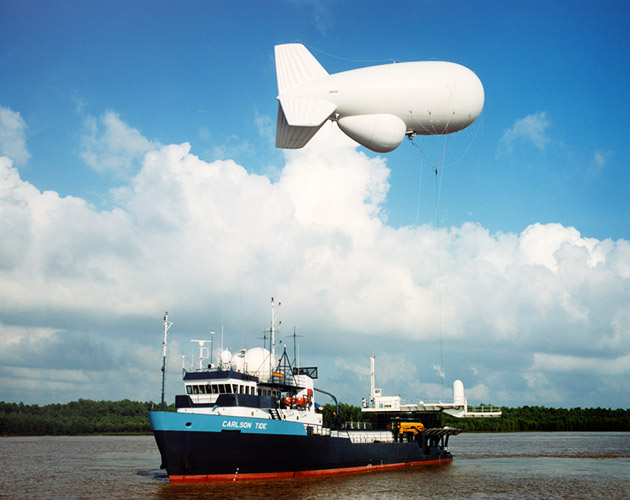}
        \caption{}
       \label{Fig.moor.a.1}  
    \end{subfigure}%
    ~ 
    \begin{subfigure}[b]{0.4\textwidth}
        \centering
        \includegraphics[scale=0.3]{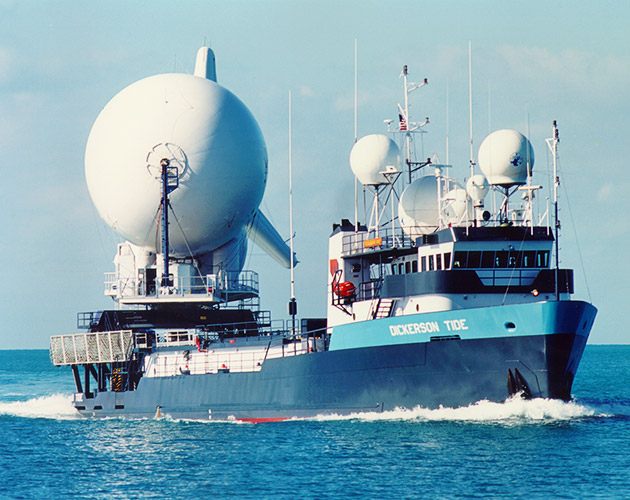}
        \caption{}
       \label{Fig.moor.a.2}  
    \end{subfigure}
    \caption{Maritime mooring station and anchor unit \cite{TCOM}. }
   \label{Fig.moor.a}  
\end{figure*}

The mooring station is the system that holds the envelope of the aerostat while it is inflated before the launching process, while it is deflated after the flight, and during maintenance. The moorings stations differ in size, form, and complexity, as depicted in Fig.\ref{Fig.moor}. For instance, large blimps require large and heavy mooring stations, while smaller platforms, such as Helikites, require lighter and simpler mooring stations. The mooring stations also depend on the environment in which the NTFPs will be used in. For instance, NTFPs can be deployed over the sea or ocean; therefore, they must have mooring stations designed for maritime applications, as depicted in Fig.\ref{Fig.moor.a}.

\paragraph{Anchor points}
The anchor point or anchor unit is the unit that the platform is anchored into, and it maintains the platform in place while aloft.
There are different types and sizes of anchor units, which we will present in more detail in section \ref{Anchor}. Mooring stations can also serve as anchor points.

\subsubsection{Winches}

\begin{figure}
 \centering
 \begin{tabular}{@{}cc}
   \includegraphics[height=2cm,width=8cm]{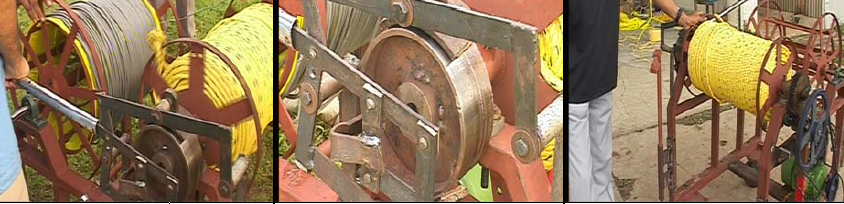} &
   \\ 
   (a) & \\
   \includegraphics[height=5cm,width=8cm]{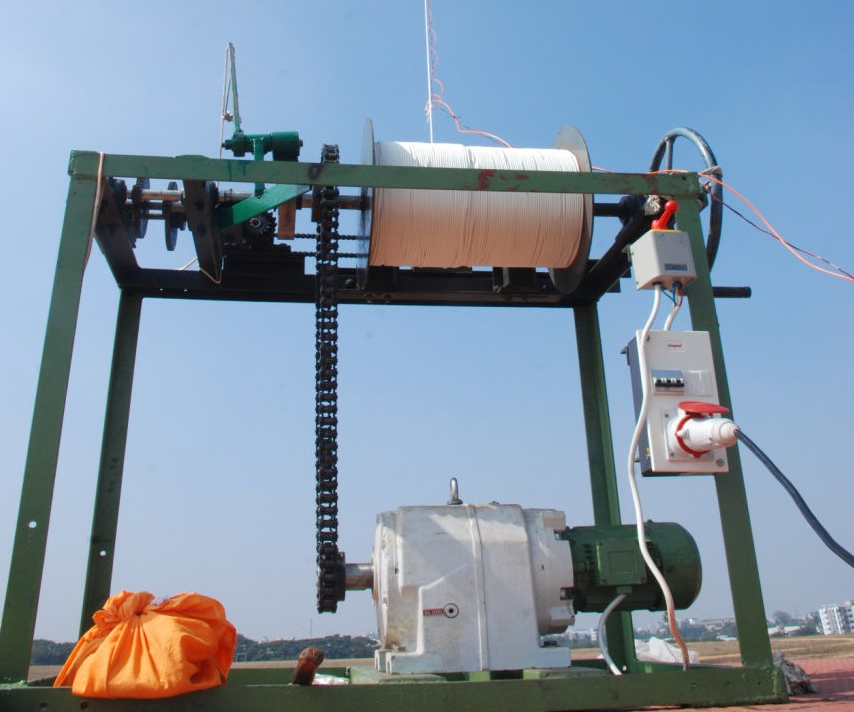}&
   \\
   (b) &  \\
    \includegraphics[height=4.5cm,width=8cm]{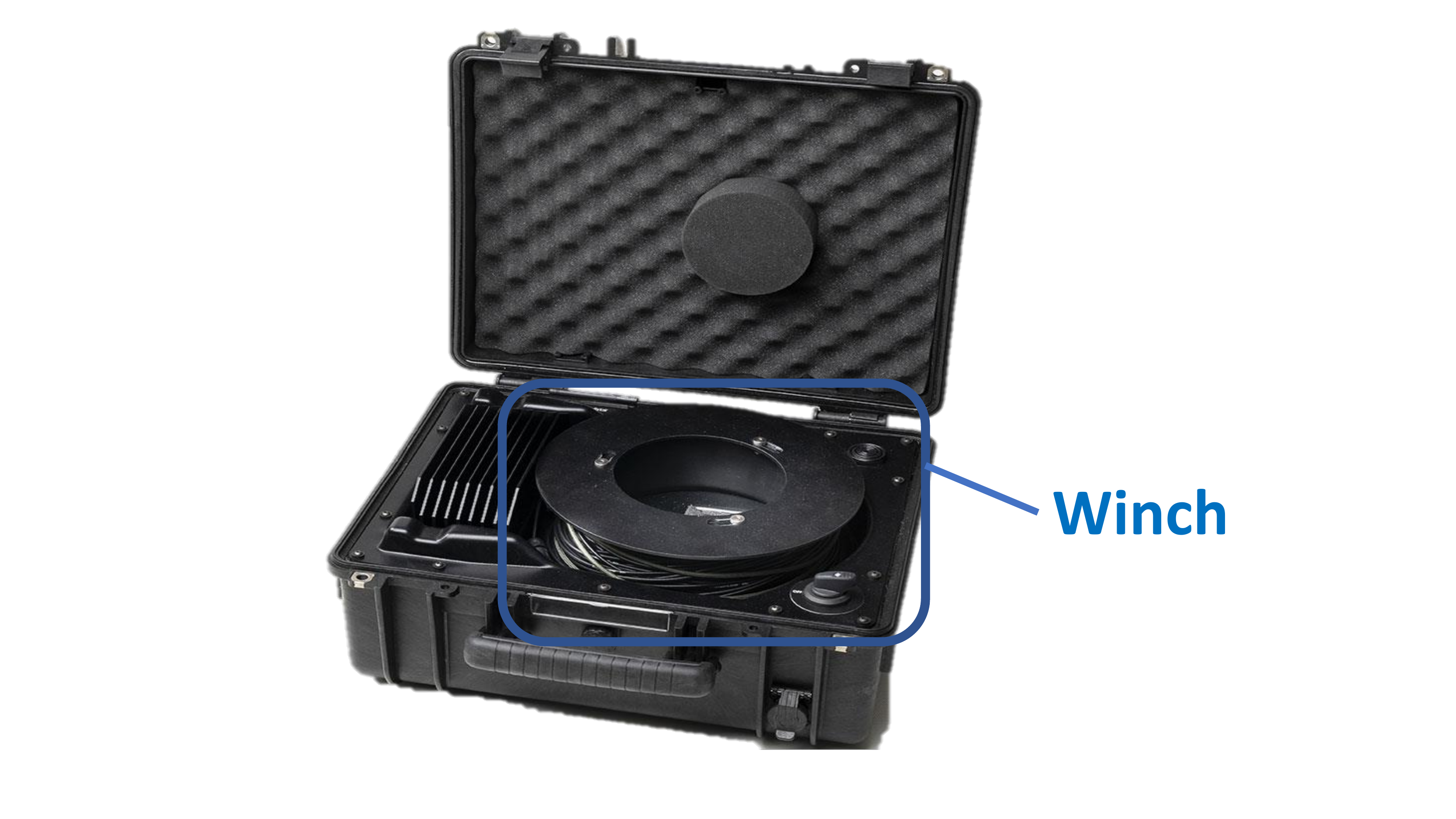}&
   \\
   (c) &  \\
 \end{tabular}
 \caption{Winches for a NTFP. Source: (a) \cite{gawande2007design}; (b) \cite{chopra2011new}; (c) \cite{Elistairwebsite}.}
 \label{Fig.winch}
\end{figure}

A winch is the device used to let out the tether during the launching process, adjust its tension while the platform is aloft, or pull it  back in during the recovery process. The tether is wound around a drum called the winch drum. The size and type of winch varies. Smaller NTFPs can be winched manually using a crank, whereas larger NTFPs require a power or motored winch. The winch can be attached to a mooring station or mounted on a trailer, such as a flat bed or a truck bed. Fig.\ref{Fig.winch} shows different types of winches. 

\subsubsection{Ground Control Station}
\begin{figure}[t!]
 \centering
 \begin{tabular}{@{}cc}
   \includegraphics[height=4cm,width=8cm]{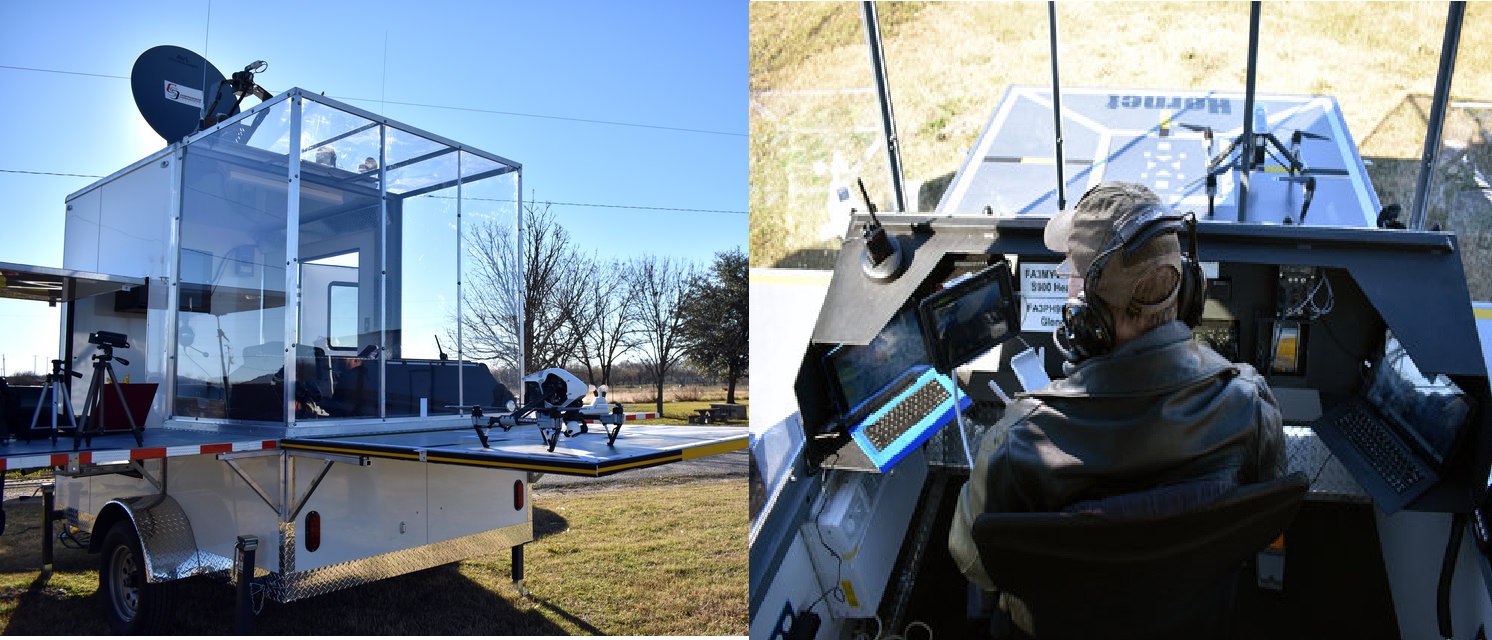} &
   \\
   (a) tUAV ground control station \cite{FortressUAV,Comprehensivecom}.& \\
   \includegraphics[height=4cm,width=8cm]{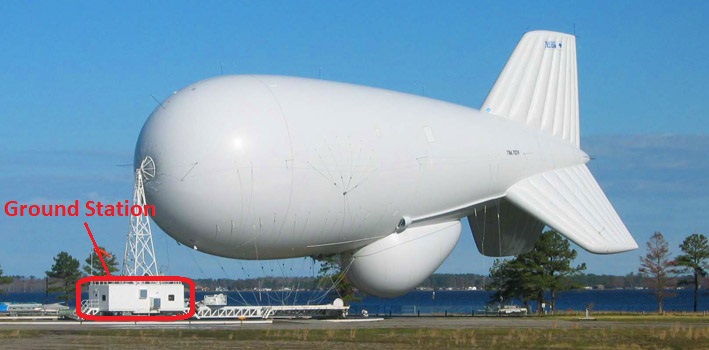}&
   \\
   (b) TCOM blimp ground control station \cite{TCOM}.&  \\
 \end{tabular}
 \caption{Ground control stations for a tUAV and blimp.}
 \label{Fig.GS}
\end{figure}

Ground control stations serve as the operation base for NTFPs, as shown in Fig.\ref{Fig.GS}. They can be used to
\begin{itemize}
    \item Control the altitude of the platforms;
    \item Monitor and control the platforms and the equipment they carry; and
    \item Store, and in some cases process, the data related to the mission, such as videos and images.
\end{itemize}
	
Depending on the mission type, a ground control station may be a building, a tent, a vehicle, a container, or any place that serves as a shelter.

\subsubsection{Transportation}
\begin{figure*}[t!]
    \centering
    \begin{subfigure}[b]{0.5\textwidth}
        \centering
        \includegraphics[scale=2.6]{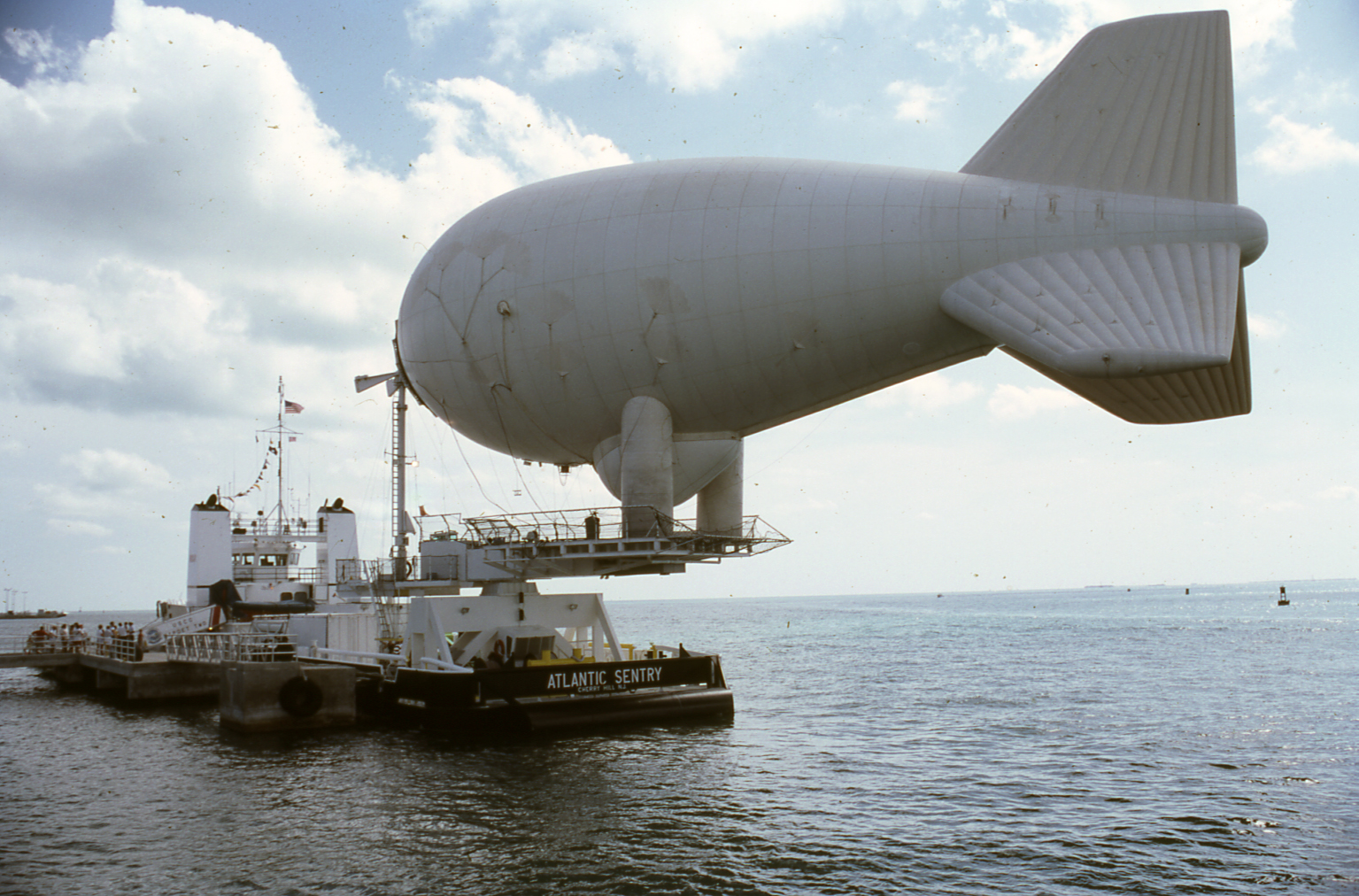}
        \caption{Transportation by sea \cite{flickr}.}
       \label{Fig.transp.1}  
    \end{subfigure}%
    ~ 
    \begin{subfigure}[b]{0.5\textwidth}
        \centering
        \includegraphics[scale=0.25]{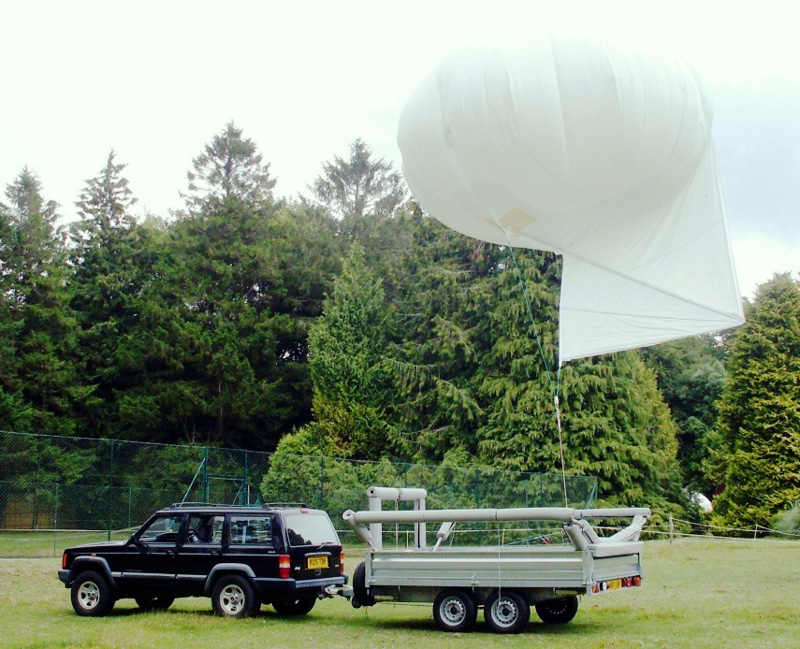}
        \caption{Transportation on land \cite{Allsoppwebsite}.}
       \label{Fig.transp.2}  
    \end{subfigure}
    \caption{Transportation of NTFPs by sea and land. }
   \label{Fig.transp}  
\end{figure*}

The components of NTFPs (mooring stations , winch, envelope, etc.) must be transported to the deployment location.
Trucks may be used to transport NTFPs on the ground, whereas for maritime applications, ships are used as transportation. Fig.\ref{Fig.transp} depicts ground and sea transportation options for NTFPs.

\subsection{Types of Anchor Units}  \label{Anchor}

\subsubsection{Ground Anchor Unit}

\begin{itemize}
    \item Mooring stations: NTFPs can be anchored to the mooring station; this is usually the case for blimps.
    \item Building: NTFPs can be anchored to a building rooftop. UAVs, which can reach a maximum altitude of 150 m, gain more altitude when placed on top of buildings \cite{bushnaq2020cellular,kishk2019capacity,teixeira2017enabling}. 
    \item Vehicles: NTFPs can be anchored to a vehicle, such as a truck, and thus benefit from the mobility of the vehicle \cite{gajra2014soptas}. 
    \item Ground: tUAVs can be placed on the ground since they are very compact and, unlike other NTFPs such as blimps
or balloons, do not require additional infrastructure \cite{kishk20203}.
\end{itemize}

\subsubsection{Sea Anchor Unit}
\begin{itemize}
    \item Ship: When NTFPs are used in a maritime context, they use the ship or another offshore floating structure as an anchor unit. The anchor point must be designed to satisfy the requirements of maritime applications \cite{carlson2018surface,teixeira2016tethered}.
    \item Buoy: When deployed over the sea or ocean, NTFPs can be anchored to oceanographic buoys \cite{campos2016bluecom+}.
    \item Drag Sail: Drag sails can be used as an anchor unit for NTFPs. For instance, the authors in \cite{akita2012feasibility} proposed a configuration where the platform, a balloon in this case, was launched with its tether wound around a reel. The tether had a drag sail at the end. After reaching the desired altitude, the tether anchored the platform into the sea. At the end of the mission, the tether was released and the balloon became a free-flying platform.
\end{itemize}

\subsubsection{Air Anchor Unit}
Although it has never been tested before, the authors in\cite{fesen2015method} proposed an interesting setup to cope with winds that blow in different directions across the stratosphere. The idea is to tether the platform to an HTA glider flying at a lower altitude than the platform itself.

\subsection{Altitudes}
The different altitudes at which NTFPs can fly are shown in Fig.\ref{Fig.altitude}. We classify NTFPs as Ultra LAPs (U-LAPs), LAPs, Medium-Altitude Platforms (MAPs), HAPs, and Ultra HAPs (U-HAP).
\subsubsection{U-LAPs (50 m--150 m)}
Generally, tUAVs, balloons, Helikites, and BATs fly at this altitude. This altitude is suitable for quick operations such as disaster relief by because NTFPs can be rapidly deployed with decent range and coverage \cite{kishk2019capacity,Elistairwebsite}. 

\subsubsection{LAPs (200 m--600 m)}
tUAVs do not reach this altitude, but balloons, Helikites and blimps can, although balloons usually do not fly higher than 400 m. This altitude is suitable for missions that last between three and seven days, and the NTFPs have more coverage than at U-LAP altitudes \cite{Altaeroswebsite,Allsoppwebsite,TCOM}.
\subsubsection{MAPs (0.7 km--5 km)}
Only Helikites and blimps can reach this altitude. Although Helikites usually do not fly beyond 1.5 km, estimates available on their website claim they can reach 3 km \cite{Allsoppwebsite}. Blimps can reach an attitude of around 4.6 km for long-duration operations and surveillance missions. At these altitudes, NTFPs benefit from higher range and coverage \cite{Allsoppwebsite,TCOM}.

\subsubsection{HAPs (15 km--22 km)}
To this day, no NTFP has reached these high altitudes. However, there are several studies that demonstrate the feasibility of high altitude NTFPs \cite{aglietti2008aerostat,chiba2017aerodynamic,fesen2015method,bely1995high,badesha2002dynamic,badesha2002sparcl,badesha1996very,badesha1996very1,aglietti2009dynamic,akita2012feasibility}.

\subsubsection{U-HAP (45 km--50 km)}
Although these altitudes seem extreme and unreachable by NTFPs, there is one paper that studied the feasibility of NTFPs flying at an ultra-high altitude \cite{izet2011low}.

\begin{table*}[h!]
\begin{center}
    \caption{Comparisons among different NTFPs. }
    \label{tab2}
\begin{tabular}{ |P{2.4cm}||P{2.2cm}|P{2.2cm}|P{2cm}|P{2.4cm}|P{2cm}| }
 \hline
Properties & UAVs & Ballons & Helikite & Blimps & BATs  \\
 \hline
Payload &  1 kg--15 kg &   5 kg--50 kg & 2 kg--25 kg &   16 kg--2600 kg &   NA\\ \hline
Altitude  &  150 m--200 m  & 150 m--700 m &  100 m--1.5 km &   100 m--5 km &   150 m--600 m\\  \hline
Wind Speed   &  40 km/h--55 Km/h  & 20 km/h--40 Km/h & 90 km/h & 75 km/h--165 km/h &   160Km/h\\  \hline
Flight Duration    & 2--4 days & 1 day--7 days & 2--4 days &   1 day--30 days  &   NA\\ \hline
Deployment Time    & Fast  & Moderate & Fast  &   Slow &   Fast\\\hline
Cost    & Low/Moderate  & Low/Moderate & Low/Moderate  &   Moderate/High &   NA\\
 \hline
\end{tabular}
\end{center}
\end{table*}

\subsection{Applications of NTFPs}

\subsubsection{Government \& Defense}

 \paragraph{Tactical Operations}
  NTFPs helps military to conduct tactical operations via accurate environment perceptions and real-time imaging \cite{chim2006employing,weiwei2015integrated,ACCwebsite,TCOM}.
 
 \paragraph{Observation and Surveillance}
 NTFPs offers continuous aerial surveillance and reconnaissance by day or night for up to several days, enabling target tracking for enemies and reduced exposure of allies \cite{dusane2017elevated,weiwei2015integrated,ACCwebsite,TCOM,sharma2014design,prior2015tethered}.
 
 \paragraph{Telecommunications}
 NTFPs play a crucial role during military operations as they extend communication in areas where cellular coverage is lacking \cite{krisztian2016military,chim2006employing}.

 \paragraph{Border Surveillance}
 NTFPs are used to track illegal aliens, unauthorized personnel, arms smugglers, and narcotics traffickers, and to detect and prevent enemy forces from crossing borders \cite{weiwei2015integrated,ACCwebsite,TCOM,sharma2014design}.

 \paragraph{Surveillance of Sensitive Sites}
 NTFPs allow the aerial surveillance of sensitive sites, such as military bases, nuclear plants, industrial sites, offshore platforms, harbors, and airports \cite{prior2015tethered,weiwei2015integrated,TCOM,sharma2014design}.

 \paragraph{Detection of Aircraft}
 Depending on their altitude, NTFPs are capable of detecting low-flying aircraft within their area of coverage \cite{ACCwebsite,TCOM}.

\subsubsection{Public Safety and Disaster Relief}

 \paragraph{Search and Rescue Missions}
 For search and rescue missions, NTFPs increase the search area. Also, they provide a cellular coverage in areas such as deserts or mountains \cite{weiwei2015integrated,hariyanto2009emergency,qiantori2012emergency,alsamhi2018tethered,valcarce2013airborne}.
 
 \paragraph{Firefighting Missions and Wildfire Monitoring}
 During firefighting missions, NTFPs not only bring cellular coverage to the area, but can also take remote-sensing aerial 
 infrared images for temperature maps in order to detect and identify a critical hot spot. This helps managers make decisions and give precise directions their the crew \cite{barrado2010wildfire,weiwei2015integrated}.

 \paragraph{Emergency Communications}
Cellphone connectivity and coverage are paramount in the aftermath of natural disasters. earthquakes, tsunamis, hurricanes, and floods can destroy cell phone towers.
Bringing cellphone coverage via NTFPs can help rescue crews communicate and make decisions. This also help identify and prioritize the areas most affected by the disaster \cite{hariyanto2009emergency,alsamhi2019performance,alsamhi2018disaster,qiantori2012emergency,alsamhi2018tethered,valcarce2013airborne}.

 \paragraph{First Operations}
 
 During rescue operations to find and help victims, first responders need to communicate with each other to coordinate their tasks. NTFPs help first responders to assess the situation by providing communication coverage and visual coverage \cite{hariyanto2009emergency,alsamhi2018disaster,qiantori2012emergency,alsamhi2018tethered}.

 \paragraph{Crowd Control and Management}
NTFPs are very useful when it comes to crowd management and riot control \cite{prior2015tethered,weiwei2015integrated}.
 
 \paragraph{Aerial Observations}
 For aerial observations and surveillance tasks, height is a huge advantage. Hence, NTFPs can be used for surveillance purposes, such as control of illegal fishing activities or for homeland security and anti-terrorism activities \cite{weiwei2015integrated,prior2015tethered,sharma2014design,kanoria2012winged}.
 
 \paragraph{Traffic Regulation, Accidents Management, and Vehicle Surveillance}
NTFPs can be useful for traffic regulations. They can also help with anticipating traffic jams, and alerting vehicles about accidents happening in their vicinity.
Finally, they can be used to locate suspect vehicles or track vehicles during car chases.

\begin{figure}[t!]
\centering
\includegraphics[scale=0.5]{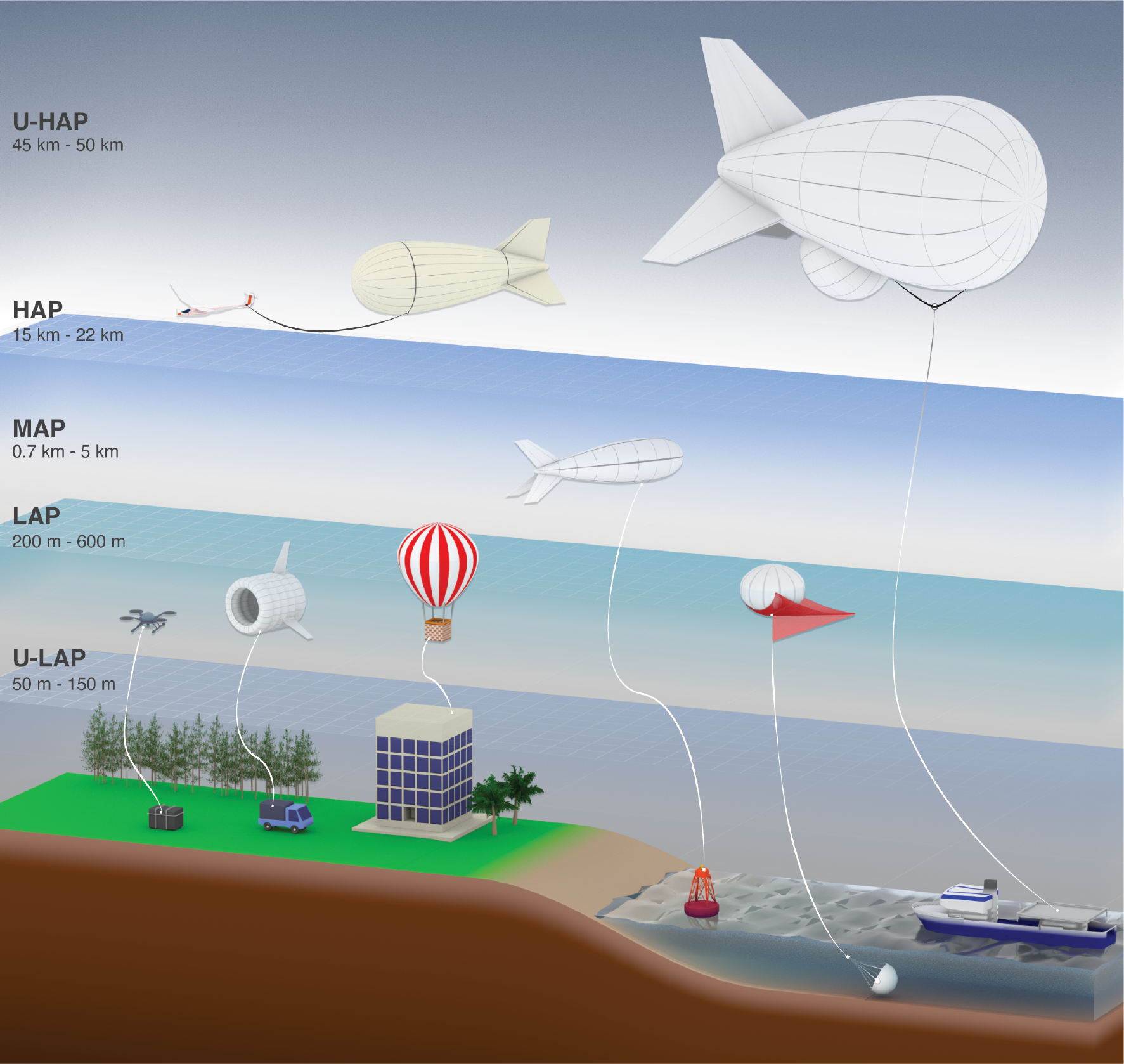}
\caption{Altitudes and anchor units of NTFPs.}
\label{Fig.altitude}
\end{figure}

\subsubsection{Communications}

 \paragraph{Cellular Coverage}
 NTFPs overcome the limitations of terrestrial communication towers. Due to their higher altitude, they have a larger coverage area and better line-of-sight (LOS). Plus, they are less costly to deploy and construct than cell towers and satellites \cite{chopra2011new,badesha2006optical,alsamhi2019tethered,alsamhi2016network,kishk2019capacity}.
 
 \paragraph{Coverage in Rural and Remote Areas}
 Almost half of the global population lives in remote or rural areas. It is not economical or profitable for phone operators to erect cell towers in these areas; consequently, most of the people living in these areas do not have access to an internet connection. NTFPs can bridge this gap by providing internet connection to these people while being economically profitable for phone operators \cite{bilaye2008low,gawande2007design,chopra2011new,badesha2006optical,alsamhi2018tethered}.

 \paragraph{Temporary Communications}
 In certain situations, there is a need for temporary communications. NTFPs can be used as temporary transmitters or as relays \cite{chim2006employing,hariyanto2009emergency,alsamhi2019performance,alsamhi2018disaster,qiantori2012emergency,alsamhi2018tethered,valcarce2013airborne}.

\paragraph{Remote Sea and Ocean Area Coverage}

Seas and oceans are lacking cellular coverage, which makes NTFPs useful in these areas to sailors, fishermen, personnel on floating structures \cite{campos2016bluecom+,teixeira2016tethered,teixeira2017enabling}.

 \begin{figure}[t!]
\centering
\includegraphics[scale=0.31]{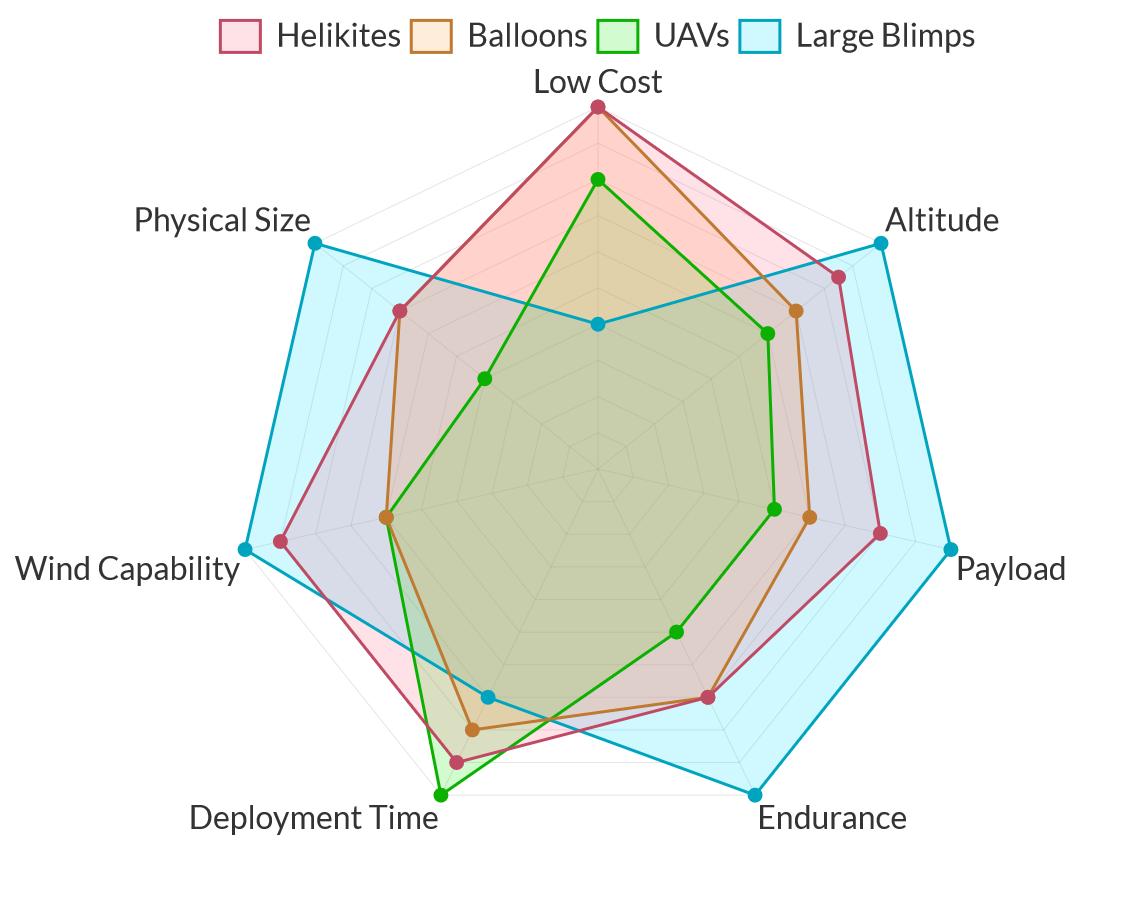}
\caption{Comparisons among Helikites, balloons, UAVs, and larger blimps.}
\label{}
\end{figure}

\subsubsection{Entertainment}

 \paragraph{Coverage of Major Sport Events}
 Major sport events, such football, baseball, rugby, etc, gather massive crowds of people, making NTFPs an excellent choice for broadcasting these events, and bringing coverage to these areas.

 \paragraph{Surveillance of Large Public Events}
 The need for surveillance of large public events cannot be overstated. NTFPs permit a wide view of those events \cite{weiwei2015integrated,sharma2014design,prior2015tethered}.
  
 \paragraph{Aerial Recording and Photography}
 NTFPs can be used to take aerial photos from a perspective that is hard to access from a regular height. They can also be used to record videos and movies \cite{weiwei2015integrated}.

 \paragraph{Advertisement}
 NTFPs are also used for advertisement purposes. The NTFP may have a written sign on it, or it can lift an advertisement sign. The NTFPs can be illuminated at night for a better visibility of the advertisement or deployed during major sport events to attract people's attention. 

\begin{figure*}[ht!]
\centering
\includegraphics[scale=0.94]{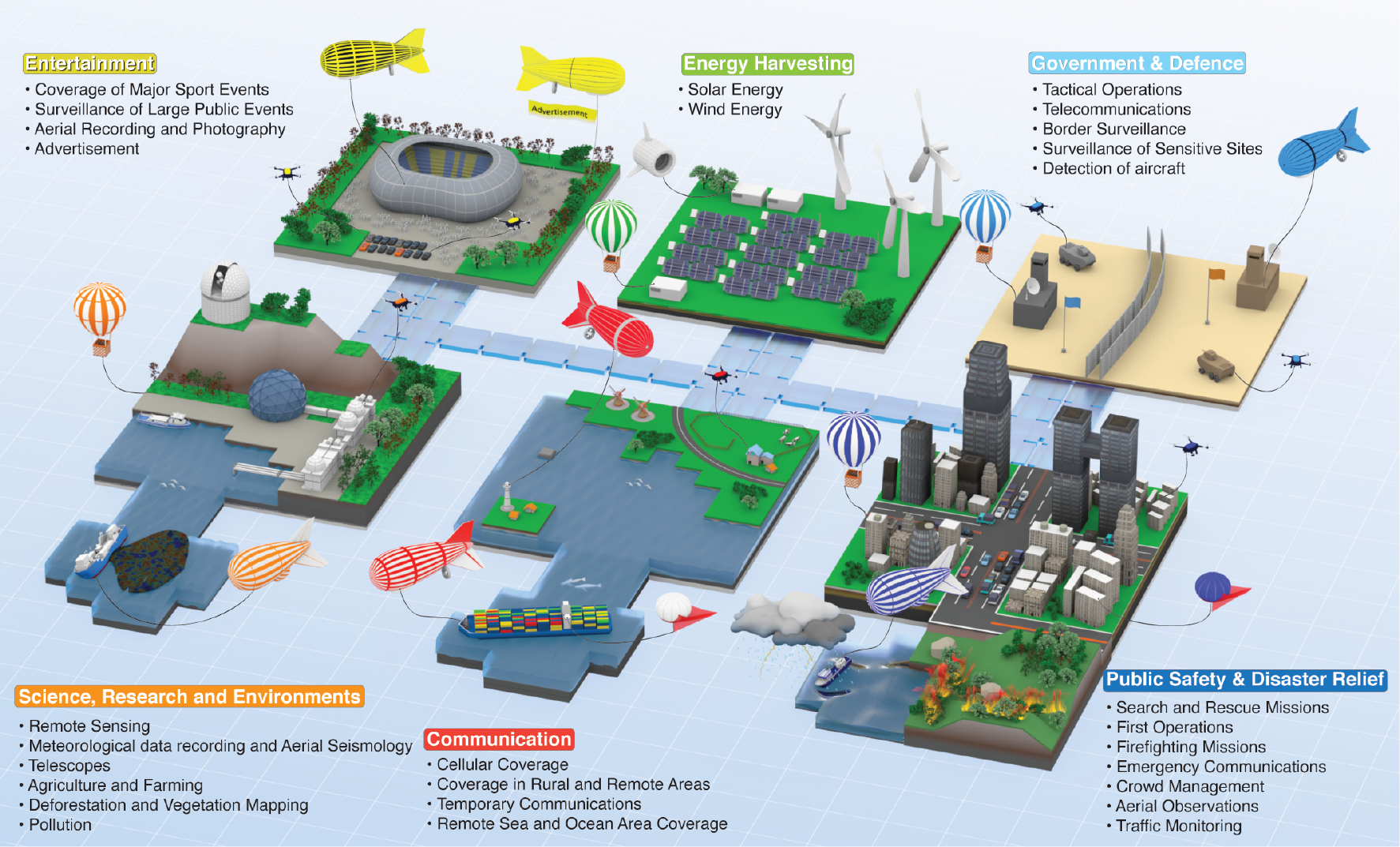}
\caption{Applications of NTFPs.}
\label{Fig.sp}   
\end{figure*}

\subsubsection{Science, Research and Environments}

  \paragraph{Remote Sensing}
  NTFPs can be equipped with remote-sensing. The sensor can be deployed to collect data for various applications, such as detecting landslides and habitat destruction \cite{chen2006spectral,wu2018experimental,inoue2000blimp,vierling2006short,carlson2018surface,jo2015mapping,lee2019estimate}.
    
  \paragraph{Education}
  NTFPs can be used as educational tool to show students phenomena that can only be see from a certain altitude.
  
  \paragraph{Meteorological Data Recording and Aerial Seismology}
   NTFPs are used to gather meteorological data, such as atmospheric temperature, wind speed, air pressure, and humidity \cite{garstang1971fluctuations,balasuriya2017development,hu2019atmospheric}. They are also used in aerial seismology to detect earthquakes \cite{krishnamoorthy,krishnamoorthy20}.

  \paragraph{Telescopes}
  Since they have a better view at higher altitudes, NTFPs are used as aerial telescopes \cite{deschesnes2005design,lambert2003design,fitzsimmons2000steady,nahon2002dynamics}.

  \paragraph{Agriculture and Farming}
   NTFPs help farmers identify several types of plants via aerial images, and may prevent farming fraud \cite{inoue2000blimp,bajoria2017design,pant2016tethered}.
   
  \paragraph{Deforestation and Vegetation Mapping}
  NTFPs can be used to capture aerial images from a high altitude, which are used to detect and prevent deforestation. These aerial images can also be used to  understand changes in biodiversity over time \cite{silva2014mapping,inoue2000blimp}.
  
  \paragraph{Pollution}
  NTFPs are used to detect oil spills and floating debris. They are also used to detect and monitor light pollution \cite{fiorentin2018minlu}.

\subsubsection{Energy Harvesting}

 \paragraph{Solar Energy}
 The power harvested by ground photovoltaic panels depends heavily on the weather. Thus, the efficiency of these panels decreases drastically in countries with fewer sunny days such as the United Kingdom. However, NTFPs can fly above the clouds, and get more solar exposure \cite{aglietti2008aerostat,aglietti2009harnessing,aglietti2008solar1,gajra2014soptas,ghosh2017power}.

 \paragraph{Wind Energy}
 Wind speed is much higher at an altitude of 150 m than at the ground. Hence, NTFPs at high altitudes can harness more powerful winds and can generate twice the energy of comparable ground-based turbines \cite{williams2008optimal,lansdorp2006laddermill,saleem2018aerodynamic}.


\subsection{Advantages of NTFPs}

\subsubsection{Cost-Efficient}
NTFPs are cost-efficient compared to free-flying platforms and other communication infrastructure such as tower masts and satellites.

NTFPs have a low operation cost overall compared to free-flying platforms. For instance, they are cheaper to purchase, maintain, and service compared to free-flying platforms. Additionally, they require less training to operate and fewer operators.
Compared with tower masts, deploying a NTFP costs ten times less than erecting a tower mast. 
Also, the cellular coverage provided by one NTFP flying at an altitude of 250 m equals the coverage provided by 14 tower masts. 
Tower masts consume large amounts of energy and fuel compared to NTFPs, which consume less energy and require no fuel.
Finally, compared to satellites, deploying a NTFP is much cheaper than launching a satellite into orbit, especially considering that satellites only operate for approximately ten years. NTFPs offer a great alternative to free-flying platforms, tower masts, and satellites.

\subsubsection{Endurance \& Persistence}
One key advantage that NTFPs have over free-flying platforms is endurance and persistence. This is more pertinent to surveillance missions and telecommunications during which the flying platform must stay aloft for a prolonged period (e.g., days or weeks) and/or in a stationary position. The persistence of free-flying platforms is only several hours, and it is very hard for free-flying platforms to remain in a stationary position.

\subsubsection{Green Credentials}
Another advantage that NTFPs have low consumption of fuel and power compared to free-flying platforms and especially tower masts. In India for instance, nearly 2 billion liters of diesel are burned each year to operate tower masts \cite{chopra2011new}. This number will increase to reach 15 billion to cover rural and remote areas in India which will dramatically increase the carbon footprint and pollution in the atmosphere.

\subsubsection{Coverage \& Backhaul Link}
NTFPs have wide coverage compared to tower masts due to their higher altitudes, which allow them to cover a larger area. When compared with satellites and terrestrial networks, NTFPs have a higher and stronger LOS than terrestrial networks and a shorter propagation delay than satellites.  Additionally, NTFPs have great backhaul link capacity compared to free-flying platforms, whose wireless backhaul is more prone to interference and higher latency. A wired (tethered) backhaul, i.e., have a wired data-link via the tether, allowing reliable and high data rates communications.

\begin{figure*}[ht!]
    \centering
    \begin{subfigure}[b]{0.5\textwidth}
        \centering
        \includegraphics[scale=0.4]{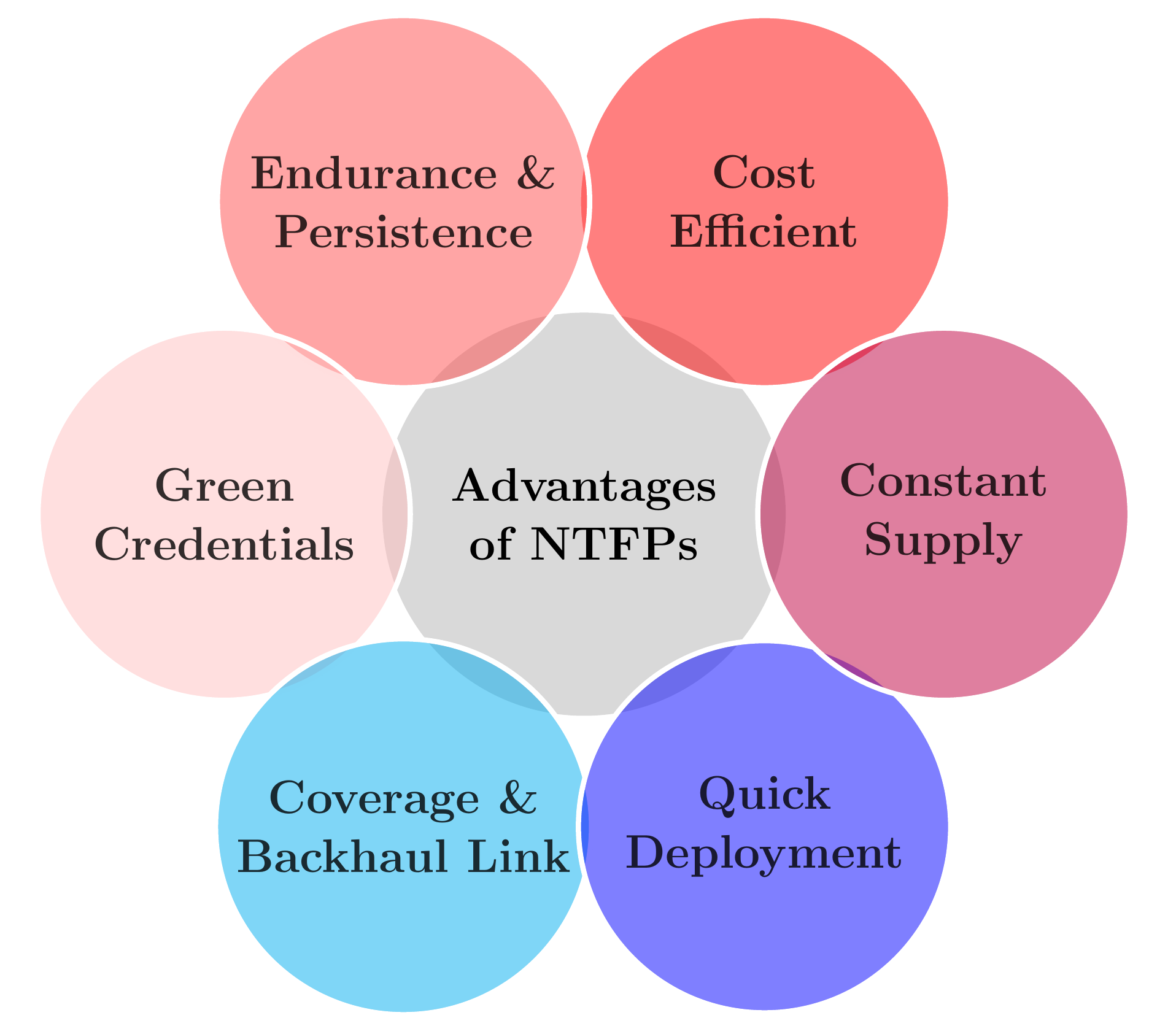}
        \caption{Advantages of NTFPs.}
       \label{}  
    \end{subfigure}%
    ~ 
    \begin{subfigure}[b]{0.5\textwidth}
        \centering
        \includegraphics[scale=0.4]{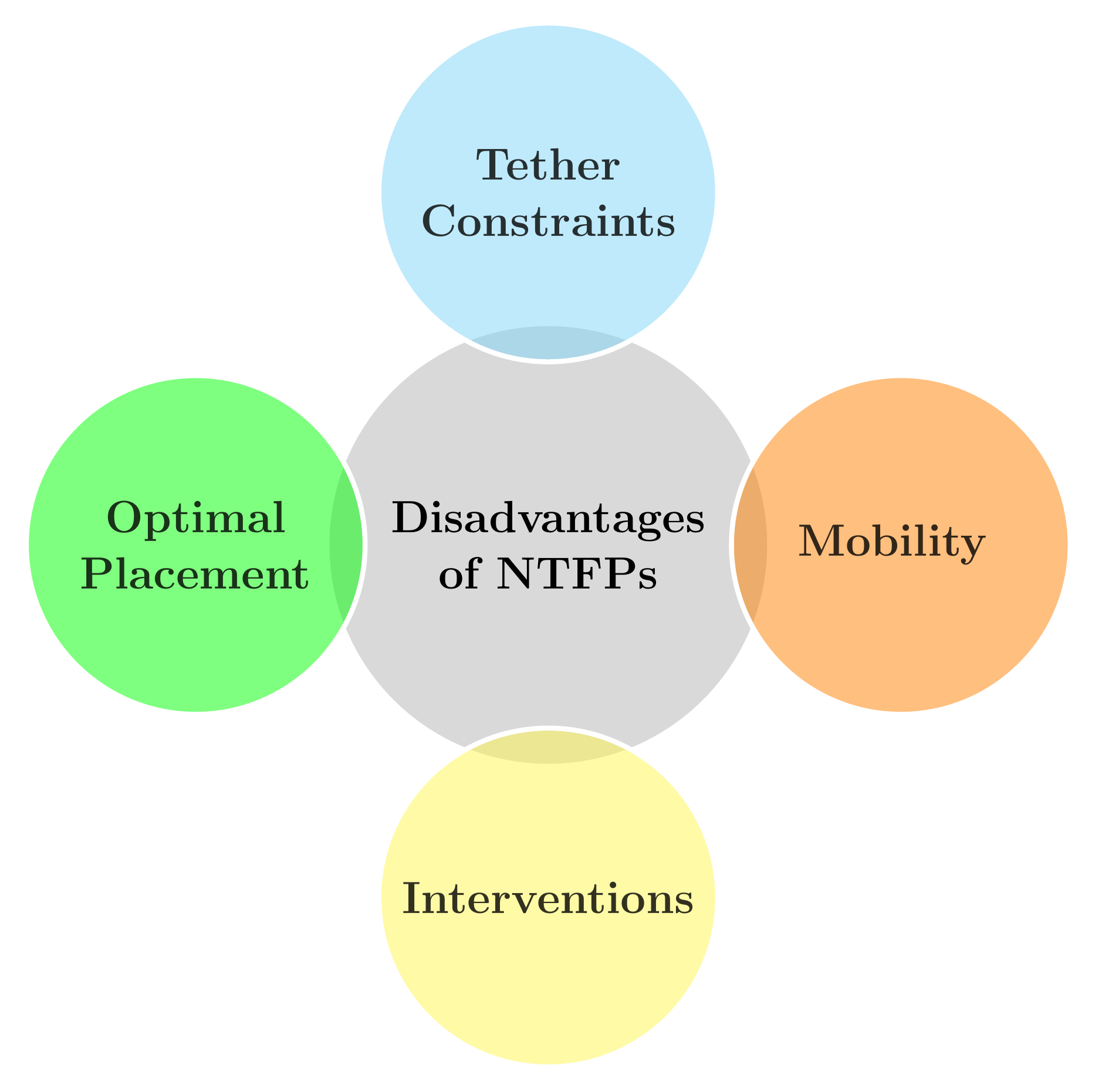}
        \caption{Disadvantages of NTFPs.}
       \label{}  
    \end{subfigure}
    \caption{Advantages and disadvantages of NTFPs.}
   \label{}  
\end{figure*}

\subsubsection{Quick Deployment}
One of NTFPs biggest advantages is their fast and quick deployment, which makes them suitable for safety missions and disaster relief operations. In addition, NTFPs can be moved and relocated. In contrast, tower masts, once erected, cannot be moved elsewhere. NTFPs can also be used in areas where it is not feasible to erect a tower mast, such as during a disaster situation where the communications infrastructure is severely damaged or destroyed or over land that is not suitable for erecting a tower mast. In summary, NTFPs can be quickly deployed, easily reconfigured, and rapidly relocated.

\subsubsection{Constant Power Supply}
The role of the tether is to supply the platform with power so it can stay aloft for days or weeks, whereas free-flying platforms have limited power supply. Also, the constant power supply allows NTFPs to carry larger payloads compared to free-flying platforms. In addition to power, the tether offers high data rate and increased backhaul link capacity via a fiber connection to the ground station compared to free-flying platforms.

\subsection{Disadvantages of NTFPs}
\subsubsection{Mobility}
Although NTFPs are rapid to deploy and have better mobility than ground stations, they are still limited in their movement due to the physical constraint imposed by the tether. The tether offers continuous power and data supply to the platforms, but at a cost to mobility. Compared to free-flying platforms, NTFPs cannot move beyond the radius of the tether’s length. 
\subsubsection{Tether Constraints}
 One constraint that we mentioned before is the limited mobility imposed by the tether. Also, the tether itself can be an issue if it sustains damages (intentionally or unintentionally), preventing the supply of data and communications. A damaged tether may hinder an ongoing operation, especially critical operations such as military missions and disaster relief operations. Hence, it is always recommended to protect and attend to the tether, and also have several tethers attached to the platform for redundancy.

\subsubsection{Optimal Placement}
Free-flying platforms can hover or fly freely in the air. Hence, they can move to optimize their positions. However, tether platforms are constrained by the tether preventing them from optimizing their position.   

\subsubsection{Intervention Operations}
NTFPs are not designed for intervention operations since they cannot move freely or quickly in the air compared to aircraft. The capabilities of NTFPs must match the requirements of the mission.

\begin{figure*}[ht!]
    \centering
    \begin{subfigure}[b]{0.5\textwidth}
        \centering
        \includegraphics[scale=0.14]{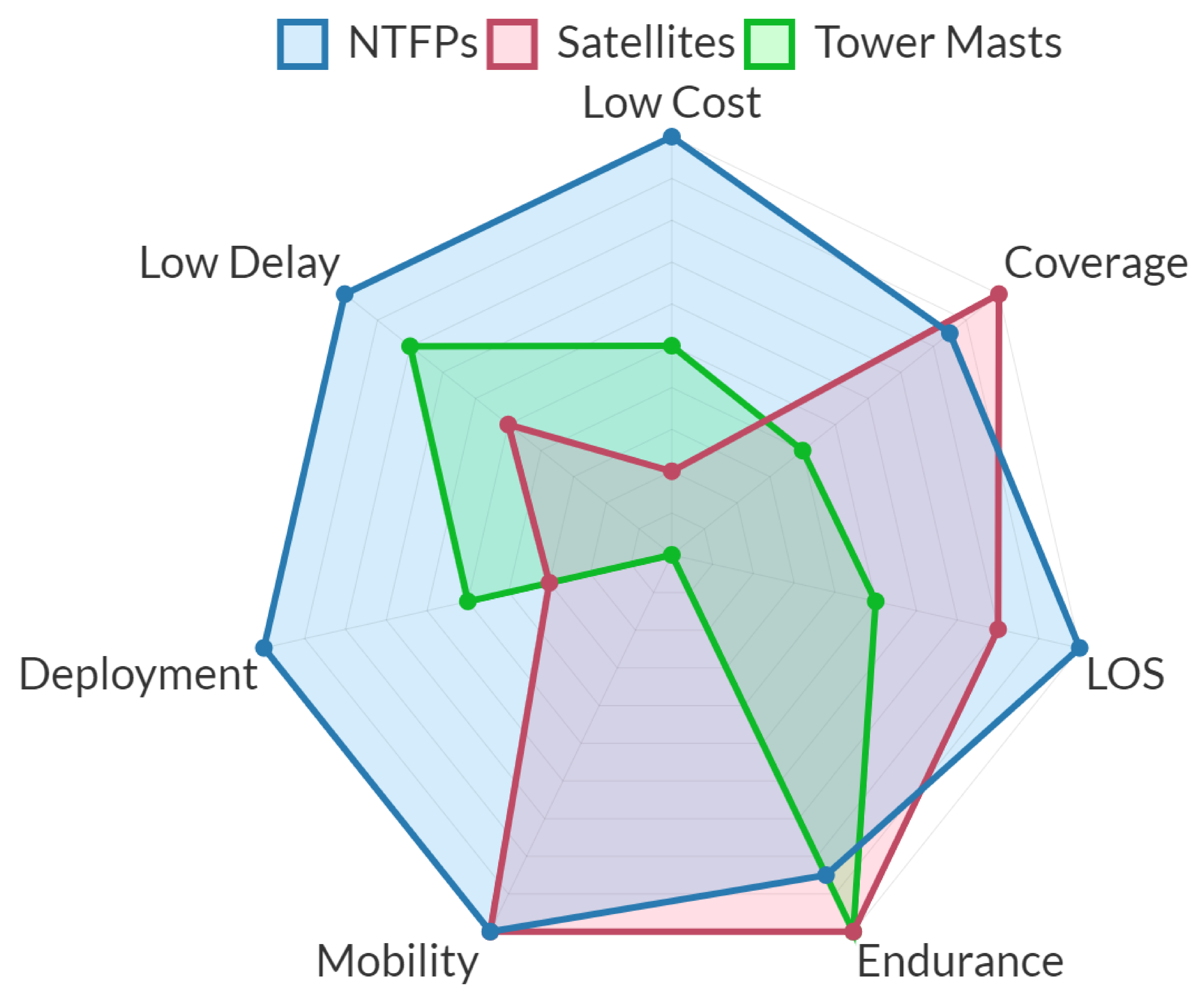}
        \caption{NTFPs, satellites, and tower masts.}
       \label{Fig.com.a}  
    \end{subfigure}%
    ~ 
    \begin{subfigure}[b]{0.5\textwidth}
        \centering
        \includegraphics[scale=0.14]{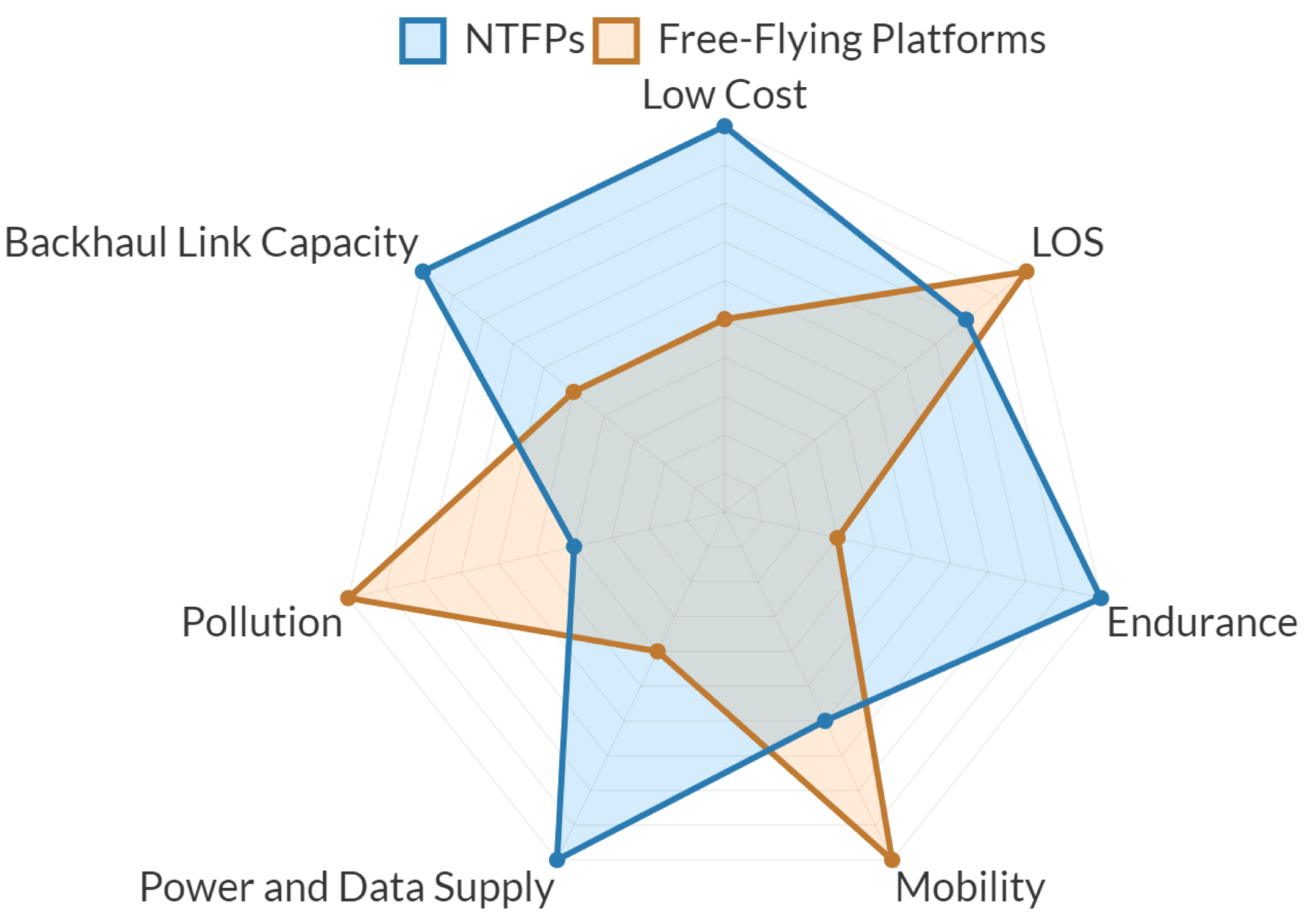}
        \caption{NTFPs and free-flying platforms.}
       \label{Fig.com.b}  
    \end{subfigure}
    \caption{Comparisons among NTFPs and other flying platforms and communications infrastructures.}
   \label{Fig.com}  
\end{figure*}

Fig.\ref{Fig.com} summarizes the advantages and disadvantages of using NTFPs. 
In Fig.\ref{Fig.com.a}, we see a comparison among NTFPs, satellites, and tower masts in terms of the communications aspect. NTFPs have more advantages in communications than disadvantages. For instance, if we compare NTFPs with satellites, we notice that satellites have more coverage and endurance than NTFPs, but with a far greater cost, greater delay, and longer time to deploy. Also, if we compare NTFPs with tower masts, we can see that NTFPs outperform tower masts in terms of mobility (tower masts are static), coverage due to their altitude, cost, and LOS probability.

From Fig.\ref{Fig.com.b}, we see the comparison between NTFPs and free-flying platforms. Free-flying platforms outperform NTFPs only in mobility and LOS probability. On the other hand, NTFPs have a lower cost, better backhaul link capacity, better endurance and persistence, constant power and data supply, and less pollution.

\subsection{Regulations} 
\subsubsection{Socio-Technical Concerns of NTFPs}
Although NTFPs offer several advantages and applications, there are several concerns that have to be taken into account regarding theses platforms such as privacy, data protection, and public safety. Before we explain the regulations related to NTFPs, we describe the socio-technical concerns related to these platforms \cite{fotouhi2019survey}:

\begin{itemize}
    \item Privacy: Since NTFPs have a great coverage over the area of interest, they can be unintentionally (or intentionally) a threat to the privacy of individuals and businesses. Therefore, legislation and regulations must be established to protect privacy.
    \item Data Protection: Due to their coverage, NTFPs can collect massive amounts of data from the public, such as images, videos, and personal data.
These data must be protected from abuse according to data protection laws, and operations of NTFPs should be subject to regulation that protects personal information \cite{finn2014study}.
    \item Public Safety: Another critical issue of NTFPs is public safety. Although NTFPs present fewer safety issues to the public than free-flying platforms, regulations must still be issued to protect the public. Some possible risks include a NTFP falling on the public, a cut in the power supply, the platform landing procedure when the tether is cut, a collision with flying aircraft, etc. 
\end{itemize}

Like free-flying platforms, NTFPs are subject to regulations.
However, tUAVs and tethered LTA aerostats are not subject to the same regulations. Tethered LTA aerostats and Helikietes fall under part 101 of the Federal Aviation Administration (FAA) aeronautics and space regulations, whereas tUAVs  fall under part 107 of aeronautics and space regulations \cite{FFAwebsite}.

We will outline below the regulations as they apply to tUAVs and tethered LTA aerostats, and highlight the main differences between tethered and free-flying regulations.

\subsubsection{tUAVs}
Whether tUAVs should be considered as kites or balloons is debated,; however, the U.S. FAA states that tUAVs fall under the same category as free-flying UAVs when it comes to regulations.
However, some countries do not classify tUAVs as free-flying UAVs, since they are tethered to an anchor point. Hence, the classification of tUAVs depends on the aviation laws of each country \cite{LOCwebsite}.

Table \ref{tab1} shows the main requirements related to UAVs.
We can see from Table \ref{tab1} that tUAVs are not under the same constraints safety-wise as their free-flying counterparts, making tUAVs easier to deploy. We can also see that from the set of requirements presented in Table \ref{tab1}, three of them have no impact on the tUAVs:

\begin{table*}[h!]
\begin{center}
    \caption{Different requirements between free-flying UAVs and tUAVs \cite{Elistairwebsite}}
    \label{tab1}
\begin{tabular}{ |p{2.2cm}||p{9cm}|p{1.7cm}|p{1.7cm}|  }
 \hline
& Requirements & Free-flying UAV & tUAV\\
 \hline
Captive system  & Have a physical connection with a mechanical strength.    & Not Required &   Compulsory\\ \hline
Max. altitude   &   90 m--152 m (see Table \ref{tab2}). & Compulsory   & Compulsory\\  \hline
Loss of data link & Restore the data line or interrupt the flight to reduce the effect on third parties in the air or on the ground. & Compulsory &  Not Required\\  \hline
Identification    & Remote direct identification system. & Compulsory &  Not Required\\  \hline
Registration &   Have a unique physical serial number.  & Compulsory & Compulsory\\  \hline
Geo-fencing & Load and update data containing information on airspace limitations in relation to UAV position and altitude imposed in relation to geographical areas.  & Compulsory   & Compulsory\\  \hline
Security & Data link system protected against unauthorized access to command and control function. & Compulsory &    Not Required\\  \hline
Endurance & Clear warning signal when the battery of the UAV or its control station reaches a low level. & Compulsory & Compulsory\\  \hline
Visibility & Equipped with positioning and navigation lights.  & Compulsory & Compulsory\\
 \hline
\end{tabular}
\end{center}
\end{table*}

\begin{table}[h!]
\begin{center}
    \caption{Regulations regarding UAV heights for commercial purposes  \cite{stocker2017review}.}
    \label{tab2}
\begin{tabular}{ |p{2.2cm}||p{2cm}|p{2cm}| }
 \hline
Country & Max altitude & Min distance to people\\
 \hline
Canada &  90 m &   150 m\\ \hline
Germany  &  100 m  & not over crowds\\  \hline
Spain   &  120 m  & not over groups \\  \hline
United States    & 122 m  & N/A\\  \hline
Chile   & 130 m & 30 m \\  \hline
Japan    & 150 m & 30 m \\  \hline 
Colombia   & 152 m & N/A \\  
 \hline
\end{tabular}
\end{center}
\end{table}

\begin{itemize}
    \item Loss of data links: In this situation, a free-flying UAVs can fly away with all the risks and safety issues involved. tUAVs, on the other hand, cannot fly away since they are tethered to the ground station. Therefore, there is no need to restore the lost data link or abort the flight.

   \item Identification: Free-flying UAVs have remote direct identification systems that allow the UAV to be identified in the case of flyaway situations. However, tUAVs are exempted from this requirement since their flying perimeter is limited by the tether length (between 90 m and 160 m).

   \item Security: In the case of free-flying UAVs, the data are transmitted though the air, which makes them vulnerable to eavesdroppers and subject to interference.
   In contrast, tUAVs transmit data via their tether, decreasing signal loss, signal attenuation, and interference. However, the tether has to be protected from physical harm and hijacking.

\end{itemize}

Another aspect worth noting about tUAVs is pilot requirements. Free-flying UAVs require a qualified pilot with a flying certificate.
But in the case of tUAVs, the pilot does not need to possess any certification. Plus, the tether makes it easier to stabilize the UAV' movements.

Furthermore, when there is a ground power cut, a safety mechanism activates a battery to keep the UAV in the air. This mechanism also triggers an alarm that alerts the pilot so they can land the UAV.

\subsubsection{Tethered LTA platforms and Helikites}

The regulations related to NTFPs can be divided into three types: regulations related to flight and aviation, regulations related to communications, and regulations related to the equipment that the platform is tethered with.

\begin{itemize}
\item Flight and Aviation Regulations: The FAA requires that all tethered aerostats have a rapid deflation device that will automatically and rapidly deflate the aerostat if the tether is cut.
If the device does not respond or does not function properly, the nearest air traffic control must be notified about the location and time of the escape of the aerostat. The deflation devices are activated when the tethered aerostat exceeds a predetermined distance from a given location monitored by a global positioning system (GPS), or when the aerostat has exceeded a predetermined altitude monitored by a barometric pressure sensor. The aerostat must be illuminated if it is flying from sunset to sunrise. For further information about the regulations concerning tethered aerostats, the reader is advised to read the electronic code on the federal regulation website \cite{FFAwebsite}. 
\item Communications Regulations: Missions or operations that require communications via tethered aerostats are regulated by the U.S. Federal Communications Commission (FCC). For further information about the regulations concerning tethered aerostats, the reader is advised to read the electronic code on the federal regulation website \cite{FCCwebsite}. 
\item Transportation Regulations: Regulations may also apply to the vehicle to which the aerostat is attached. 
\end{itemize}

\section{Case Studies, Projects, and Companies Related to NTFPs}
\label{section_3}
In this section, we explore the numerous applications of NTFPs. We present projects and case studies from real-life scenarios involving NTFPs. Also, for completeness, we present the major companies that manufacture and sell NTFPs worldwide.

\subsection{Project and Case Studies}

\subsubsection{Paris Airport Maintenance}
\begin{figure*}
        \centering
        \begin{subfigure}[b]{0.475\textwidth}
            \centering
\includegraphics[scale=0.63]{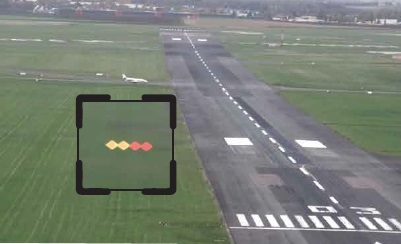}
\caption{PAPI lights beside the runway.}
\label{Fig.PAPI.1}  
        \end{subfigure}
        \hfill
        \begin{subfigure}[b]{0.475\textwidth}  
            \centering 
\includegraphics[scale=0.6]{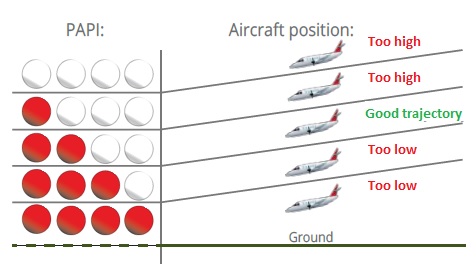}
\caption{The PAPI red and white lights.}
\label{Fig.tUAVElistair.1}
        \end{subfigure}
        \vskip\baselineskip
        \begin{subfigure}[b]{0.475\textwidth}   
            \centering 
\includegraphics[scale=0.72]{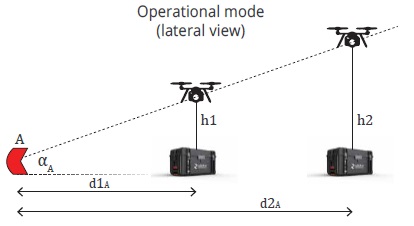}
\caption{Two tUAVs at two different distances to calibrate the PAPI.}
\label{Fig.tUAVElistair.2}
        \end{subfigure}
        \quad
        \begin{subfigure}[b]{0.475\textwidth}   
            \centering 
\includegraphics[scale=0.65]{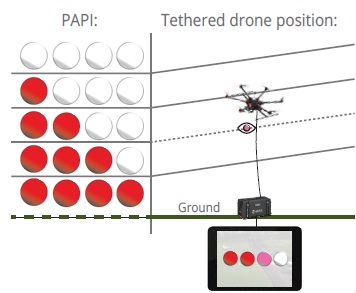}
\caption{The tUAV calibrating the transition between each color.}
\label{Fig.tUAVElistair.3}
        \end{subfigure}
\caption{The PAPI mode of operation, and its calibration by a tUAV \cite{elistair_SC_1}.}
\label{Fig.tUAVElistair}
    \end{figure*}

The Precision Approach Path Indicator (PAPI) is a system with four lights (two white lights and two red lights) placed beside landing runways that helps pilots assess their landing slope (as shown in Fig.\ref{Fig.PAPI.1}). 
If the PAPI displays two white lights and two red lights, then the airplane has the correct slope. If the PAPI shows three white lights or more, then the slope is too high; if it shows three red lights or more, then the slope is too low (Fig.\ref{Fig.tUAVElistair.1}).

To calibrate the precision of this system, maintenance is carried out using elevating work platforms, which block access to the runway. Closing the runway causes time loss, complicated maintenance logistics, and risks associated with the ground operators on the runway and the operators on the elevating work platforms.

To solve this problem, the French airport authority Groupe ADP used a tUAV at the Paris airport Charles-de-Gaulle.
An Elistair hexacopter was used to perform the maintenance task and calibrate the PAPI. The tUAV allows a clear view of the Paris airport runways. The advantages of using a tUAV to perform this task are: 1) it can stay in a stationary position for prolonged periods of time, 2) the tUAV can take off and land on a very small surface (1,2 m diameter), 3) the tUAV can detect precisely the boundary of each color, which is depicted in pink (a mix of white and red) as shown in Fig.\ref{Fig.tUAVElistair.3}.
To further increase the accuracy of the PAPI, two tUAVs are placed at different distances and altitudes; hence, by using the required threshold angle $\alpha_A$, the threshold can be verified by checking the accurate altitude of the pink color detected by the tUAVS at distances $d1_A$ and $d2_A$, as shown in Fig.\ref{Fig.tUAVElistair.2}.

The use of a tUAV prevents any risk to the operators, such as falling down from the elevated work platforms. Since the UAV is tethered, it is directly linked to the control tower, which permits secure and interference-free communications. Also, the tUAV can be deployed over the runway with the necessary safety measures, while the rest of the airport continues running without interruption. Finally, the tUAV calibrates all the thresholds in one hour.

\subsubsection{Road Traffic Monitoring in Lyon}

To monitor traffic in Lyon, Elistair proposed a solution using their tUAV
to continuously monitor a roundabout.
Lyon, which is the second largest urban area in France, has a big concentration of traffic flow, especially in the suburban areas.

\begin{figure*}[ht!]
    \centering
    \begin{subfigure}[b]{0.62\textwidth}
        \centering
        \includegraphics[scale=0.3]{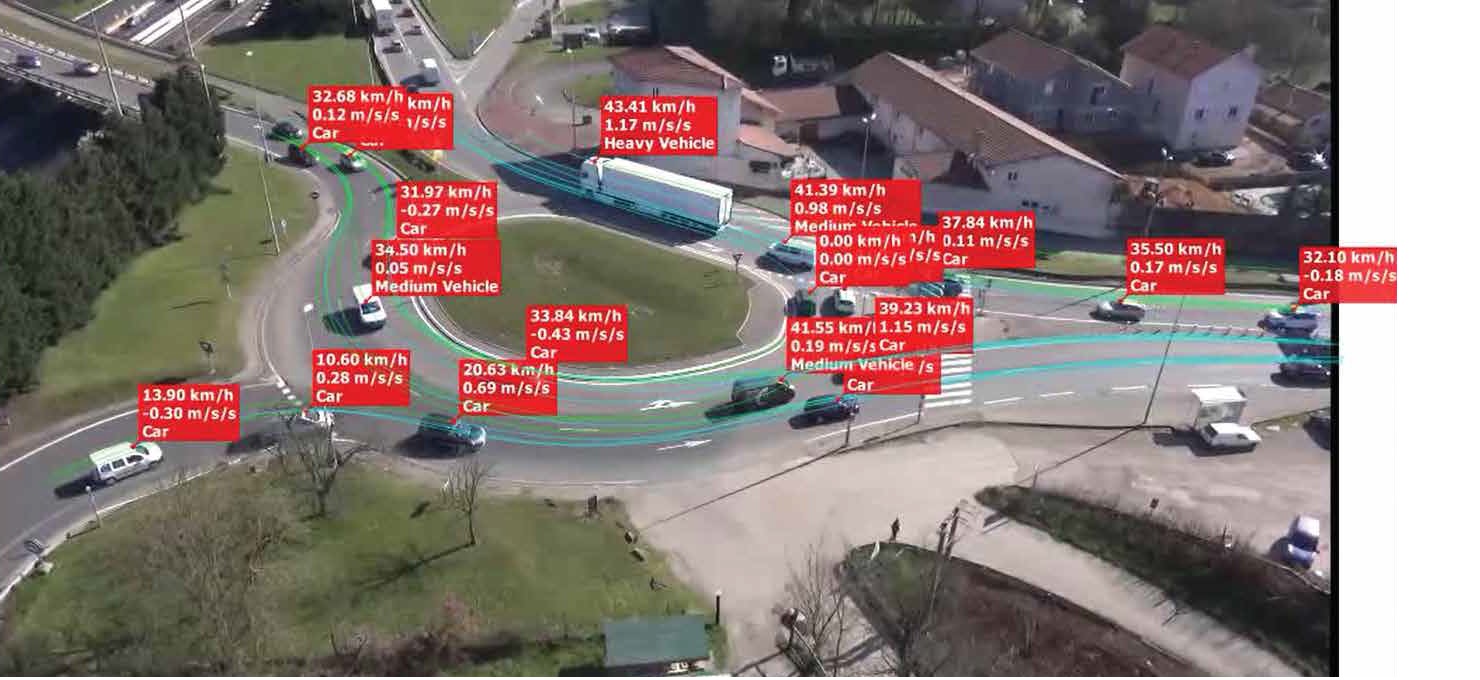}
        \caption{A snapshot of the traffic video after DataFromSky processing.}
       \label{Fig.datafromsky.1}  
    \end{subfigure}%
    ~ 
    \begin{subfigure}[b]{0.4\textwidth}
        \centering
        \includegraphics[scale=0.4]{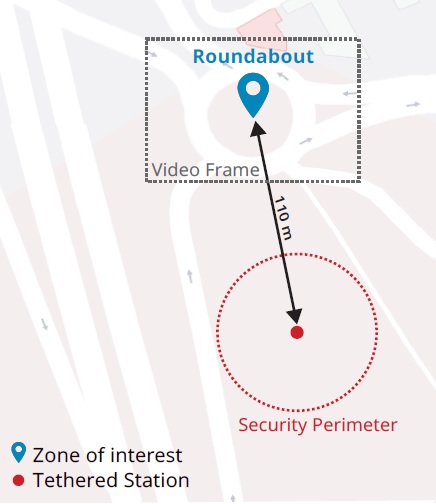}
        \caption{The tUAV placement.}
       \label{Fig.datafromsky.2}  
    \end{subfigure}
    \caption{An aerial view of the roundabout and the placement of the tUAV \cite{elistair_SC_2}. }
   \label{Fig.datafromsky.}  
\end{figure*}

The tUAV was equipped with a Full HD camera, and the operation lasted 3 hours during rush hours. The tUAV was placed 110 m away from the roundabout. For security reasons, the tUAV was located within a 50 m secured radius, as depicted in Fig.\ref{Fig.datafromsky.2}, to ensure that the operation would not be interrupted or jeopardized.

Using the tUAV offered several advantages. For example, the tether cable allowed the tUAV to maintain a steady position when controlling the camera. Also, the tether allowed safe data transfer and data display in real-time. The cloud-based platform DataFromSky was used to analyze the road-traffic data collected from the tUAV with artificial intelligence and machine learning tools \cite{datafromsky}.

The tUAV recorded the traffic flow in the roundabout, then the videos were uploaded to DataFromSky. After processing and analyzing the data, DataFromSky sent a video and metrics, such as speed, acceleration, and trajectory of the vehicles, as shown in Fig.\ref{Fig.datafromsky.1}. It also provided the quantity, categories, and the types of the vehicles: cars, motorcycles, trucks or buses.  


Thus, using a tUAV for traffic monitoring has an easy configuration and fast deployment.
The tUAVs also outperformed the traditional traffic monitoring method in terms of mobility and coverage. By using a tether cable: 1) the UAV maintained a persistent and steady position to record the traffic; 2) the communications and data were secured; 3) the tUAV could stay aloft for several hours (days if needed) thanks to the constant supply of power.

\subsubsection{Border Security in Southern Texas}

Rio Grande Valley, located in southern Texas along the Mexican border, has accounted for the highest number of apprehended illegal immigrants in the U.S. since 2016 \cite{CBP}.
The U.S. Department of Homeland Security (DHS) and Customs and Border Protection (CBP) are facing a huge challenge in monitoring this very long border. Also, the terrain in this area is challenging, limiting the capabilities of ground-based surveillance systems, and pushing the DHS and CBP to seek an elevated or aerial solution.

\begin{figure}[]
\centering
\includegraphics[scale=0.25]{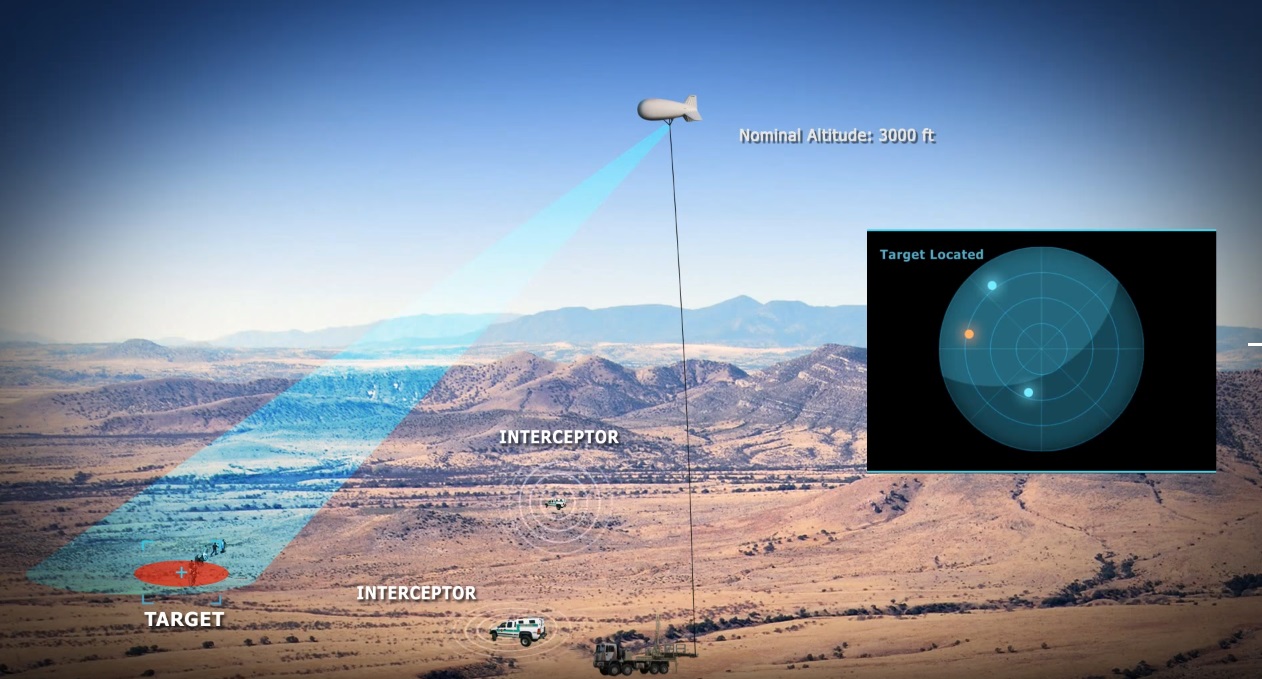}
\caption{Borders control and illegal immigrants tracking with a tethered TCOM blimp \cite{TCOM_CS1}.}
\label{}
\end{figure}

To deal with this challenge, the DHS and CBP used TCOM tethered blimps. The  tethered blimps provided large coverage and continuous surveillance of the Rio Grande Valley to the border authorities. The advantage of tethered blimps is that they can stay aloft for several weeks while providing real-time videos and monitoring, which helps the authorities make better decisions. Also, the blimps have a high degree of mobility, can be rapidly deployed, are battle-proven, and low-cost.

After using the tethered blimps, the border authorities have witnessed a decrease in illegal immigrant crossings, thanks to the tethered blimps’ wide coverage and long persistence.

\subsubsection{Oil-Spill Detection in the Arctic Ocean, Norway}

The Norwegian Clean Seas Association for Operating Companies, also called NOFO, is an  oil-spill response organization. NOFO works with  30 offshore operators,  providing and managing oil-spill  preparedness plans.
An aerial solution can detect oil spills better than at sea level.
Using an aerial camera offers wide coverage for assessing the extent as well as the thickness of the oil spill.
Although aircraft and UAVs can be a solution, they lack persistence and steadiness in the air, especially for long oil sweeping missions. 

\begin{figure}[]
\centering
\includegraphics[scale=0.4]{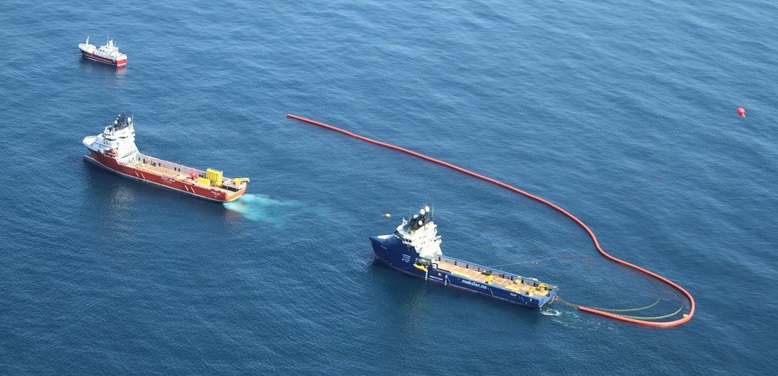}
\caption{NOFO using Helikite to detect and clean up oil spills \cite{NOFO}.}
\label{Fig.oilspill}
\end{figure}

To overcome these limitations, NOFO used a Helikite system called "The Ocean Eye" to detect oil spilling.
Helikites have the advantage of being small and compact; they can be easily handled, rapidly deployed, and can sustain harsh sea weather.
The Helikite can be anchored to the cleaning ship or nearby boats, as shown in Fig.\ref{Fig.oilspill}. The Helikite provides real-time video of the oil spill, which helps the cleaning boat locate accurate positioning and, thus, extract more oil in a shorter time from the sea.

\begin{figure}[h!]
\centering
\includegraphics[scale=0.13]{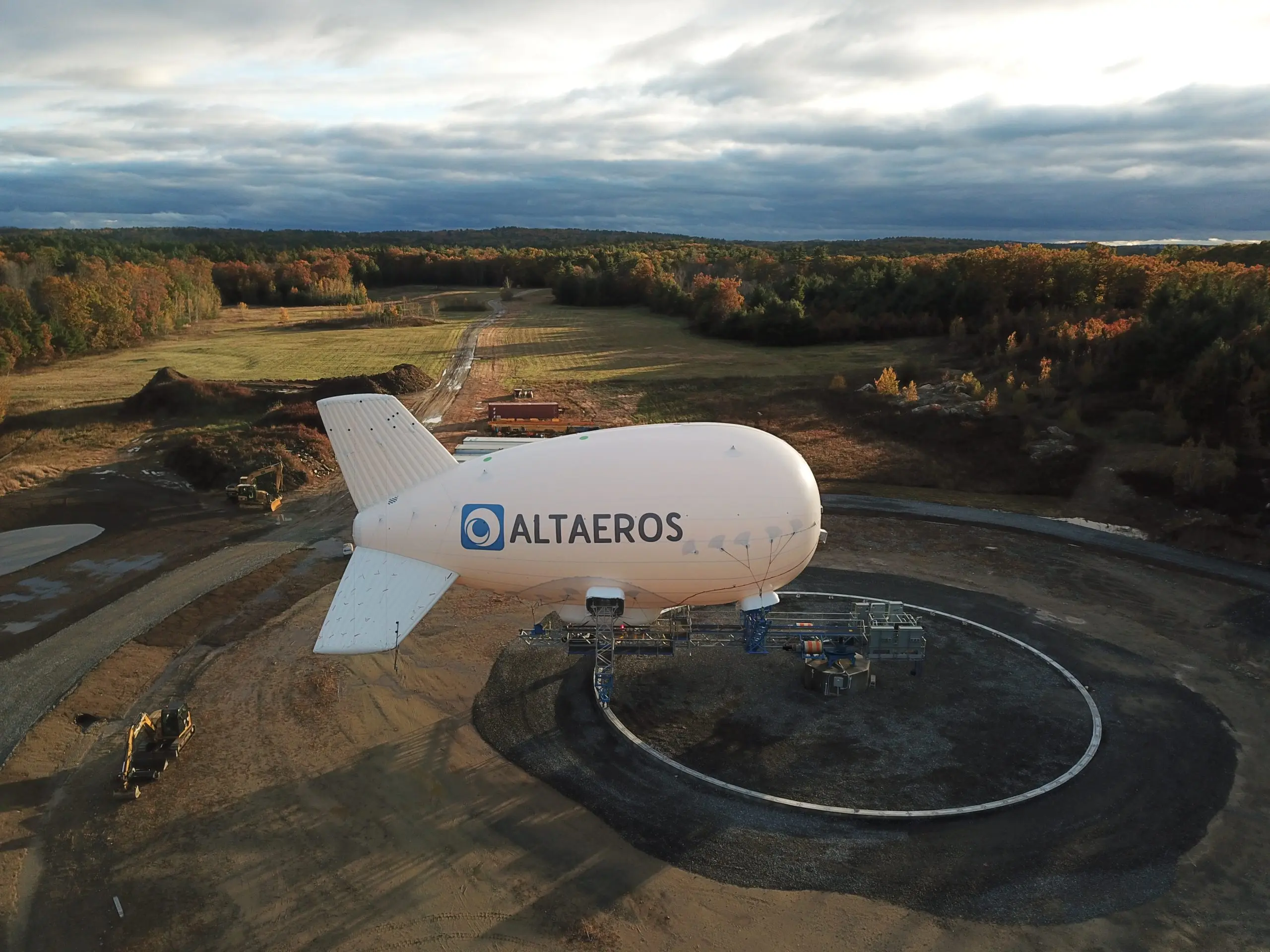}
\caption{Altaeros SuperTower setup \cite{Altaeroswebsite}.}
\label{Fig.Altaerosimage1}
\end{figure}

\subsubsection{Aerial Photographic Survey of Armarna, Egypt}

A Cambridge University research group wanted to conduct an archaeological photography of the ancient site Armarna in Egypt, where the famous Tutankhamen was born. They wanted to take still stereo-images of the site. They used Helikites equipped with two 35 mm single-lens reflex (SLR) cameras.
The Helikites provided ultra-sharp, still stereo-images of the whole area.

The Helikite is a low-cost solution, allowing to get closed to the site with low vibrations, and with a steady position to take still images. Additionally, the Helikite performed very well under the challenging hot weather conditions in Egypt.

\subsubsection{Aerostats All Australia (AAA) Mobile Coverage}

According to \cite{AAA}, nearly $70\%$ of Australia does not have mobile coverage.
To deal with that problem, a project named AAA has offered a plan to extend mobile coverage throughout Australia and the surrounding sea areas. In order to do so, AAA project proposes the use of NTFPs to bring wide coverage at a lower cost than other alternative solutions.
The AAA envisions mobile coverage with low latency for all mobile users in remote areas. In the short term (5 years), AAA proposes doubling Australian coverage from one-third to two-thirds of the total land areas. Over the long term, AAA aims to provide mobile coverage to all Australia.

\begin{table*}[ht!]
\begin{center}
    \caption{LTA NTFP companies. }
    \label{Company_LTA}
\begin{tabular}{ |p{0.9cm}||p{3.5cm}|p{3.3cm}| }
 \hline
Type & Company & Contact   \\
 \hline
Blimps & 
    • TCOM \cite{TCOM}\newline
    • Lindstrand Technologies \cite{LT} \newline
    • CNIM Air Space \cite{CNIM}  \newline
    • ADASI \cite{ADASI} \newline
    • Altaeros \cite{Altaeroswebsite}
&  
• NA \newline
• sales@lindstrandtech.com \newline
• +33 5 34 43 04 09  \newline
• NA \newline
• info@altaeros.com
\\ \hline
Balloons  &  
    • Vigilance \cite{Vigilance}   \newline
    • Drone Aviation Corp \cite{DroneAC}  \newline
    • SkyDoc \cite{SkyDoc} 
  & 
• +31 402 340 600 \newline
• info@droneaviationcorp.com \newline
• charlie@skydoc.com
\\  
 \hline
 BATs  &  
  • Altaeros \cite{Altaeroswebsite}
  & 
 • info@altaeros.com 
\\  
 \hline
\end{tabular}
\end{center}
\end{table*}

\begin{table*}[ht!]
\begin{center}
    \caption{HTA NTFP companies. }
    \label{Company_HTA}
\begin{tabular}{ |p{2.4cm}||p{4.7cm}|p{4.4cm}| }
 \hline
Type & Company & Contact   \\
 \hline
tUAVs & 
    • Elistair \cite{Elistairwebsite} \newline
    • Equinox Innovative Systems \cite{Equinox}  \newline
    • Tethered Drone Systems \cite{TDS} \newline
    • Hoverfly tech \cite{Hoverfly}  \newline
    • Drone Aviation Corp \cite{DroneAC} \newline
    • Fotokite Sigma \cite{Fotokite}
&  
• +33 9 83 57 06 39 \newline
• NA      \newline
• info@tethereddronesystems.co.uk  \newline
• info@hoverflytech.com  \newline
• info@droneaviationcorp.com \newline
• contact@fotokite.com
\\ \hline
Air Wind Turbines  &  
    • Makani \cite{Makaniwebsite}\newline
    • Airborne Wind Europe \cite{AWE}
  & 
• NA \newline
• +32 27396212

\\  
 \hline
\end{tabular}
\end{center}
\end{table*}

\begin{table*}[ht!]
\begin{center}
    \caption{Hybrid NTFP companies. }
    \label{Company_hyb}
\begin{tabular}{ |p{1.8cm}||p{3cm}|p{2.8cm}| }
 \hline
Type & Company & Contact   \\
 \hline
Helikite  & 
   •Allsopp Helikite \cite{Allsoppwebsite}
&  
• info@helikites.com

\\ \hline
Hybrid Airship  &  
    • Hybrid Air Vehicles \cite{HAVwebsite} \newline
    • Lockheed Martin \cite{Lockheedwebsite}
  & 
  • +44 (0) 1234 336400 \newline
  • NA
\\  
 \hline
\end{tabular}
\end{center}
\end{table*}

\begin{table}[ht!]
\begin{center}
    \caption{tUAV solution companies. }
    \label{Company_tUAV}
\begin{tabular}{ |p{2.4cm}||p{2.6cm}| }
 \hline
Company & Contact  \\
 \hline
Spooky Action \cite{spooky} & 
+ 952-649-1637
\\ \hline
AeroMana \cite{aeromana}  &  
info@aeromana.com
\\  
 \hline
\end{tabular}
\end{center}
\end{table}

\subsubsection{Altaeros SuperTower}

The Altaeros SuperTower is a solution proposed and developed by Altaeros to provide cellular coverage in rural areas (Fig.\ref{Fig.Altaerosimage1}). Standard infrastructure solutions have the disadvantage of being expensive and not lucrative in areas with few subscribers. The SuperTower is a tethered blimp that flies at an altitude of 240 m. The coverage gained by flying at this high altitude allows one SuperTower to replace 15 cell towers, reducing costs by 60\%. Hence, using this solution can accelerate the implementation of a mobile network more quickly and efficiently than standard cell towers, with significantly less cost \cite{6GSummit}.

\subsection{Companies Related to NTFPs}

Here we present the major companies that manufacture and sell NTFPs. Table \ref{Company_LTA} shows the companies that manufacture LTA platforms, i.e., blimps, balloons, and BATs. 

Table \ref{Company_HTA} shows the companies that manufacture HTA platforms, such as tUAVs and airwind turbines.

Finally, Table \ref{Company_hyb} shows the companies that manufacture hybrid platforms, such as Helikites and hybrid airship. We recall that hybrid airships have the possibility to be tethered, but are not systematically NTFPs.

For the sake of completeness, we present two companies that propose solutions for UAVs, namely, Spooky Action and AeroMana. These companies propose a tether configuration that can be plugged into existing UAVs without any modifications. This will give an additional freedom of choice to free-flying UAVs, by allowing them to be tethered when needed.

\section{NTFPs from a Wireless Communications Perspective}
\label{section_4}

\subsection{Geometric Analysis}

\begin{figure}[h!]
\centering
\includegraphics[scale=0.43]{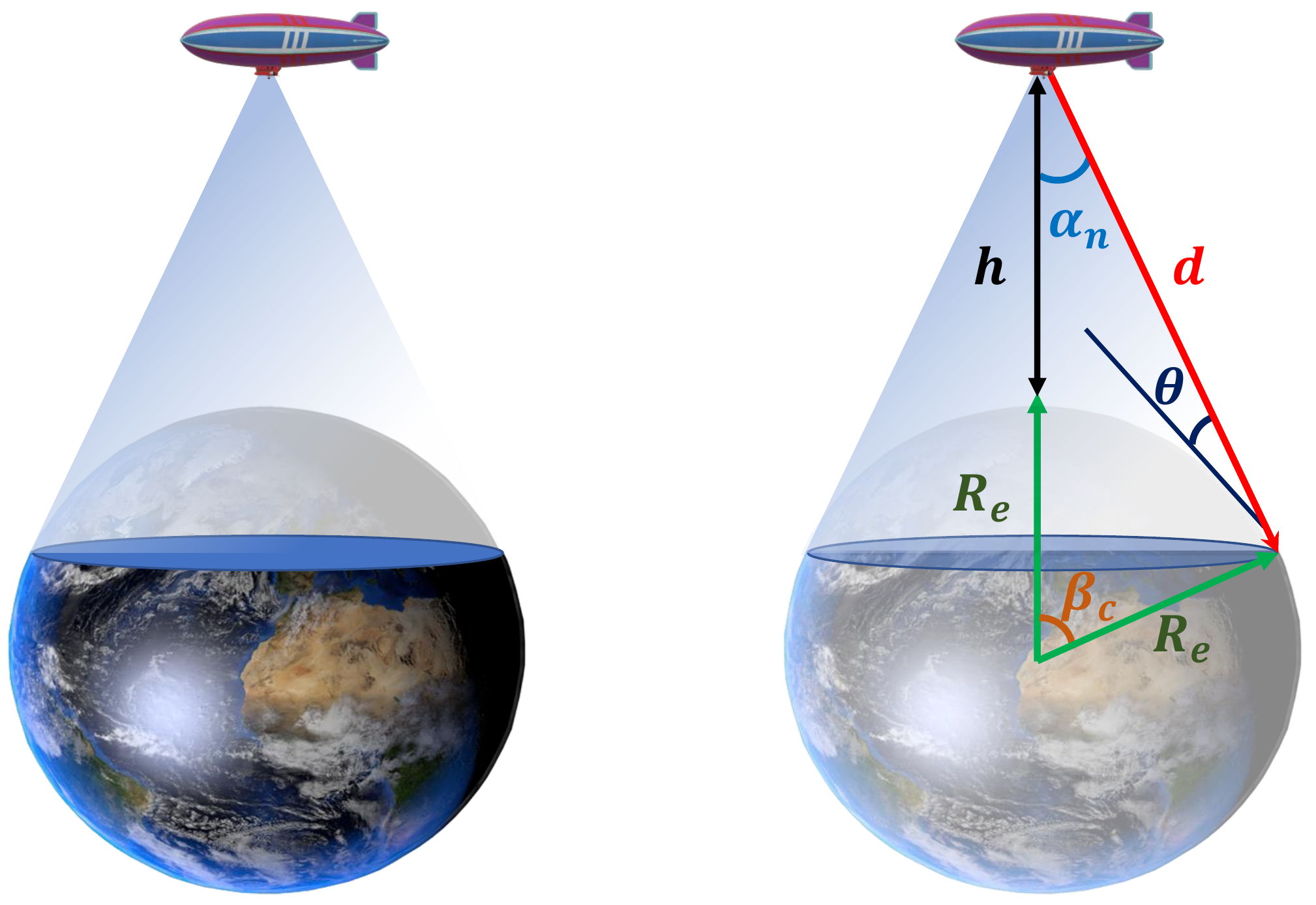}
\caption{ Coverage of NTFPs from a geometric perspective. }
\label{Cov_TP}
\end{figure}

In order to evaluate the performance of NTFPs, we have to investigate the geometrical aspect between a given NTFP (e.g., blimp) and the Earth. This can be carried out following an approach similar to that used for LEO satellites \cite{cakaj2011range,geyer2016earth}. Fig.\ref{Cov_TP} shows the geometrical aspect and the coverage surface between a blimp and a user located on the ground. In Fig.\ref{Cov_TP}, $R_e=6378$ km  denotes the Earth's radius at sea level, $h$ is the height of the blimp above the Earth, $d$ is the distance between the blimp and the ground user, also known as the slant range, $\alpha_n$ is the nadir angle, that is, the angle under which the blimp views the ground user,
$\beta_c$ is central angle, that is, the geocentric angle between the user and blimp nadir, and 
$\theta$ is elevation angle, that is, the angle between the slant range and the horizon plane.

Following the same approach as \cite{cakaj2011range,geyer2016earth}, we obtain the following equations:

\begin{equation}
    \alpha_n + \beta_c + \theta = \frac{\pi}{2},
    \label{eqa.1}
\end{equation}

\begin{equation}
    d \cos(\theta)=\left(h+R_e\right) \sin(\beta_c),
    \label{eqa.2}
\end{equation}
and
\begin{equation}
    d \sin(\alpha_n)=R_e \sin(\beta_c).
    \label{eqa.3}
\end{equation}

When distance $d$ is required, to compute path loss, we apply the law of cosines for the triangle in Fig.\ref{Cov_TP}, which yields 

\begin{equation}
    (h+R_e)^2=R_{e}^2+d^2+2R_ed\cos\left(\frac{\pi}{2}+\theta\right).
    \label{eqa.4}
\end{equation}

Solving equation (\ref{eqa.4}) with respect to $d$ yields the following solution:

\begin{equation}
d=R_e\Bigg[ \sqrt{\Big(\frac{R_e+h}{R_e}\Big)^2-\cos^2(\theta)}-\sin(\theta)\Bigg].
\label{eqa.5}
\end{equation}

Hence, $d$ reaches its maximum value when $\theta=0$, which is given by
\begin{equation}
d_{\max}=d \left( \theta=0 \right)=\sqrt{h^2+2hR_e}.
\label{eqa.6}
\end{equation}

Also, from applying the law of sines in Fig.\ref{Cov_TP}, we get

\begin{equation}
    \frac{\sin(\alpha_n)}{R_e}=\frac{\sin\left(\frac{\pi}{2}+\theta\right)}{R_e+h}
    \label{eqa.7}
\end{equation}
Then, we have
\begin{equation}
\sin(\alpha_n)=\frac{R_e\cos(\theta)}{R_e+h}.
\label{eqa.8}
\end{equation}
Finally, the angle $\alpha_n$ is given by
\begin{equation}
\alpha_n=\sin^{-1}\Bigg(\frac{R_e\cos(\theta)}{R_e+h}\Bigg).
\label{eqa.9}
\end{equation}

\noindent To calculate $\beta_c$, we use equation (\ref{eqa.2}):
\begin{equation}
\beta_c=\sin^{-1}\Bigg(\frac{d}{R_e+h}\cos(\theta)\Bigg).
\label{eqa.10}
\end{equation}

\noindent Also, we can use equations (\ref{eqa.1}) and (\ref{eqa.9}) to calculate $\beta_c$; hence,

\begin{equation}
\beta_c=\frac{\pi}{2}-\theta-\sin^{-1}\Bigg(\frac{R_e}{R_e+h}\cos(\theta)\Bigg).
\label{eqa.11}
\end{equation}

\noindent Finally, the surface coverage achieved by the blimp can be computed as follows:

\begin{equation}
S_{cov}=2\pi R_{e}^2 \left[1-\cos(\beta_c)\right].
\label{eqa.12}
\end{equation}

\begin{figure}[ht!]
\centering
   \begin{subfigure}[b]{0.5\textwidth}
        \includegraphics[scale=0.6]{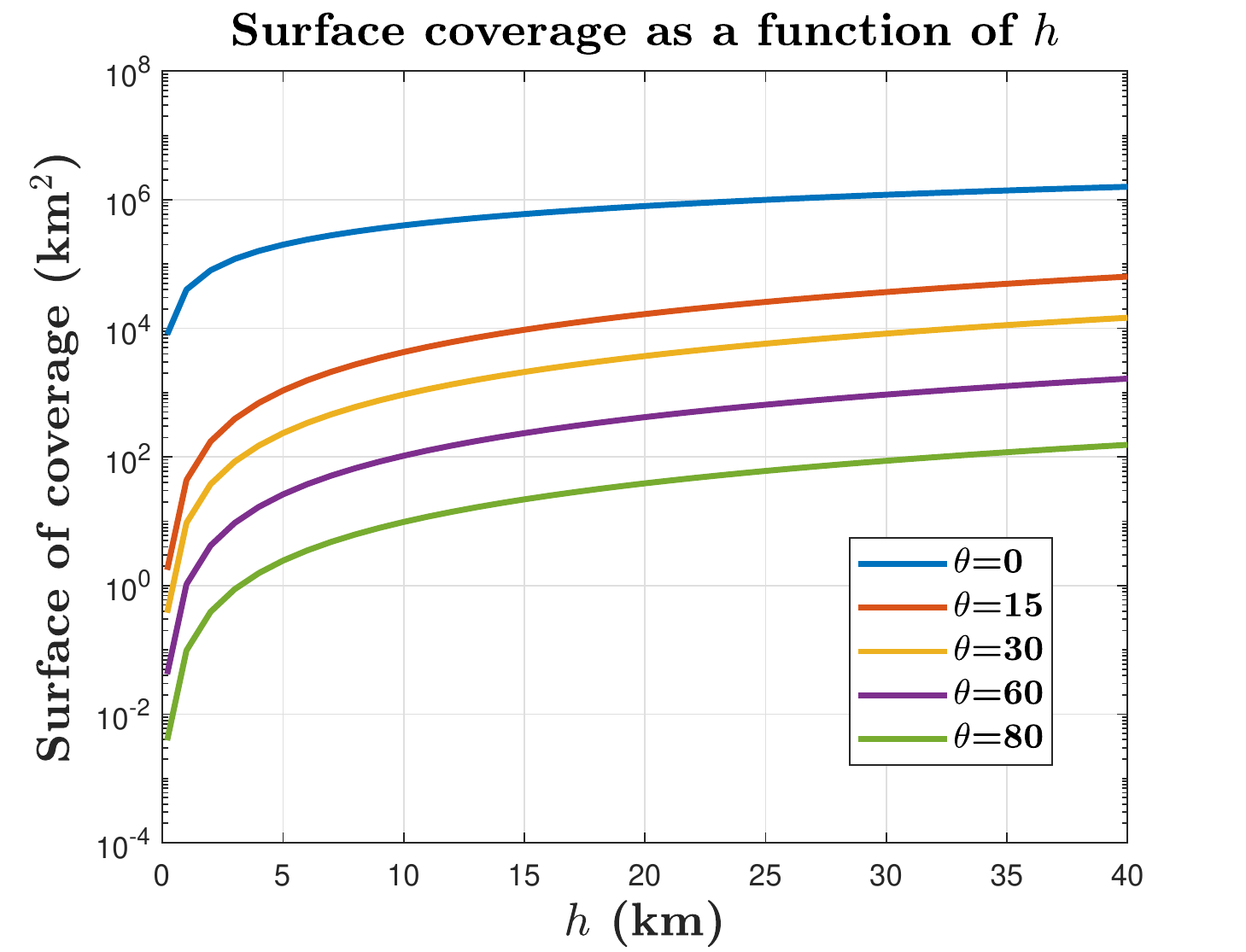}
        \caption{Surface coverage as a function of $h$ for several values of $\theta$.}
       \label{Fig.SC.t}  
\end{subfigure}
\begin{subfigure}[b]{0.5\textwidth}
        \includegraphics[scale=0.6]{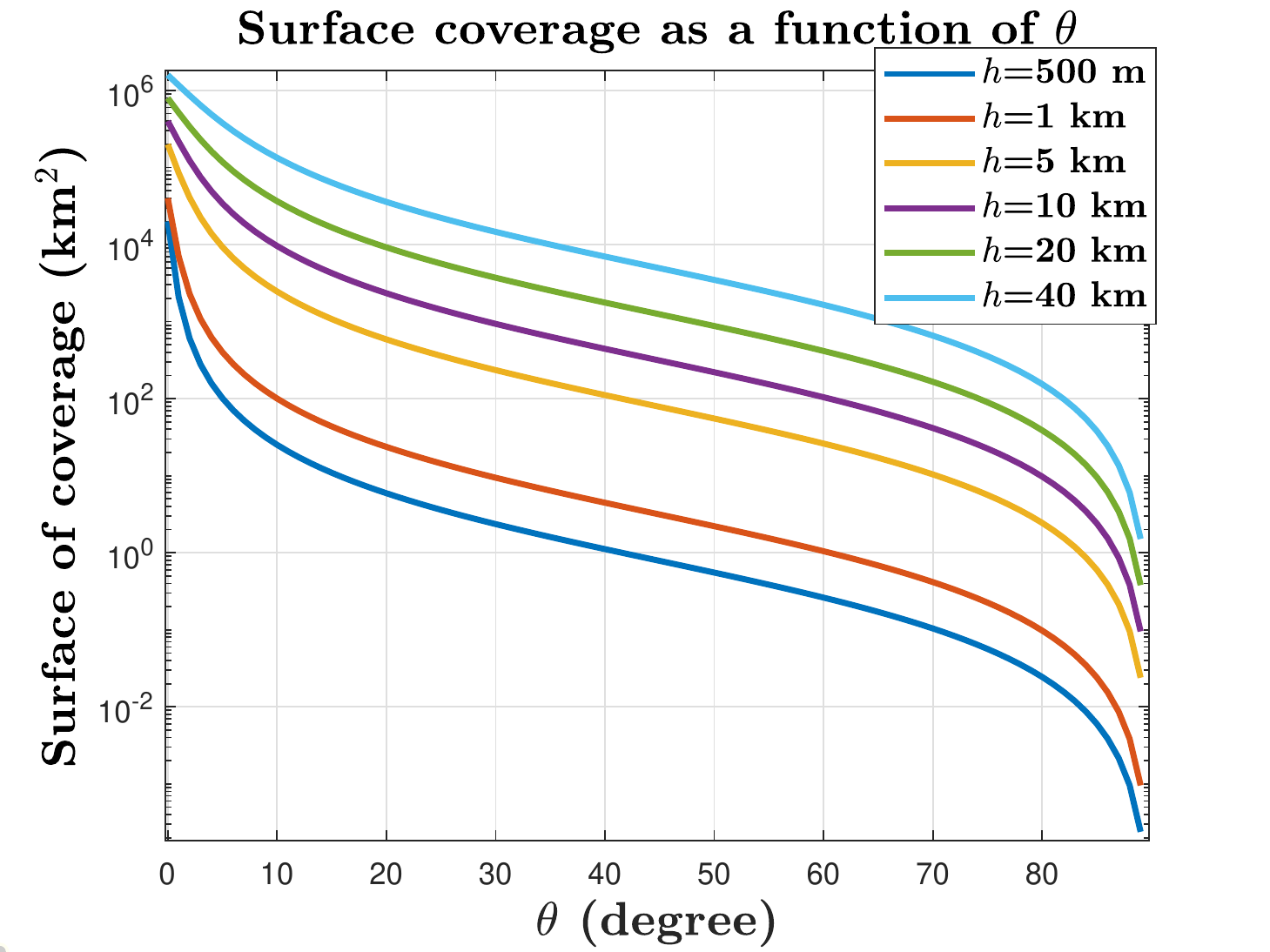}
        \caption{Surface coverage as a function of $\theta$ for several values of $h$.}
       \label{Fig.SC.h}  
\end{subfigure}
    \caption{Surface coverage as a function of $\theta$ and $h$.}
   \label{Fig.SC}  
\end{figure}

\begin{figure}[ht!]
\centering
   \begin{subfigure}[b]{0.5\textwidth}
        \includegraphics[scale=0.6]{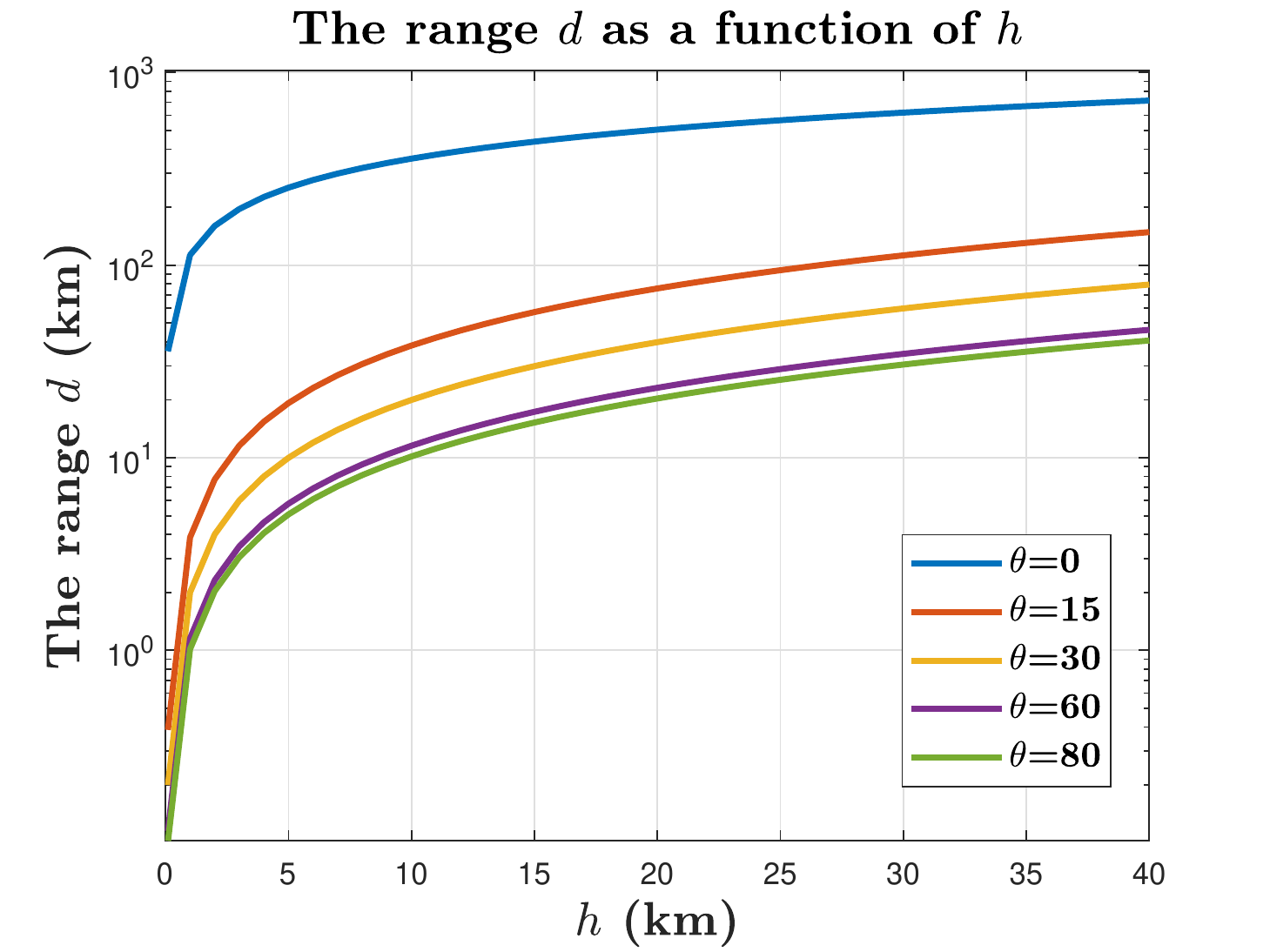}
        \caption{The range $d$ as a function of $h$ for several values of $\theta$.}
       \label{Fig.d.t}  
\end{subfigure}
\begin{subfigure}[b]{0.5\textwidth}
        \includegraphics[scale=0.6]{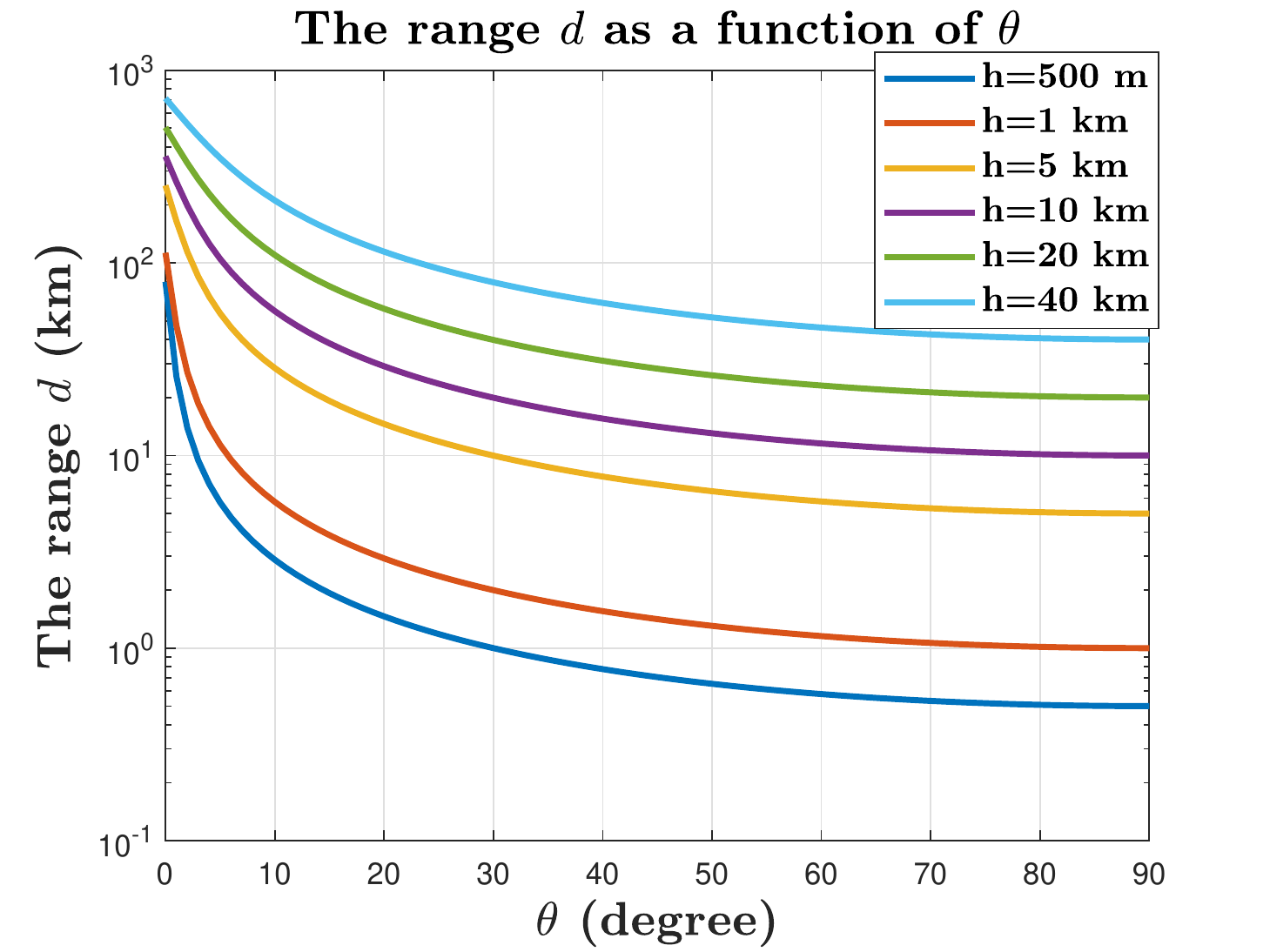}
        \caption{The range $d$ as a function of $\theta$ for several values of $h$.}
       \label{Fig.d.h}  
\end{subfigure}
    \caption{The range $d$ as a function of $\theta$ and $h$.  }
   \label{Fig.d}  
\end{figure}

To better assess the impact of $h$ and $\theta$ on different metrics, such as surface coverage and the rang $d$, we plotted different curves in Fig.\ref{Fig.SC} and  Fig.\ref{Fig.d}.

In Fig.\ref{Fig.SC.t}, we plotted the surface coverage as a function of $h$ for several values of $\theta$. We can see that, as the altitude of the blimp increases ($h$ increases), the surface coverage increases as well. This remark is intuitive, because as altitude of the blimp increases, the blimp can cover a greater surface. We can also notice that, the difference between the surface coverage when $h=100$ m and when $h=40$ km is four orders of magnitude.   

The range $d$ is plotted in Fig.\ref{Fig.d.h} as a function of the elevation angle $\theta$ for several values of $h$. We can see that the range $d$ decreases as the elevation increases. For instance, when $\theta=90$, the user is exactly below the blimps, which corresponds to the shortest distance possible between the blimp and the user. Inversely, when $\theta=0$, the user is the farthest from the blimp, which corresponds to $d_{\max}$.

\subsection{Performance Analysis}

Very few works have investigated the performance of NTFPs in wireless communications. Most focus on free-flying platforms, such as HAPs and UAVs. For example, in the absence of terrestrial infrastructure, UAVs are used to assist
cellular networks. But two major issues limiting the connection between UAVs and the core network are backhaul constraints and limited energy.

To address this issue, the work in \cite{li2018placement} proposed a multihop connection using several UAVs to alleviate the backhaul constraint. However, by increasing the hops, the latency increased and the spectral efficiency of communications was reduced.

\begin{figure}[h!]
\centering
\includegraphics[scale=0.27]{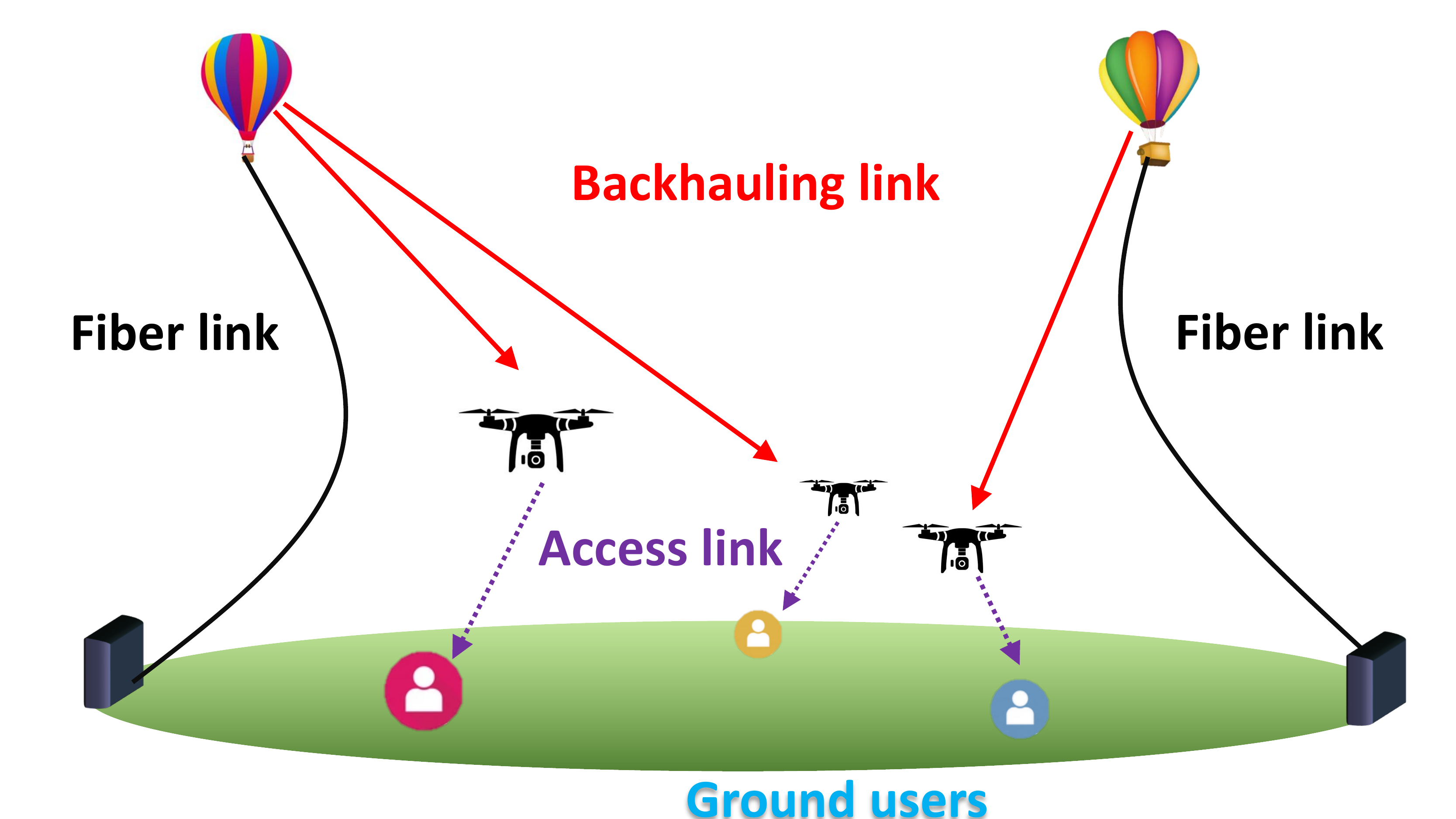}
\caption{Backhaul connection between UAVs and tethered balloons.}
\label{Fig.uav_tb}
\end{figure}

Hence, the authors in \cite{alzidaneen2019resource} proposed a configuration with a tethered balloon connected to the core network via fiber, as shown in Fig.\ref{Fig.uav_tb}. The tethered balloon acted as a “flying base station” located at higher altitude than the UAVs, which created a strong LOS backhaul connection. The proposed configuration showed an increase in the achievable end-to-end data rate of the users. Plus, they proposed a framework that optimized the transmit power, placement and association of the UAVs.

Regarding the use of tUAVs, the authors in \cite{selim2018post} proposed a hybrid solution to overcome these challenges. The solution consisted of three different types of UAVs: UAVs that acted as communication drones, tUAVs that provided a backhaul connection to the  communication drones via a RF/FSO
hybrid link \cite{zedini2016performance,alzenad2018fso}, and UAVs that powered the communication UAVs by providing on-the-fly battery charging. The authors showed that unlimited cellular communications could be provided while guaranteeing a minimum rate for all users.

Using tUAVs as airborne base stations has huge potential to extend network capacity and coverage for 6G \cite{kishk2020aerial}. In \cite{kishk20203}, the authors investigated the optimal placement of a tUAV tethered to a rooftop to minimize the path loss between the tUAV and a ground user; constraints included the limited length of the tether and the inclination angle of the tUAV for safety issues. The authors in \cite{bushnaq2020optimal} compared the performance of UAVs and tUAVs under heavy traffic conditions. Their results showed that tUAVs outperformed UAVs.

NTFPs can also be used to connect other flying platforms, such as future flying cars,  with the core network. Indeed, flying cars require a reliable aerial wireless communications network. The communications technologies currently used in vehicular communications \cite{belmekki2019cooperative,2020outagecopMW} are ill-suited for flying cars due to their lack of aerial coverage \cite{saeed2020wireless}.

Finally, NTFPs can be used as a relay between airborne platforms and ground stations. For instance, in \cite{sudheesh2017sum}, the authors considered communications between free-flying HAPs and ground-based stations. However, in this case, channel state information (CSI) were hard to obtain due to the high altitude and mobility of the HAPs. Thus, without CSI knowledge, the performance in terms of sum-rate was significantly degraded.
To overcome the absence of CSI and maximize the sum-rate, the authors proposed  an interference
alignment scheme that used a tethered balloon as a relay between the HAPs and the ground-based stations. The proposed scheme achieved the maximum sum-rate without CSI. Additionally, they showed that there is an optimal altitude for the tethered balloon that maximized the achievable sum-rate.

\subsection{NTFP Communications for 6G}

\begin{figure}[ht!]
\centering
\includegraphics[scale=0.35]{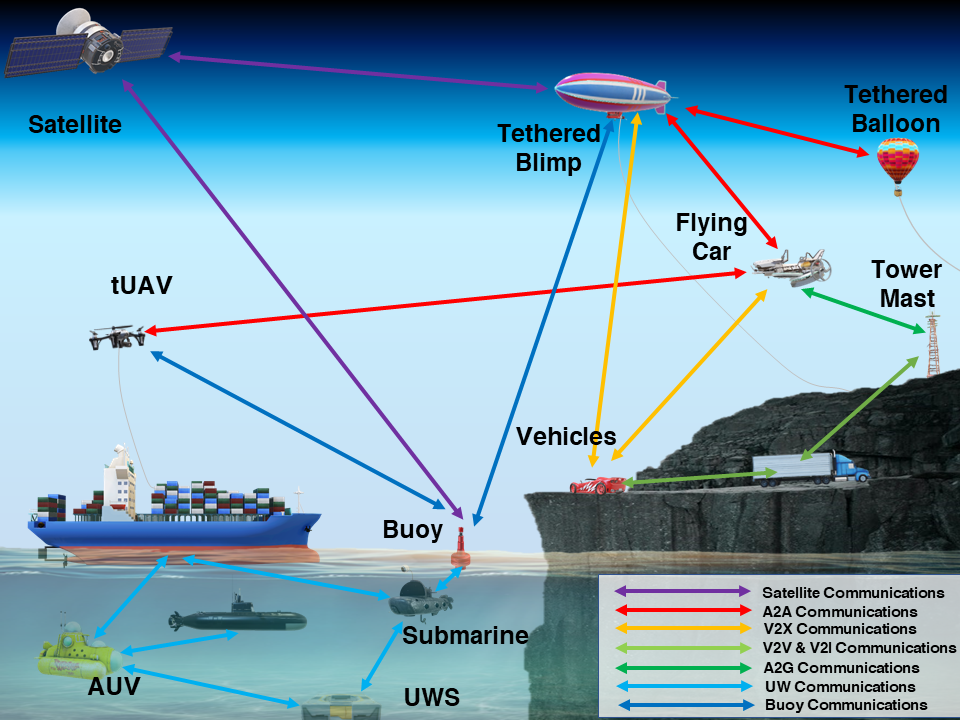} 
\caption{Various type of communications used by NTFPs.}
\label{exosystem}
\end{figure}

Now we show how NTFPs will be the key enablers to interconnect heterogeneous networks and communications for 6G. 
To this end, they will be the bridge between land, air, sea, and space by being connected to ground communications (e.g., cellular communications and vehicular communications), air communications (e.g., UAV communications and HAP communications), maritime communications (underwater communications, and remote-sea communications) and satellite communications, as shown in Fig.\ref{exosystem}.
One of key aspect of 6G is that everything will be connected. 
In that context, more than 4 billion do not have access to an internet connection, NTFPs will help bridge the digital divide and bring connection to rural and remote areas. 
They will also provide last mile connectivity, ensuring seamless connection to vehicles and flying cars, interconnecting airborne platforms, providing coverage remote sea areas, and serving as relay between satellite communications and ground/sea communications. 

\subsubsection{Remote communications and last mile connectivity}
To connect the unconnected, NTFPs are a cost-efficient solution to bridge the digital divide and provide high data rate connection to remote areas. They are used as aerial base stations with a huge backhaul capacity to provide last mile connectivity for hard-to-reach areas \cite{kishk2020aerial}. Additionally, this solution is ready to be used immediately, can be quickly deployed, and costly efficient.

\subsubsection{Vehicular communications}
With the upcoming rise of autonomous vehicles, the need for seamless connectivity, high reliability, and low latency communications is of utmost importance \cite{belmekki2020performance1,belmekki2020performance2}. Autonomous and driverless vehicles need to be constantly connected and aware of their surrounding. NTFPs allow these vehicles to be connected all the time with their wide mobile coverage and high data rate connection. They also guarantee seamless connectivity at the edge and under-connected areas.

\subsubsection{Flying cars}

Flying cars in general, and electric vertical take-off landing  (eVTOL) in particular, are expected to be part of our daily lives in the next decade and 6G envisions to serve the eVTOLs as of 2030 \cite{pan2021flying}. 
Flying cars need to be constantly connected, in order to avoid collisions, to acquire other vehicles' positions, or to change their itinerary  when necessary.
However, terrestrial networks, as they are now, are not suitable for eVTOL communications since flying car fly at an altitude of 300 m, which is way above tower masts. In that context, NTFPs are the perfect solution since they have great coverage and can provide aerial connectivity for eVTOLs \cite{saeed2020wireless}.

\subsubsection{Airborne communications}
Airborne platforms will be ubiquitous in 6G forming a three dimensional networks. Hence, they need to be connected with other platforms (such as vehicles), but also connected among themselves whether they are tethered or free-flying. NTFPs stay aloft for a prolonged period of time at the same position, which allows them to be used as relay to interconnect different airborne solutions flying at different altitudes, such as, HAPs and UAV \cite{saeed2020point}.

\subsubsection{Maritime communication}
Maritime economy and maritime activities have witnessed a significant growth with various activities and applications, such as fisheries, maritime
transportation, sea monitoring, deep sea mining, surveillance, and inspection missions (e.g., oil and gas facilities). 
However, as important as these  activities are, they still lack broadband communications capable
of withstanding the needs of theses activities and applications.
For instance, underwater missions rely heavily on a massive number of autonomous underwater
vehicles (AUVs), remote operated vehicles (ROVs), and underwater sensors (UWSs). These vehicles need to transmit large amounts of data among themselves and between the shore. Another example is that fishers or travelers need an internet connection in their cruise ship. NTFPs are a suitable solution to connect the sea (above and under the surface) with the shore by providing broadband connectivity and mobile coverage. They are also less costly compared to satellite solutions, and with lower latency \cite{campos2016bluecom+,teixeira2016tethered,teixeira2017enabling}.

\subsubsection{Satellite communications}
Finally, NTFPs help connecting satellite with ground platforms. In fact,
satellite communications suffer attenuation and delay. Hence, NTFPs can act as relay by
converting a long-range transmission using one single hop into a short-range transmission using multiple hops. 
This will alleviate the overall propagation delay and improve the data rate \cite{saeed2020point}.

\subsection{Economics of NTFPs}
One of the several advantages of NTFPs is that they are cost-efficient. Therefore, it is relevant to investigate the economic aspects of using NTFPs in a wireless communications context.
To this end, we present two analyses: the first one was conducted in the AAA project \cite{AAA}, and the second one was done by Altaeros for their SuperTowers \cite{Altaeroswebsite,6GSummit}. The analysis carried out in the AAA project provide a detailed deployment plan for 250 NTFPs over four years. The analysis done by Altaeros compared the cost of one single NTFP with several cell towers during their respective lifetime.

\subsubsection{AAA Analysis}
As mentioned in Section \ref{section_3}, the AAA project aimed to extend mobile coverage in Australia by using tethered aerostats, especially in remote areas \cite{AAA}. One of the main advantages of using tethered aerostats is that they cost less than alternative solutions.
We will show the costs of using tethered solutions from a CAPEX/OPEX perspective, as demonstrated in \cite{AAA}. Also, we review the different stages proposed in the AAA project and the level of support.

The AAA project proposed a three-stage deployment strategy spread over four years!
\begin{itemize}
    \item Stage 1: Aerostat center for excellence,
    \item Stage 2: Aerostat co-location with existing remote Australia cell towers, and
    \item Stage 3: Public Safety Agency points-of-presence (POPs) and point-to-point links.
\end{itemize}

\paragraph{Stage 1} The tasks and responsibilities of the aerostat center for excellence include designing and testing the aerostats, carrying out trials across Australia, designing radio access network (RAN), obtaining regulation approvals, and training the operators and the support personals, etc. For more details, we invite the reader to read the AAA report \cite{AAA}.

\paragraph{Stage 2}
The advantage behind co-locating tethered aerostats with an existing cell tower is that there are existing access roads, fiber optic backhaul and electricity, which, according to \cite{AAA}, reduce the cost of a tethered aerostat from \$2 million  to \$500k. However, they recommend some security measures such as the length of the tether has to respect a given distance in case the tethered aerostat cannot be taken down (for instance, due to winch malfunction), hence protecting the communications structure in its vicinity. Also, they recommend that the tethered aerostat have a dedicated radio spectrum in order to avoid interference with the already existing cell site.

\paragraph{Stage 3} New tethered aerostat sites will act as new Public Safety Agency POPs. They should be erected on the major Australian islands and external territories.
The sites on islands and territories that have high strategic importance should be erected jointly with the
defense department and border surveillance. In that context, larger tethered aerostats can carry heavier payloads. Therefore, they can carry both communication payloads as well as surveillance payloads, such cameras, sensors, etc., allowing the tethered aerostats to be used for both communications and surveillance.

\begin{table}[ht!]
  \begin{center}
    \caption{Total OPEX from year 1 to year 4.}
    \label{tab:OPEX}
    \begin{tabular}{l|S}
      \textbf{OPEX} & \textbf{Annual Cost} \\
      \toprule 
      L1 Support & \$ 25 M  \\
      L2 Support & \$ 15 M \\
      L3 and L4 Support & \$ 10 M \\
      \toprule 
       \textbf{Total L1 to L4 Support} & \$ 50 M \\
      \bottomrule
      Center for Aerostat Excellence & \$ 10 M \\
      Amortization & \$ 10 M   \\
      \toprule
      \textbf{OPEX} & \$ 75 M \\
      \bottomrule 
    \end{tabular}
  \end{center}
\end{table}

\begin{table*}[ht!]
  \begin{center}
    \caption{Combined CAPEX and OPEX for Year 1–4 }
    \label{tab:CAPEXOPEX}
    \begin{tabular}{l|l|l|S|S|S}
      \textbf{Year}  &  \textbf{Stage} & \textbf{Description} & \textbf{CAPEX} & \textbf{OPEX} & \textbf{Total}\\
      \toprule 
      Year 1  & Stage 1 & Setup center of excellence &  \$ 10 M &  \$ 10 M & \$ 20 M\\
      \midrule
      Year 2 & Stage 2 & 80 aerostats cell-sites co-located x \$500K  &  \$ 40 M  &  \$ 25 M & \$ 65 M \\
      \midrule
      Year 3 & Stage 2 + Stage 3 & 70 aerostats x \$500K + 50 aerostats x\$2M &  \$ 135 M  &  \$ 50 M & \$ 185 M\\
      \midrule
      Year 4  &  Stage 3 & 50 aerostats x \$2M &  \$ 100 M &  \$ 75 M & \$ 175 M\\
      \bottomrule 
      \textbf{Year 1-4} & All stages & Combined CAPEX and OPEX & \textbf{\$ 160 M} & \textbf{\$ 285 M} & \textbf{\$ 445 M}  \\
      \bottomrule 
    \end{tabular}
  \end{center}
\end{table*}
\begin{figure}[ht!]
\centering
\includegraphics[scale=0.25]{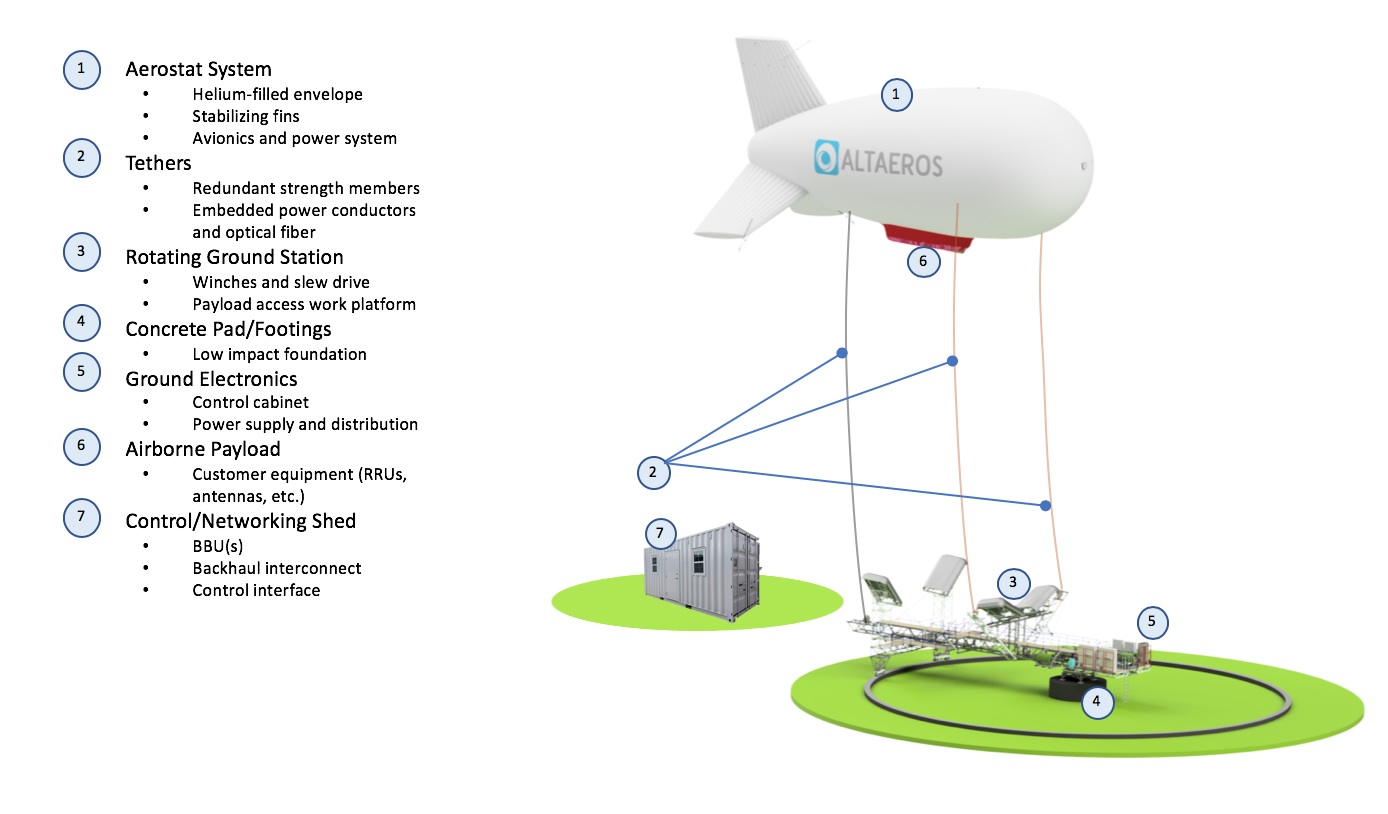}
\caption{Components Altaeros SuperTower System \cite{AltaerosST}.}
\label{Fig.ssystem3}
\end{figure}

\begin{table}[h!]
  \begin{center}
    \caption{CAPEX/OPEX analysis between cell towers and a single NTFP.}
    \label{table_compa}
    \begin{tabular}{l|S|S}
      \textbf{} & \textbf{16 x } & \textbf{1 x}\\
      \textbf{} & \textbf{Cell Towers} & \textbf{SuperTower}\\
      \toprule 
      RAN/Site Setup & \$ 1280 K & \$ 105 K\\
      Backhaul Setup & \$ 560 K & \$ 35 K \$\\
      \midrule
      \textbf{CAPEX} & \$ 1840 K & \$ 140 K\\
      \midrule
      Tower Rent & \$ 240 K & \$ 400 K\\
      Site Operations \& Maintenance & \$ 112 K & \$ 10 K\\
      Backhaul Service & \$ 256 K & \$ 60 K\\
      Power \& Other & \$ 128 K & \$ 25 K\\
      \midrule
      \textbf{OPEX} & \$ 736 K & \$ 495 K\\
      \bottomrule 
      \textbf{NPV during their Lifetime} & \$ 9.6 M & \$ 4.7 M\\
      \bottomrule 
    \end{tabular}
  \end{center}
\end{table}

They also plan to offer four levels of support, depending on the task and requirements:

\begin{itemize}
    \item Level 1: Weekly on-site inspections and recovery of aerostats before/after cyclones
    \item Level 2: Support for monthly service and setting up new aerostat sites
    \item Level 3: Major service, technology upgrades, and major repair performed annually
    \item Level 4: Joint collaboration with vendors requiring expertise
\end{itemize}

\paragraph{Level 1 - Support to Inspect and Protect Aerostat} 
Each site requires two people, called the L1 crew, for a weekly inspection that includes winching the aerostat down, visually inspecting it, applying the necessary adjustments, and then winching it back. These weekly inspections extend the operation life of an aerostat to ten years. Although tethered aerostats perform better in adverse weather than other airborne solutions, such as HAPs, aircraft, UAVs, and free-flying platforms, the L1 crew have to winch down the aerostat when there are cyclones and strong winds, check if there are any damages, apply the necessary repairs if needed, and winch the aerostat up. The L1 crew may be regular contractors or trusted locals. While the aerostat is winched down, the L1 crew will have access to satellite communication.

\paragraph{Level 2 - Monthly Service and Setup of New Aerostats} 
Ten L2 crews of two people each will provide basic repairs and refill the aetostats with helium. They will also erect new aerostat sites, train the L1 crews, and prepare the aerostats for the L3 crews.

\paragraph{Level 3 - Support for Aerostat Annual Overhaul and Technology Upgrade} 
The L3 crew offers necessary technological and physical upgrades and an annual overhaul. They also perform repairs and on-demand facilities.

\paragraph{Level 4 - Expert Support} The L4 expert support will provided by vendors and the Australian Center for Aerostat Excellence.\\

In total, the cost estimates provided by \cite{AAA} for 250 aerostats are \$25 million per annum (PA) L1 support,
\$15 million PA for L2 support, and \$10 million PA for L3 plus L4 support. Hence, the total cost of all support is \$50 million PA.
The Australian Center for Aerostat Excellence requires a budget of \$10 million PA for administration fees and for research and development. 
A further amortization budget of \$10 million is dedicated to refurbishing the tethered aerostats every five to ten, which is 20\% PA of the total support budget. The total OPEX is shown in Table \ref{tab:OPEX}, and the total CAPEX and OPEX for year 1 through year 4 (until a full fleet operation is completed) is shown in Table \ref{tab:CAPEXOPEX}.

It is noted in \cite{AAA} that the funding allocated to the AAA project would save the government and National Broadband Network \$1 billion in a Sky Muster satellite which is twice the cost of the AAA project. Also, the AAA bandwidth is 250 Gbps, whereas the satellite bandwidth is only 135 Gbps, that is double the aggregate bandwidth \cite{SNH}.

\begin{figure}[ht!]
\centering
\includegraphics[scale=0.4]{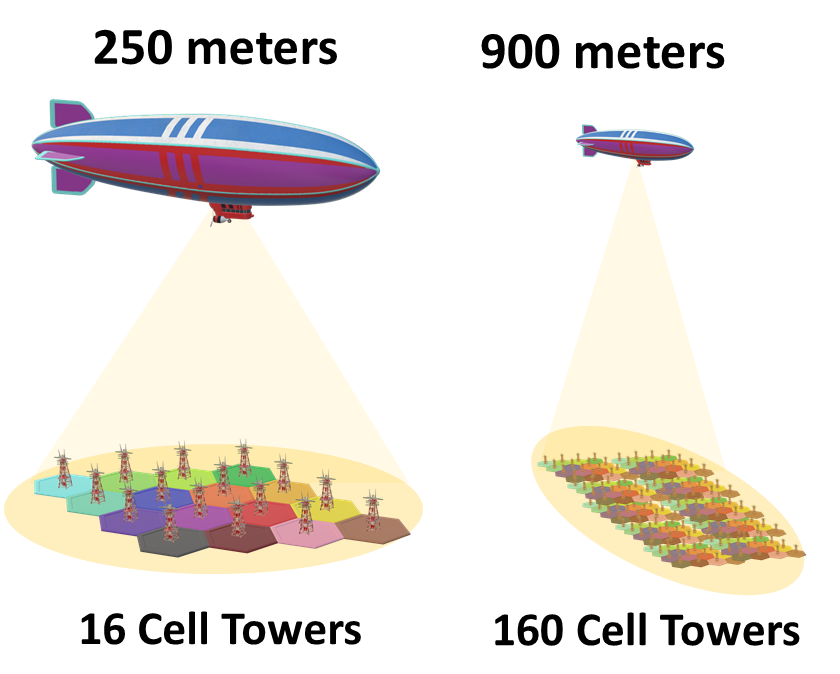}
\caption{Comparison between the coverage of NTFPs and cell towers.}
\label{cov}
\end{figure}

\subsubsection{Altaeros SuperTower Analysis}
The SuperTower made by Altaeros is a high capacity and long-endurance NTFP, that delivers data and provides coverage \cite{Altaeroswebsite}. The components of the Altaeros SuperTower are shown in Fig.\ref{Fig.ssystem3}. Altaeros conducted a cost analysis with CAPEX/OPEX comparisons between NTFPs (SuperTowers) and cell towers \cite{6GSummit}. The aim of this comparison was to investigate whether NTFPs are cost-effective compared to cell towers over their lifetime. 
A single cell tower is 40--60 m tall with a coverage radius of 10--15 km. This means that the tower has a service area ranging between 300--700 km$^2$. On the other hand, a NTFP that flies at a low altitude, for instance 250 m, has radius of 40--60 km with a service area ranging between 5,000--10,000 km$^2$. Hence, a single NTFP has a coverage equivalent of 16 cell towers, as shown in Fig.\ref{cov}. Table \ref{table_compa} shows a CAPEX/OPEX analysis between 16 cell towers and a single NTFP and their Net Present Value (NPV) during their lifetime. We can see from Table \ref{table_compa} that one NTFP flying at an altitude of 250 m is equivalent to 16 cell towers in terms of coverage at 50\% of the cost. Furthermore, we can see that the NTFP has 90\% lower CAPEX and 30\% lower OPEX compared to 16 cell towers. This analysis considers that the NTFP flies at altitude of 250 m. If we consider that the NTFP flies at 900 m such as the platforms described in the AAA project, the coverage of the NTFP will be the equivalent of 160 cell towers in terms of coverage (as depicted in Fig.\ref{cov}).

\section{Channel Modeling of NTFPs}\label{section_5}

In this section, we provide a comprehensive channel modeling framework for NTFPs. Furthermore, for this section, the term LAP refers to all the platforms that fly at ultra low-altitudes (50 m--150 m), low-altitudes (200 m--600 m), and medium-altitudes (0.7 km--5 km).
Also, we do not include channel modeling for U-HAP since there is only one paper that considers the feasibility of NTFPs flying at such altitudes \cite{izet2011low}.
For the sake of completeness, we consider both RF links and FSO links. Indeed, FSO links can provide a high speed connection via air-to-air (A2A) communications between NTFPs at different altitudes, and with ground stations or ground users via air-to-ground (A2G) communications \cite{mynaric}. Finally, at the end of this section, we summarize all channel models in Table \ref{table_channel} with the relevant references.

\subsection{LAPs Channel Modeling using RF Links}
For LAPs channel modeling considering RF links, we present both large-scale fading and small-scale fading.

\begin{figure}[]
\centering
\includegraphics[scale=0.25]{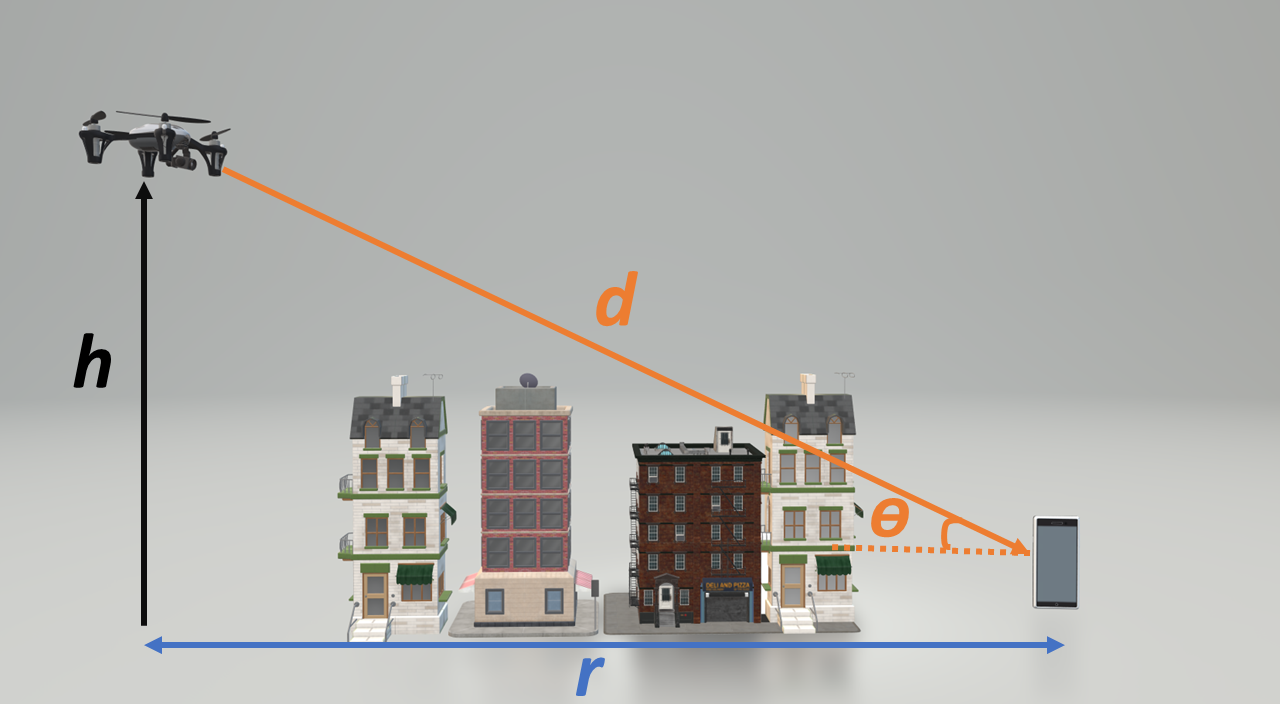}
\caption{Connection between a LAP and a mobile user in an urban environment.}
\label{los_diagrame}
\end{figure}

\subsubsection{Large Scale Path Loss}
The International Telecommunication Union (ITU) proposed a model for deriving 
LOS probability in an urban environment between a transmitter at $h_{TX}$
and a receiver at elevation $h_{RX}$ \cite{data2003prediction}. 
The LOS probability is a function of the following parameters:

\begin{itemize}
    \item $p_\alpha$: the ratio of the built-up land area to
the total land area,
\item $p_\beta$: the mean number of buildings per km$^2$,

\item $p_\gamma$: which is a scale parameter that models the buildings’
heights. The distribution of the buildings’
heights follows a Rayleigh distribution given by 
$f(H) = \frac{H}{\gamma^2}  \exp\Big(\frac{H^2}{2\gamma^2}\Big) $,
where $H$ is the building height in meters. 
\end{itemize}

Hence, according to \cite{data2003prediction}, the LOS probability denoted by $ \mathrm{P}(\mathrm{LOS})$ is expressed as

\begin{equation}\label{equ1}
    \mathrm{P}(\mathrm{LOS})=\prod_{n=0}^m 
    \left[ 
            1-\exp\left(- \frac{{\left[h_{TX}-
            \frac{-(n+\frac{1}{2})( h_{TX}-h_{RX} )}{m+1}
             \right]}^2}{2\gamma^2} \right)   
                                                   \right],
\end{equation}
where $m=\mathrm{floor}(r\sqrt{p_\alpha p_\beta -1})$ and $r$ is the ground projection of the distance between the transmitter and the receiver as shown in Fig.\ref{los_diagrame}.
In the case of LAPs, $h_{RX}$ can be neglected since the height of the receiver is much lower than the heights of buildings and the LAP altitude.
In \cite{al2014optimal}, the authors showed that (\ref{equ1}) can be expressed as a function of $\theta$ and the environment parameters as follows:

\begin{equation}
    \mathrm{P}(\mathrm{LOS},\theta)=
    \frac{1}{1+a_1\exp(-b_1(\theta-a_1))},
\end{equation}
where the parameters $a_1$ and $b_1$ are a function of the environment variables $p_\alpha$, $p_\beta$, and $p_\gamma$. We note that the non-line-of-sight (NLOS) probability equals $\mathrm{P}(\mathrm{NLOS},\theta)=1-\mathrm{P}(\mathrm{LOS},\theta)$.

The expression of path loss is then expressed as 

\begin{align}
&\mathrm{PL} (\alpha_{\mathrm{PL}},\theta,d)= \nonumber\\
&\left[\mathrm{Loss}_{\mathrm{LOS}} \cdot \mathrm{P}(\mathrm{LOS},\theta)+\mathrm{Loss}_{\mathrm{NLOS}} \cdot \mathrm{P}(\mathrm{NLOS},\theta) \right]\:d^{-\alpha_{\mathrm{PL}}(\theta)}, \nonumber
\end{align}
where $\mathrm{Loss}_{\mathrm{LOS}}$ and $\mathrm{Loss}_{\mathrm{NLOS}}$ denote
the mean additional losses for the LOS and NLOS transmissions, respectively, $d$ is the distance between the transmitter and the receiver, and $\alpha_{\mathrm{PL}}(\cdot)$ is the path loss exponent.

The path loss exponent is a function of the density of buildings and an obstacle between the transmitter and the receiver. For instance, larger values of $\alpha_{\mathrm{PL}}$ are assumed in dense urban areas, whereas lower values of $\alpha_{\mathrm{PL}}$ are assumed for rural areas. Hence, the path loss exponent $\alpha_{\mathrm{PL}}(\theta)$ can be modeled as a function of $\mathrm{P}(\mathrm{LOS},\theta))$, which, in turn, is a function of $\theta$.
Consequently, the path loss exponent is defined  as \cite{azari2017ultra}

\begin{equation}
    \alpha_{\mathrm{PL}}(\theta)=a_2 \mathrm{P}(\mathrm{LOS},\theta) + b_2,
\end{equation}
where the expression of $a_2$ and $b_2$ are given by
\begin{equation}
a_2 = \frac{\alpha_{\frac{\pi}{2}}-\alpha_0}
           {\mathrm{P}(\mathrm{LOS},\frac{\pi}{2})- \mathrm{P}(\mathrm{LOS},0)}    \cong, \alpha_{\frac{\pi}{2}}-\alpha_0
\end{equation}
and
\begin{equation}
    b_2 = a_0 - a_2 \mathrm{P}(\mathrm{LOS},0) \cong a_0,
\end{equation}
where $\alpha_{\frac{\pi}{2}}$ and $\alpha_0$ are the path loss exponent values when $\theta=\pi/2$ and $\theta=0$, respectively. The values of $\alpha_{\frac{\pi}{2}}$ are usually smaller since the transmitter is in LOS with the receiver, whereas the values of $\alpha_0$ are usually larger since there are more obstacles when $\theta=0$.
Also, $\mathrm{P}(\mathrm{LOS},0)\rightarrow 0$ and $\mathrm{P}(\mathrm{LOS},\frac{\pi}{2})\rightarrow  1$.
For more comprehensive path loss models of LAPs, we direct the readers to reference \cite{khuwaja2018survey}.

\subsubsection{Small Scale Fading}
\paragraph{Rayleigh distribution}
This distribution is widely and extensively used in the literature. It is used to model a transmission when the path between the transmitter and the receiver is heavily obstructed, and when LOS is not available. 
It can be used when the cooperative transmission between LAPs is considered \cite{abualhaol2010performance}.
Also, it has been shown in \cite{newhall2003wideband} that the fading model in urban environments with large elevation angles follows the Rayleigh distribution. We recall the Rayleigh distribution is given by
\begin{equation}
    f(y)=\frac{y}{\sigma^2}\exp\left(\frac{-y^2}{2\sigma^2}\right).
\end{equation}

\paragraph{Rician distribution}

Rician distribution is used to model fading of transmission with LOS for LAPs \cite{cai2017low,ye2017air,goddemeier2015investigation,azari2017ultra}. We recall the Rician distribution is given by
\begin{equation}
    f(y)=\frac{y}{\sigma^2}
    e^{\frac{-(y^2 + a_{\textrm{LOS}}^2)}{2\sigma^2}} 
    I_{0} \left( \frac{y\:a_{\textrm{LOS}}}{\sigma^2}\right),
\end{equation}
where $y \geq 0$, $a_{\textrm{LOS}}$ and $\sigma$ denotes the magnitude of the LOS and the diffuse multipath components, respectively, and $I_0(\cdot)$ is the zeroth–order modified Bessel function.

Also, the Rician parameter $K$ is defined as
\begin{equation}
    K=\frac{a_{\textrm{LOS}}^2}{2\sigma^2}.
\end{equation}

In Rician fading, the Rician factor, commonly noted by $K$, is a parameter that measures the severity of the 
fading. For instance, $K = 5.29$ dB when the LAP is in the takeoff and landing phase, whereas $K = 19.14$ dB when the LAP flies at an altitude of 20--30 m o\cite{cai2017low}. In \cite{ye2017air}, the authors proposed a piece-wise model of $K$ as a function of the altitude. 
The authors in \cite{azari2017ultra} proposed a model of $K$ as a function of the elevation angle: 

\begin{equation}
    K(\theta)=a_3 \cdot \exp(b_3\theta),
\end{equation}

where 
\begin{equation}
    a_3=k_0\:, \:\:\:b_3=\frac{2}{\pi}\ln\left(\frac{k_{\frac{\pi}{2}}}{k_0}\right),
\end{equation}
where $k_{\frac{\pi}{2}}$ and $k_0$ are the values of $K$ at $\theta=\frac{\pi}{2}$ and $\theta=0$, respectively.
Hence, larger values of $\theta$ lead to higher values of $K$, which characterizes fewer multipath scatters. In contrast low values of $\theta$ lead to low values of $K$, which characterizes severe multipath scatters \cite{azari2017ultra}.
Also, the A2A channel for LAPs can also be modeled as a Rician fading distribution \cite{goddemeier2015investigation}.

\paragraph{Nakagami-$m$ distribution}
Nakagami-$m$ distribution offers great flexibility to model LAP channels, since it embodies several other different distributions thanks to its Nakagami shape and spread-controlling parameters defined by $m$ and $\Omega$, respectively \cite{abualhaol2010performance,frew2008airborne}. 
The Nakagami-$m$ distribution is given by
\begin{equation}
f(y)=\frac{2m^m}{\Gamma(m) \Omega^m}y^{2m-1}e^{\frac{-my^2}{\Omega}},\nonumber
\end{equation}
where  $\Gamma(\cdot)$ denotes the Gamma function.
The authors in \cite{yanmaz2013achieving} showed that the Nakagami–$m$ distribution fits the empirical measurement better than the Rayleigh distribution.

\paragraph{Loo Model (Rice+Log-Normal) distribution} Loo Model is composed of the Rician and log–normal distributions. 
The authors in \cite{simunek2013uav} showed that the Loo model fits the empirical data of A2G narrowband channels in urban areas. The Loo model is given by

\begin{align}
    &f(y)=\frac{y}{\sigma^2 \sqrt{2\pi\Sigma^2_A}} \times \nonumber\\
    & \int^{\infty}_{a_{\textrm{LOS}}=0} \frac{1}{a_{\textrm{LOS}}}
    e^{\frac{-(20\log a_{\textrm{LOS}} - M_A)^2}{2\Sigma^2_A}  } 
    e^{\frac{-(y^2 + a_{\textrm{LOS}}^2)}{2\sigma^2}} 
    I_{0} \left( \frac{y\:a_{\textrm{LOS}}}{\sigma^2} \right) d a_{\textrm{LOS}},
\end{align}
where $M_A$ and $\Sigma$ denote the mean and standard deviation of the Gaussian distribution for the direct LOS signal, respectively, and  $a_{\textrm{LOS}}$ is the amplitude of LOS signal.

\subsection{HAPs Channel Modeling using RF Links}
For HAPs channel modeling considering RF links, we present both large-scale fading and small-scale fading.
\subsubsection{Large-Scale Path Loss}
The large-scale fading of a communication between a HAP and a given receiver located on the ground is mainly caused by free-space path loss. The path loss equation is given by

\begin{equation}
    \mathrm{PL}_{\mathrm{Free\:Space}}= {\left(\frac{c}{4\pi d f} \right) }^2,
\end{equation}

\noindent where $c$ is the speed of light and $f$ is the operating frequency. We recall that $d$ is the distance between the transmitter and the receiver.

\subsubsection{Small-Scale Fading}

\paragraph{Rayleigh}
Although the Rayleigh distribution is not often used to characterize small-scale fading for HAPs, some works have considered Rayleigh fading for HAPs \cite{karapantazis2005broadband, cuevas2004channel}. For instance, the authors in \cite{cuevas2004channel} used a two-states channel model with the assumption that when the channel is considered "good", the Rician model is used to model the fading, whereas when the channel is considered "bad", the Rayleigh model is used.

\paragraph{Rician}
The Rician model is commonly used in the literature to model small-scale fading for HAPs \cite{cuevas2004channel,saeed2020wireless,karapantazis2005broadband}.

\paragraph{Loo Model}
In \cite{zhao2020ka}, the authors provided a multi-states model considering a Ka-band channel for HAPs.
First, they modeled the effect of tropospheric weather on the amplitude of the transmitted signal as a Gaussian distribution defined by
\begin{equation}
    f_{w,r}(r)= \frac{1}{\sqrt{2\pi \sigma_{w,r}^2}} 
    \exp\left( -\frac{ \left( r-m_{w,r} \right)^2}{2 \sigma_{w,r}^2}   \right),
\end{equation}
where $m_{w,r}$ and $\sigma^2_{w,r}$ respectively denote the mean and variance of the Gaussian distribution of amplitude.

Then, to model impairment related to the ground environment, they provided a three-states model, with $S\in\{s_1, s_2, s_3\}$ defining the different states. The first state, $s_1$, describes a LOS state, the second one, $s_2$, describes a moderate shadowing state, and the last one, $s_3$, describes a deep shadowing state.
The Loo model for these three states is given by

\begin{align}
    f_{S,r}(r)=& \frac{8.686}{\sigma_S^2 \Sigma_{dB,S} \sqrt{2\pi}}
    \nonumber\\ &\int_{0}^{\infty} \frac{1}{a_{\textrm{LOS}}} 
    \:\exp\left( - \frac{ \left(20\log(a_{\textrm{LOS}})-M_{dB,S} \right)^2 }{2\Sigma_{dB,S}^2}   \right)\nonumber\\
    &\times\:\exp\left(- \frac{r^2 + a_{\textrm{LOS}}^2}{2\sigma_{S}^2} \right)
    \:I_0 \left(\frac{r a_{\textrm{LOS}}}{\sigma_{S}^2} \right) d a_{\textrm{LOS}},
\end{align}

\noindent where $a_{\textrm{LOS}}$ is the amplitude of LOS signal, and
$M_{dB,S}$ and $\Sigma_{dB,S}$ denote the mean and standard deviation, respectively, of the
log-normal distribution for the state $S$.

\subsection{Channel Modeling using FSO Links}
An FSO signal undergoes different types of fluctuation and attenuation due to atmospheric turbulence, pointing errors, path loss attenuation, and atmospheric attenuation \cite{trichili2020roadmap}.
Atmospheric turbulence is caused by random fluctuations of the refractive index, causing,  in turn, fluctuation in the intensity and phase of the received signal.
Pointing errors arise from the misalignment of the transmitter and the receiver. Finally, path loss and atmospheric attenuation depend on the atmospheric and weather conditions, such as, fog, rain, snow, and dust.

\subsubsection{Atmospheric Turbulence}
\begin{figure*}
        \begin{subfigure}[b]{0.25\textwidth}
                \includegraphics[height=4cm,width=5cm]{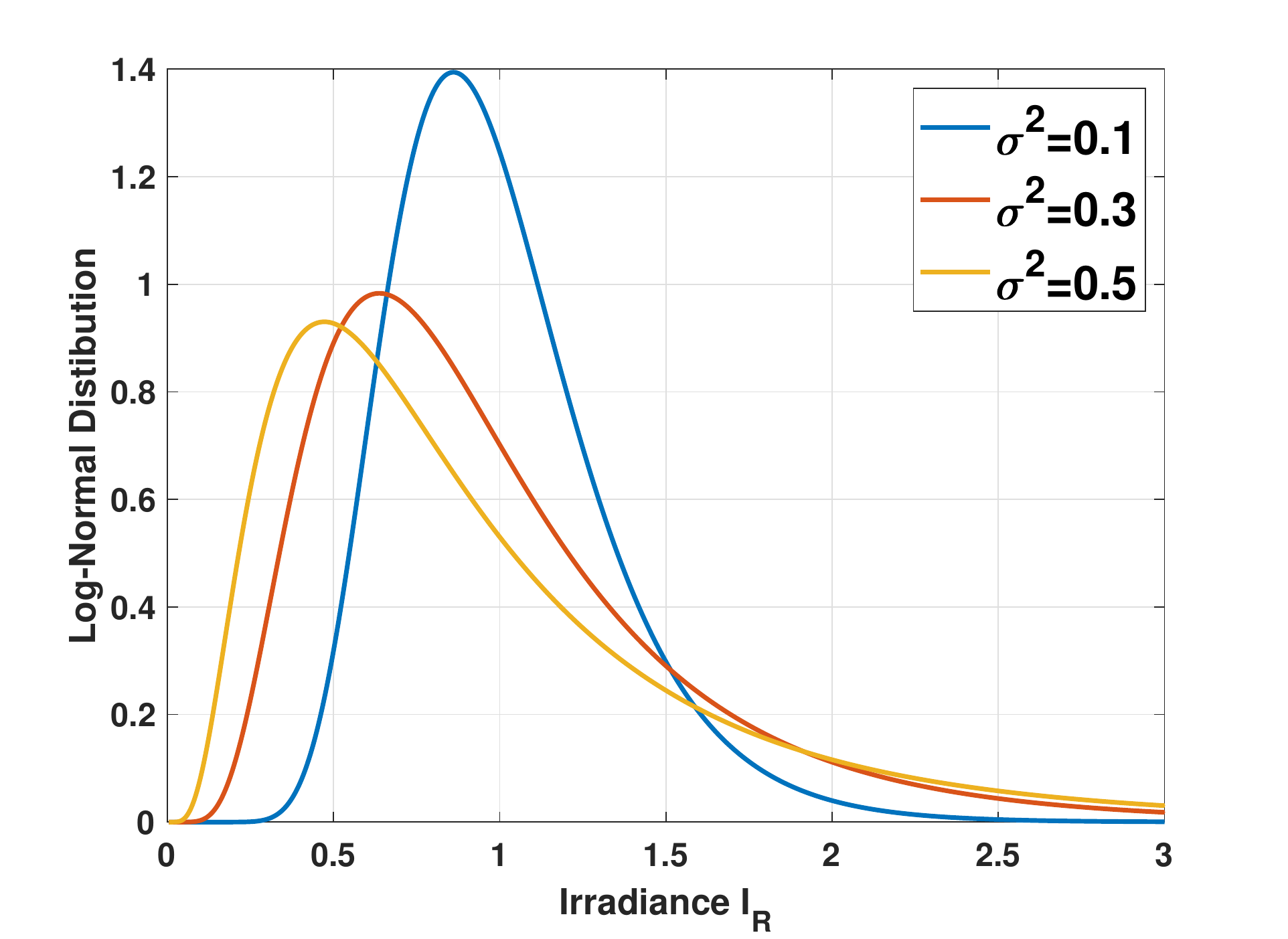}
                \caption[]{Log-normal distribution }    
                \label{Fig9aFigdis}
        \end{subfigure}%
        \begin{subfigure}[b]{0.25\textwidth}
                \includegraphics[height=4cm,width=5cm]{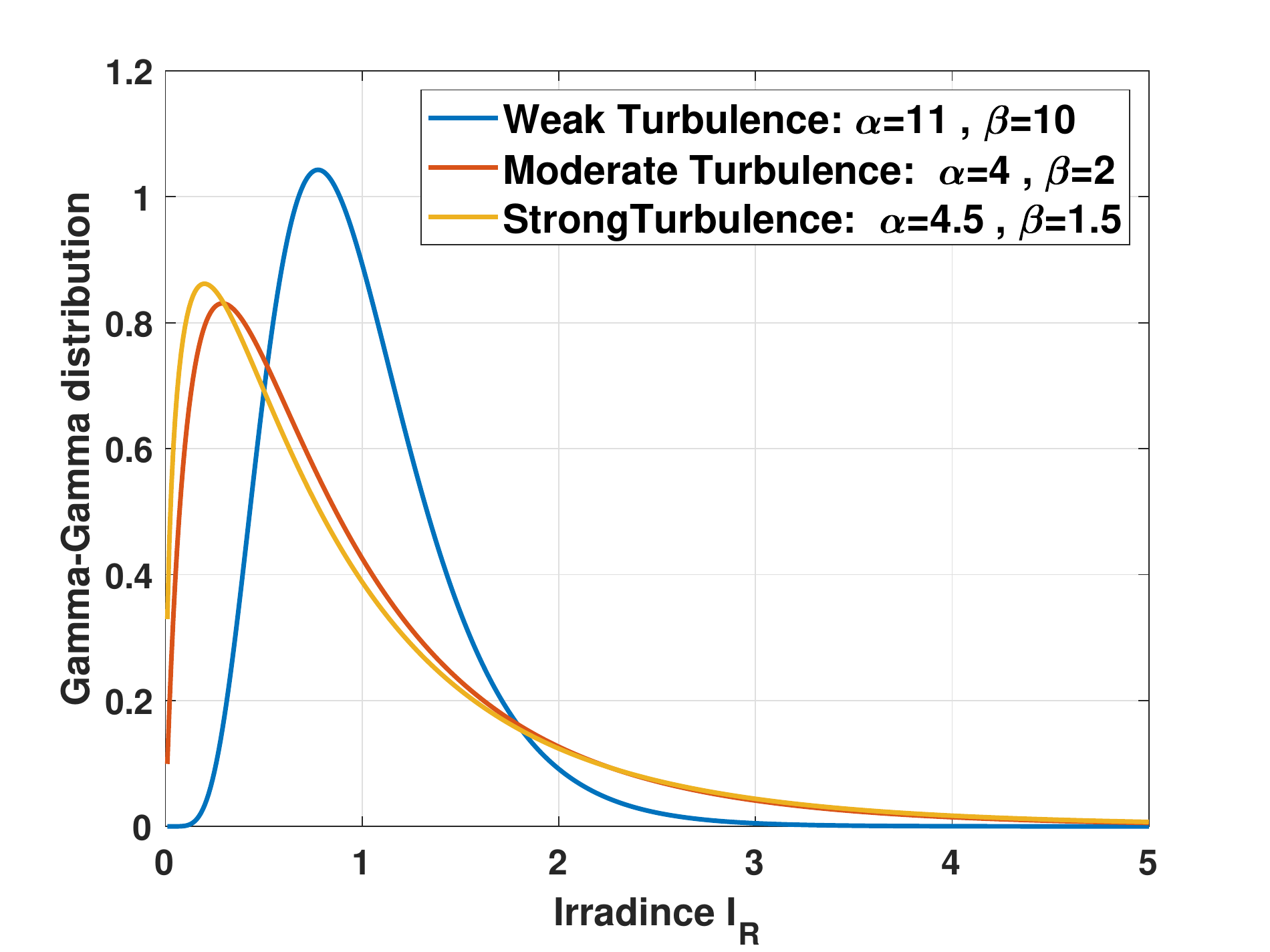}
                \caption[] {Gamma-Gamma distribution}    
                \label{Fig9bFigdis}
        \end{subfigure}%
        \begin{subfigure}[b]{0.25\textwidth}
                \includegraphics[height=4cm,width=5cm]{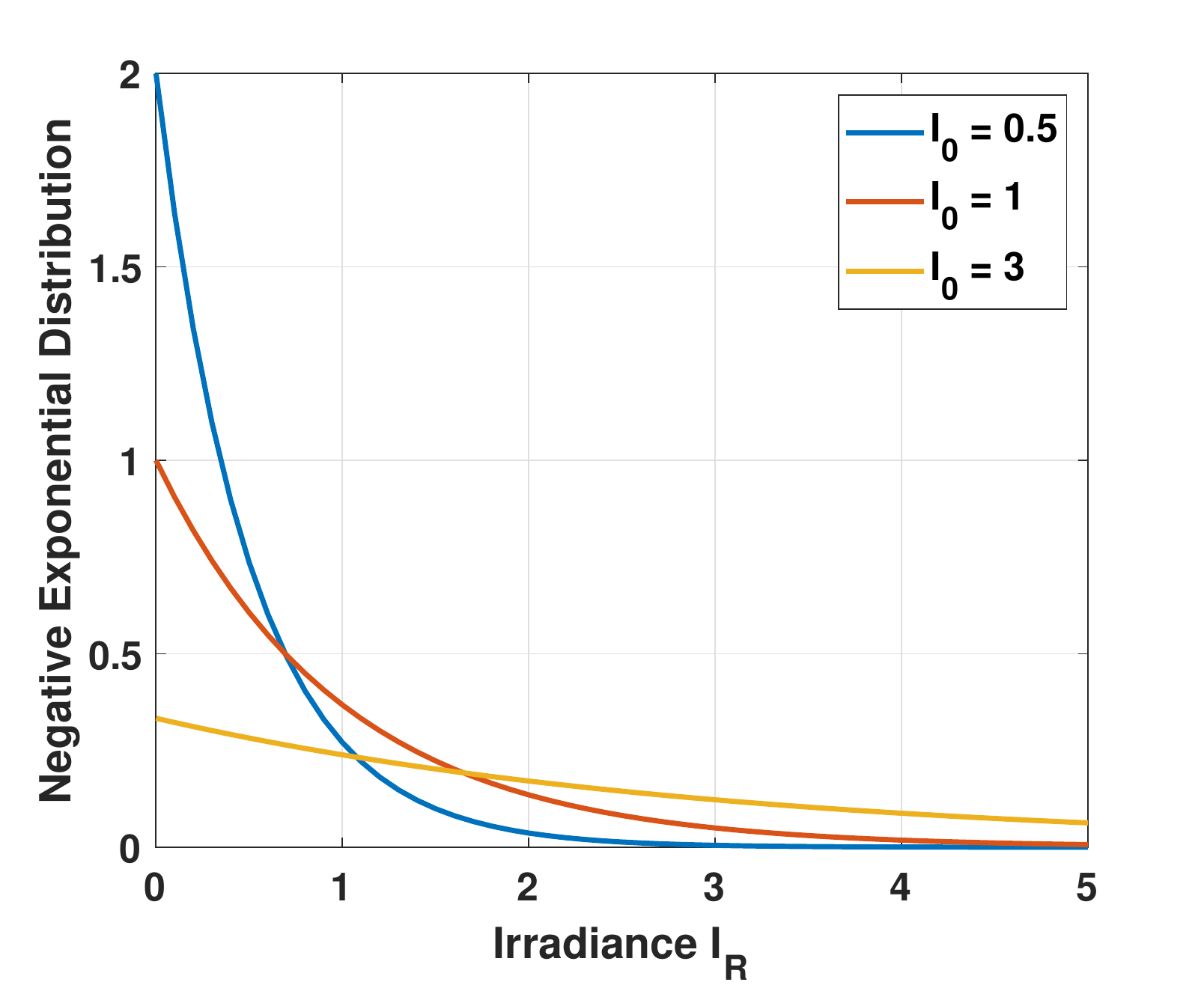}
                \caption[]{Negative exponential distribution}    
                \label{Fig9cFigdis}
        \end{subfigure}%
        \begin{subfigure}[b]{0.25\textwidth}
                \includegraphics[height=4cm,width=4.8cm]{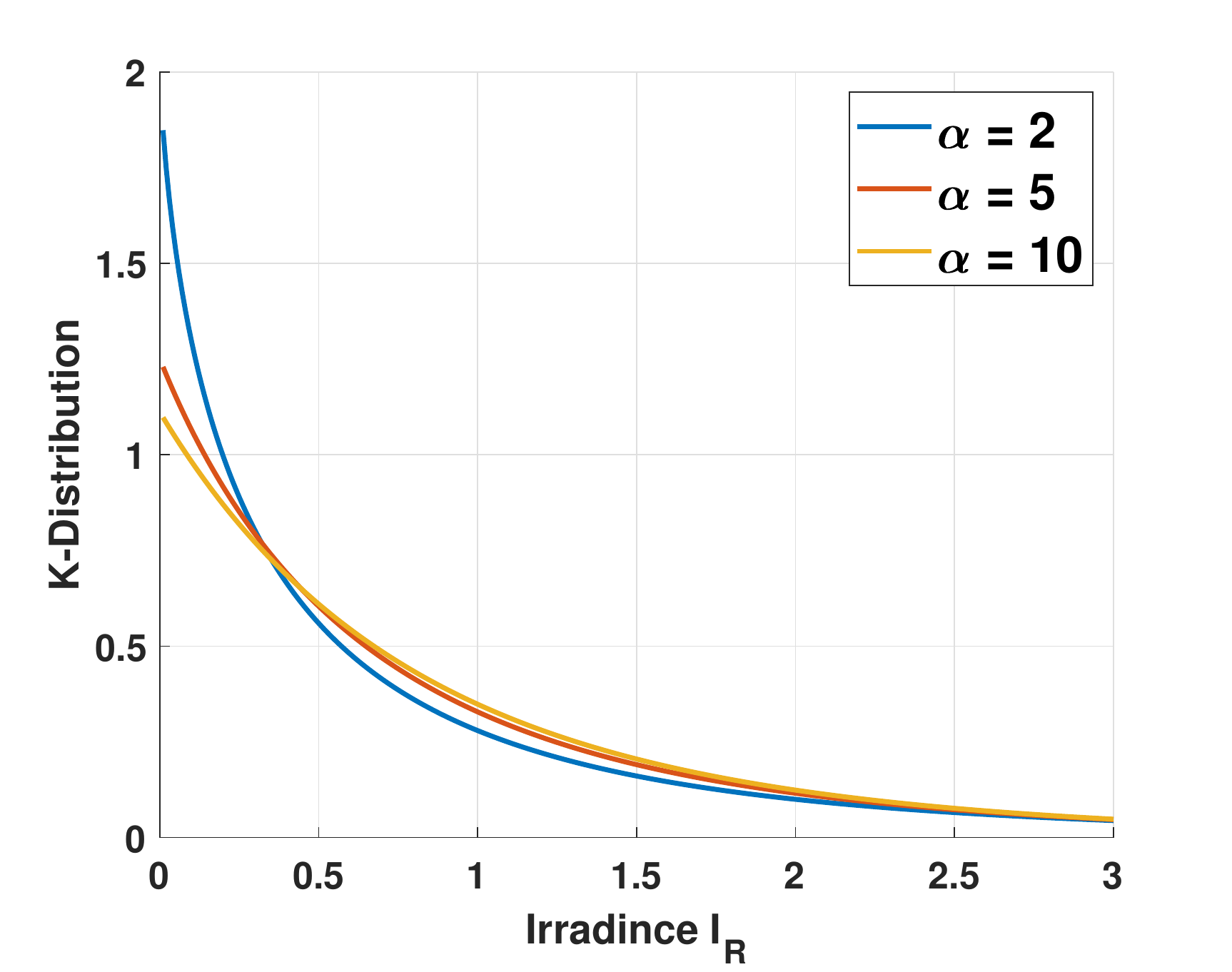}
                 \caption[]{$K$-distribution}    
                \label{Fig9dFigdis}
        \end{subfigure}
        \caption{Different atmospheric-turbulence distributions.}\label{Figdis}
\end{figure*}

The FSO signal undergoes atmospheric turbulence causing fluctuations in the intensity
and phase of the received signal. Several models are used to characterize atmospheric turbulence for an FSO link. For instance, in weak-to-moderate turbulence, the log-normal distribution is used \cite{nistazakis2009average}. In moderate-to-strong turbulence, the Gamma-Gamma distribution \cite{al2001mathematical} or the $K$-distribution \cite{tsiftsis2009optical} is used. A negative exponential distribution is used when the turbulence effect is very strong \cite{nistazakis2011performance}. Finally, the Málaga distribution, also called the $M$-distribution, embodies all the aforementioned distributions \cite{jurado2011unifying}. Fig.\ref{Figdis} shows the probability density function (PDF) of different atmospheric turbulence distributions.

\paragraph{Log-Normal Distribution}
Considering weak turbulence conditions, a log-normal distribution is used. The PDF of the log-normal distribution modeling atmospheric turbulence, denoted by $I_a$, is given by
\begin{equation}
     f_{I_a}(I_a)=\frac{1}{2 I_a \sqrt{2\pi\sigma^2}} 
    \exp\left(-\frac{(\log(I_a) + 2 \sigma^2)^2}{8\sigma^2}  \right),
\end{equation}
where $\sigma^2$ is the variance, which is given by
\begin{equation}
\sigma^2= \frac{\sigma_{\textrm{R}}^2}{4} =0.3 k^{\frac{7}{6}}C_n^2(h) d^{\frac{11}{6}},
\end{equation}
where $\sigma_{\textrm{R}}^2$ is the Rytov variance for a plane wave propagation ($\sigma_{\textrm{R}}^2=1.23 k^{\frac{7}{6}}C_n^2(h) d^{\frac{11}{6}}$), $C_n^2(h)$ is the index of refraction structure parameter at an altitude $h$, and $k=2\pi/\lambda$ is the optical wavenumber. The expression of $C_n^2(h)$ is given by
\begin{align}
    C_n^2(h)&=0.00594 \left( \frac{v_{\textrm{wind}}}{27} \right)^2 (10^{-5} h)^{10}
   \nonumber\\ & \times \exp \left[- \frac{h}{1000}\right] + 2.7\times 10^{-16} \exp\left[-\frac{h}{1500} \right]+
    C_n^2(0),
\end{align}
where $v_{\textrm{wind}}$ is the wind speed and $ C_n^2(0)$ is the structure constant at level ground.

\paragraph{Gamma-Gamma Distribution}
For moderate-to-strong turbulence, the Gamma-Gamma fading is used to model atmospheric turbulence. Hence, the PDF of the irradiance, $I_a$, is given by 

\begin{equation}
    f_{I_a}(I_a)=\frac{2(\alpha \beta)^{(\alpha+\beta)/2}}{\Gamma(\alpha)\Gamma(\beta)} 
     I_a^{(\alpha+\beta)/2-1}  
     K_{\alpha-\beta}(2\sqrt{\alpha \beta I_a} ),
\end{equation}
where $K_v(\cdot)$ is the modified Bessel function of the second kind and order $v$. $\alpha$ and $\beta$ are the turbulence fading parameters, and under the plane wave approximation, are expressed as follows:

\begin{equation}
\alpha = \Bigg[  \exp\Bigg( \frac{ 0.49 \sigma_R^2}{{(1+1.11 \sigma_R^{12/5})}^{7/6} }  \Bigg) - 1 \Bigg]^{-1},
\end{equation}

\begin{equation}
\beta = \Bigg[  \exp\Bigg( \frac{ 0.51 \sigma_R^2}{{(1+0.69 \sigma_R^{12/5})}^{5/6} }  \Bigg) - 1 \Bigg]^{-1}.
\end{equation}


\paragraph{$K$-Distribution}
The $K$-distribution can be achieved as a multiplication of a Gamma distribution and an exponential distribution. The PDF is therefore given by

\begin{equation}
    f_{I_a}(I_a)=\frac{2\alpha}{\Gamma(\alpha)} 
     {(\alpha I_a)}^{(\alpha-1)/2}  
     K_{\alpha-1}(2\sqrt{\alpha I_a}).
\end{equation}

\paragraph{Negative Exponential Distribution}
The negative  exponential  distribution is used to model strong turbulence conditions. The PDF of its distribution is given as
\begin{equation}
    f_{I_a}(I_a)=  \frac{1}{I_a(0)} \exp(-I_a/I_a(0)),  \:\: I_a(0)>0,
\end{equation}
where $\mathbb{E}[I_a]=I_a(0)$ is the  mean  of  receiver  optical  irradiance.

\paragraph{Málaga Distribution ($M$-Distribution)}
Hence, we employ an FSO link that experiences $M$ turbulence, for which the PDF of the irradiance $I_a$ is given by 
\begin{equation}
    f_{I_a}(I_a)=  A \sum_{m=1}^\beta a_m I_a 
    K_{(\alpha - m)} \Bigg( 2 \sqrt{ \frac{\alpha \beta I_a}{g \beta + \Omega^{'}} } \Bigg), I_a>0,
\end{equation}
where
\begin{equation}
 A \triangleq \frac{2 \alpha^{\alpha/2}}{g^{1+\alpha/2} \Gamma(\alpha)} 
 { \Bigg( \frac{g \beta}{ g \beta +\Omega^{'}} \Bigg)}^{\alpha+\beta /2},
\end{equation}

\begin{equation}
 a_m \triangleq   \binom{\beta-1}{m-1}
 \frac{ {(g\beta+\Omega^{'})}^{1-m/2} }{(m-1)!}
 { \bigg(\frac{\Omega^{'}}{g}\bigg) }^{m-1}
{ \bigg(\frac{\alpha}{\beta}\bigg) }^{m/2},
\end{equation}
where $g$ is the average power of the scattering component and $\Omega^{'}$ is the average power from
the coherent contributions.
\begin{table*}[h!] 
\begin{center}
    \caption{Channel modeling for NTFPs considering RF and FSO links.}
    \label{crouch}
    \begin{tabular}{   p{6cm} | p{6cm} }
        \toprule
    \textbf{RF Links}   
& \textbf{FSO Links} \\\midrule
\underline{\textbf{LAPs Large-Scale Path Loss}} 
\begin{itemize}
    \item LOS probability model \cite{data2003prediction} 
    \item Simplified LOS probability model \cite{al2014optimal}
    \item Path loss exponent model   \cite{azari2017ultra} 
\end{itemize}
 \underline{\textbf{LAPs Small-Scale Fading}}  
\begin{itemize} 
\item Rayleigh distribution \cite{newhall2003wideband,abualhaol2010performance}
\item Rician distribution \cite{cai2017low,ye2017air,goddemeier2015investigation,azari2017ultra}
\item Nakagami-$m$ distribution \cite{abualhaol2010performance,frew2008airborne,yanmaz2013achieving}
\item (Rice + Log-Normal) Loo Model \cite{simunek2013uav}    
\end{itemize}
\underline{\textbf{HAPs Large-Scale Path Loss}} 
\begin{itemize} 
\item Free Space Path loss \cite{saeed2020wireless}
\end{itemize}
\underline{\textbf{HAPs Small-Scale Fading}}      
\begin{itemize} 
\item Rayleigh distribution \cite{karapantazis2005broadband, cuevas2004channel}
\item Rician distribution \cite{cuevas2004channel,saeed2020wireless,karapantazis2005broadband}
\item Loo Model \cite{zhao2020ka}   
\end{itemize}

& \underline{\textbf{Atmospheric Turbulence}}
\begin{itemize} 
\item Log-normal distribution \cite{nistazakis2009average}  
\item Gamma-Gamma distribution  \cite{al2001mathematical} 
\item $K$-distribution \cite{tsiftsis2009optical}
\item Negative exponential distribution \cite{nistazakis2011performance}
\item Málaga distribution \cite{jurado2011unifying} 
\end{itemize}
\underline{\textbf{Pointing Errors}} 
\begin{itemize} 
\item Beckmann distribution \cite{alquwaiee2016asymptotic}
\item Rayleigh distribution \cite{farid2007outage} 
\item Rician distribution \cite{yang2014free}
\item Hoyt distribution \cite{gappmair2011ook}
\end{itemize} 
\underline{\textbf{Large-Scale Path Loss}} 
\begin{itemize} 
\item Path loss \cite{vavoulas2012weather}
\end{itemize} 
\underline{\textbf{Atmospheric Attenuation}} 
\begin{itemize} 
\item Fog  and  haze attenuation \cite{nadeem2009weather}
\item Rain attenuation \cite{data2007prediction} 
\item Snow attenuation \cite{data2007prediction}
\item Dust attenuation \cite{esmail2016experimental}
\end{itemize} 

\\
\bottomrule
    \end{tabular}\label{table_channel}
    \end{center}
\end{table*}

\subsubsection{Pointing Errors}
The pointing errors originate from a misalignment between the transmitter and the receiver due to errors in tracking, vibrations in the system, or building sway.
We assume that a Gaussian beam, with a beamwidth denoted by $w_z$, propagates in a photo-detector.
We denote the radial displacement by $r_{dep}$. Hence, the collected power at a given distance $z$ is approximated by the following formula:

\begin{equation}
I_{p} \approx A_0 \exp \left(- \frac{2r_{dep}^2}{w_{z_{eq}}^2}\right),
\end{equation}

where we denote by $w_{z_{eq}}$ the equivalent beamwidth given by
\begin{equation}
  w_{z_{eq}}= w_z^2 \frac{\sqrt{A_0 \pi}}{2v\exp(-v^2)},
\end{equation}

where $ A_0 = {[\textrm{erf}(v)]}^2$ is the maximum fraction of the collected power,
and $v=\sqrt{  \frac{a^2 \pi}{w_z^2}  }$ is the ratio between the aperture
radius denoted by $a$ and the beamwidth $w_z$.

The radial displacement at the receiver is expressed as $\textrm{r}_{dep}=[x\: y]^T$, where $x$ and $y$ are the vertical and horizontal displacement on the plane.

Consequently, the distribution of $r_{dep}=|\textrm{r}_{dep}|=\sqrt{x^2+y^2}$ depends on
the distribution of $x$ and $y$. Assuming independent Gaussian displacements on the horizontal and elevation axes, $r_{dep}$ can have several distributions depending on the Gaussian parameters. For instance, several distributions are used in the literature for modeling pointing errors, such as the Beckmann distribution \cite{alquwaiee2016asymptotic}, Rayleigh distribution \cite{farid2007outage}, Rician distribution \cite{yang2014free}, and Hoyt distribution \cite{gappmair2011ook}.

\paragraph{Beckmann distribution}
When both the $x$ and $y$ displacements are nonzero mean Gaussian random variables with
$x \sim \mathcal{N}(\mu_x,\sigma_x)$ and $y \sim \mathcal{N}(\mu_y,\sigma_y)$, then $r_{dep}$ follows a Beckmann distribution with a PDF given by \cite{alquwaiee2016asymptotic}

\begin{align}
  &f_{r_{dep}}(r_{dep}) = \frac{r_{dep}}{2\pi\sigma_x \sigma_y} \nonumber \\
          & \int_0^{2\pi} 
           \exp\left(-\frac{(r_{dep}\cos(\theta)-\mu_x)^2}{2\sigma_x^2}- \frac{(r_{dep}\sin(\theta)-\mu_y)^2}{2\sigma_y^2}\right) d\theta.
\end{align}

The $n$th moment of the pointing error effect, denoted by $I_p$, is given by
\begin{equation}
    \mathbb{E}[I_p^n]=\frac{A_0^2 \xi_x \xi_y}{\sqrt{(n+\xi_x^2)(n+\xi_y^2)}} 
     \exp\left(-\frac{2n}{w_{z_{eq}}^2} \left[\frac{\mu_x^2}{1+\frac{n}{\xi_x^2}} + \frac{\mu_y^2}{1+\frac{n}{\xi_y^2}}\right]\right),
\end{equation}

where $\xi_x=\frac{w_{z_{eq}}}{2\sigma_x}$ and $\xi_y=\frac{w_{z_{eq}}}{2\sigma_y}$.

\paragraph{Rayleigh distribution}
If both $x$ and $y$ have
zero mean and the same variance, that is, $\mu_x =\mu_y=0$ and $\sigma_x = \sigma_y = \sigma$, then  $r_{dep}$ follows a Rayleigh distribution with the PDF given by \cite{farid2007outage}
\begin{equation}
  f_{r_{dep}}(r_{dep}) = \frac{r_{dep}}{\sigma^2} \exp\left(-\frac{r_{dep}^2}{2\sigma^2}\right),
\end{equation}

and the PDF of the pointing error $I_p$ given by
\begin{equation}
f_{I_p}(I_p)=\frac{\xi^2}{A_0^{\xi^2}} I_p^{\xi^2-1},    
\end{equation}
where $\xi=\frac{w_{z_{eq}}}{2\sigma^2}$.

\paragraph{Rician  distribution}
In case both displacements have different
non-zero means and the same variance, that is, $\mu_x \neq \mu_y$ and $\sigma_x = \sigma_y = \sigma$, then $r_{dep}$ follows a Rician distribution with the PDF given by \cite{yang2014free}
\begin{equation}
  f_{r_{dep}}(r_{dep}) = \frac{r_{dep}}{\sigma^2} \exp\left(-\frac{(r_{dep}^2+s^2)}{2\sigma^2}\right) I_0 \left(\frac{r_{dep} s}{\sigma^2}\right),
\end{equation}
where $s=\sqrt{\mu_x^2 + \mu_y^2}$. The PDF of the pointing error $I_p$ is given by
\begin{equation}
f_{I_p}(I_p)=\frac{\xi^2 \exp(\frac{-s^2}{2\sigma^2})}{A_0^{\xi^2}} I_p^{\xi^2-1}  
I_0 \left(\frac{s}{\sqrt{2}\sigma^2} \sqrt{-w_{z_{eq}} \log\left(\frac{I_p}{A_0}\right)}\right).
\end{equation}

\paragraph{Hoyt distribution}
When $x$ and $y$ have zero mean and different variances, that is, $\mu_x=\mu_y=0$ and $\sigma_x \neq \sigma_y$, then $r_{dep}$ follows a Hoyt distribution with the PDF given by \cite{gappmair2011ook}
\begin{equation}
f_{r_{dep}}(r_{dep})= \frac{r_{dep}}{q \sigma_y^2}\exp\left(-\frac{r_{dep}^2(1+q^2)}{4q^2 \sigma_y^2}\right)
        I_0 \left(\frac{r_{dep}^2(1-q^2)}{4q^2 \sigma_y^2}\right),
\end{equation}
where $q=\frac{\sigma_x}{\sigma_y}=\frac{\xi_x}{\xi_y}$. The PDF of the pointing error $I_p$ is given by

\begin{equation}
f_{I_p}(I_p) = \frac{\xi_x \xi_y}{A_0}
      {\bigg(\frac{I_p}{A_0}\bigg)}^{\frac{\xi_x^2(1+q^2)}{2}-1}
      I_0 \bigg( \frac{\xi_x^2 (1-q^2)}{2}   \log\left(\frac{I_p}{A_0}\right)\bigg), 
\end{equation}
where $ 0\leq I_p \leq A_0$.
\subsubsection{Atmospheric Attenuation}
The FSO signal is affected by path
loss, which depends on the distance between the transmitter and receiver, but also on the atmospheric conditions given by \cite{vavoulas2012weather} 

\begin{equation}
    I_{Att}=\frac{A_r}{(\Theta d)^2} \textrm{Att},
\end{equation}
where $A_r$ is the receiver effective area and $\Theta$ is the beam divergence. The parameter $\textrm{Att}\in\{\textrm{Att}_{\textrm{fog}}, \textrm{Att}_{\textrm{rain}}, \textrm{Att}_{\textrm{snow}}, \textrm{Att}_{\textrm{dust}} \}$ represents the atmospheric attenuation, and it depends on atmospheric conditions, such as fog, rain, snow, and dust, denoted by  $\textrm{Att}_{\textrm{fog}}$ \cite{nadeem2009weather}, $\textrm{Att}_{\textrm{rain}}$ \cite{data2007prediction}, $\textrm{Att}_{\textrm{snow}}$ \cite{data2007prediction}, and $\textrm{Att}_{\textrm{dust}}$ \cite{esmail2016experimental}, respectively.

\section{Conclusion} \label{section_conclusion}
In this survey, we provided an extensive and comprehensive overview of NTFPs. 
We provided a general overview of this solution for all readers interested irrespective of their background by reviewing all the existing types of NTFPs, including their components, their characteristics, their applications, advantages, challenges, and regulations.
We also have detailed some case studies that have been used and the major companies related to NTFPs. Then, we addressed platforms from a wireless communications perspective. 
We carried out different type of analysis: geometrical analysis, performance, and economic analysis. Finally, we proved comprehensive channel modeling for NTFPs and free-flying platforms for different altitudes and type of links. In this survey, we highlighted the effectiveness of NTFPs as a future solution solution for 6G.

\section{Acknowledgment}
Figure \ref{Fig.altitude} and Figure \ref{Fig.sp} were created by Heno Hwang, scientific illustrator at King Abdullah University of Science and Technology (KAUST). 

\bibliographystyle{ieeetr}
\bibliography{0_bibnoma}

\begin{thebibliography}{100}

\bibitem{dang2020should}
S.~Dang, O.~Amin, B.~Shihada, and M.-S. Alouini, ``What should 6g be?,'' {\em
  Nature Electronics}, vol.~3, no.~1, pp.~20--29, 2020.

\bibitem{2david20186g}
K.~David and H.~Berndt, ``6g vision and requirements: Is there any need for
  beyond 5g?,'' {\em IEEE Vehicular Technology Magazine}, vol.~13, no.~3,
  pp.~72--80, 2018.

\bibitem{3raghavan2019evolution}
V.~Raghavan and J.~Li, ``Evolution of physical-layer communications research in
  the post-5g era,'' {\em IEEE Access}, vol.~7, pp.~10392--10401, 2019.

\bibitem{4yastrebova2018future}
A.~Yastrebova, R.~Kirichek, Y.~Koucheryavy, A.~Borodin, and A.~Koucheryavy,
  ``Future networks 2030: Architecture \& requirements,'' in {\em 2018 10th
  International Congress on Ultra Modern Telecommunications and Control Systems
  and Workshops (ICUMT)}, pp.~1--8, IEEE, 2018.

\bibitem{5saad2019vision}
W.~Saad, M.~Bennis, and M.~Chen, ``A vision of 6g wireless systems:
  Applications, trends, technologies, and open research problems,'' {\em IEEE
  network}, vol.~34, no.~3, pp.~134--142, 2019.

\bibitem{6strinati20196g}
E.~C. Strinati, S.~Barbarossa, J.~L. Gonzalez-Jimenez, D.~Ktenas, N.~Cassiau,
  L.~Maret, and C.~Dehos, ``6g: The next frontier: From holographic messaging
  to artificial intelligence using subterahertz and visible light
  communication,'' {\em IEEE Vehicular Technology Magazine}, vol.~14, no.~3,
  pp.~42--50, 2019.

\bibitem{7tariq2020speculative}
F.~Tariq, M.~R. Khandaker, K.-K. Wong, M.~A. Imran, M.~Bennis, and M.~Debbah,
  ``A speculative study on 6g,'' {\em IEEE Wireless Communications}, vol.~27,
  no.~4, pp.~118--125, 2020.

\bibitem{sharma2018itu}
A.~Sharma, ``Itu calls to connect almost 4 billion unconnected individuals
  globally,'' {\em The National (UAE), October}, vol.~28, 2018.

\bibitem{yaacoub2020key}
E.~Yaacoub and M.-S. Alouini, ``A key 6g challenge and opportunity—connecting
  the base of the pyramid: A survey on rural connectivity,'' {\em Proceedings
  of the IEEE}, vol.~108, no.~4, pp.~533--582, 2020.

\bibitem{6GSummita}
``6g summit, connecting the unconnected.'' http://6gsummit.org/.

\bibitem{saeed2020wireless}
N.~Saeed, T.~Y. Al-Naffouri, and M.-S. Alouini, ``Wireless communication for
  flying cars,'' 2020.

\bibitem{lynk}
lynk, ``Lynk global.'' https://lynk.world/.

\bibitem{ast}
A.~S. T, ``Ast \& science, space mobile, transforming connectivity.''
  https://ast-science.com/.

\bibitem{han2021abstracted}
B.~Han, W.~Jiang, M.~A. Habibi, and H.~D. Schotten, ``An abstracted survey on
  6g: Drivers, requirements, efforts, and enablers,'' {\em arXiv preprint
  arXiv:2101.01062}, 2021.

\bibitem{akhtar2020shift}
M.~W. Akhtar, S.~A. Hassan, R.~Ghaffar, H.~Jung, S.~Garg, and M.~S. Hossain,
  ``The shift to 6g communications: vision and requirements,'' {\em
  Human-centric Computing and Information Sciences}, vol.~10, no.~1, pp.~1--27,
  2020.

\bibitem{saad2020wireless}
W.~Saad, M.~Bennis, M.~Mozaffari, and X.~Lin, {\em Wireless Communications and
  Networking for Unmanned Aerial Vehicles}.
\newblock Cambridge University Press, 2020.

\bibitem{kurt2020vision}
G.~Kurt, M.~G. Khoshkholgh, S.~Alfattani, A.~Ibrahim, T.~S. Darwish, M.~S.
  Alam, H.~Yanikomeroglu, and A.~Yongacoglu, ``A vision and framework for the
  high altitude platform station (haps) networks of the future,'' {\em arXiv
  preprint arXiv:2007.15088}, 2020.

\bibitem{mahmood2020tethered}
K.~Mahmood, N.~Ismail, and N.~M. Suhadis, ``Tethered aerostat envelope design
  and applications: A review,'' in {\em AIP Conference Proceedings}, vol.~2226,
  p.~050003, AIP Publishing LLC, 2020.

\bibitem{TCOM}
TCOM, ``Tcom web site,.'' {https://tcomlp.com/}.

\bibitem{Altaeroswebsite}
Altaeros, ``Altaeros website.'' {http://www.altaeros.com/}.

\bibitem{kalabic2013reference}
U.~Kalabi{\'c}, C.~Vermillion, and I.~Kolmanovsky, ``Reference governor design
  for computationally efficient attitude and tether tension constraint
  enforcement on a lighter-than-air wind energy system,'' in {\em 2013 European
  Control Conference (ECC)}, pp.~1004--1010, IEEE, 2013.

\bibitem{kehs2017online}
M.~Kehs, C.~Vermillion, and H.~Fathy, ``Online energy maximization of an
  airborne wind energy turbine in simulated periodic flight,'' {\em IEEE
  Transactions on Control Systems Technology}, vol.~26, no.~2, pp.~393--403,
  2017.

\bibitem{samson2015adaptive}
J.~Samson and R.~Katebi, ``Adaptive envelope control design for a buoyant
  airborne wind energy system,'' in {\em 2015 American Control Conference
  (ACC)}, pp.~2395--2400, IEEE, 2015.

\bibitem{saleem2018aerodynamic}
A.~Saleem and M.-H. Kim, ``Aerodynamic analysis of an airborne wind turbine
  with three different aerofoil-based buoyant shells using steady rans
  simulations,'' {\em Energy Conversion and Management}, vol.~177,
  pp.~233--248, 2018.

\bibitem{Elistairwebsite}
Elistair, ``Elistair-the tethered drone company.'' {https://elistair.com//}.

\bibitem{williams2008optimal}
P.~Williams, B.~Lansdorp, and W.~Ockesl, ``Optimal crosswind towing and power
  generation with tethered kites,'' {\em Journal of guidance, control, and
  dynamics}, vol.~31, no.~1, pp.~81--93, 2008.

\bibitem{lansdorp2006laddermill}
B.~Lansdorp and P.~Williams, ``The laddermill: Innovative wind energy from high
  altitudes in holland and australia,'' {\em Windpower 2006}, 2006.

\bibitem{loyd1980crosswind}
M.~L. Loyd, ``Crosswind kite power (for large-scale wind power production),''
  {\em Journal of energy}, vol.~4, no.~3, pp.~106--111, 1980.

\bibitem{Kitewinderwebsite}
Kitewinder, ``Kitewinder website.'' {https://kitewinder.fr/}.

\bibitem{Makaniwebsite}
M.~P, ``Makani power website.'' {https://makanipower.com/}.

\bibitem{Allsoppwebsite}
Allsopp, ``Allsopp helikites website.'' {http://www.allsopphelikites.com/}.

\bibitem{HAVwebsite}
H.~A. Vehicles, ``Hybrid air vehicles.'' {https://www.hybridairvehicles.com/}.

\bibitem{Lockheedwebsite}
L.~M, ``Lockheed martin website.'' {https://www.lockheedmartin.com/}.

\bibitem{zhang2010flight}
K.-s. Zhang, Z.-h. Han, and B.-f. Song, ``Flight performance analysis of hybrid
  airship: revised analytical formulation,'' {\em Journal of Aircraft},
  vol.~47, no.~4, pp.~1318--1330, 2010.

\bibitem{meng2019aerodynamic}
J.~Meng, M.~Li, L.~Zhang, M.~Lv, and L.~Liu, ``Aerodynamic performance analysis
  of hybrid air vehicles with large reynolds number,'' in {\em 2019 IEEE
  International Conference on Unmanned Systems (ICUS)}, pp.~403--409, IEEE,
  2019.

\bibitem{Drone_Aviation}
D.~A. C, ``Drone aviation corp..'' {https://droneaviationcorp.com/}.

\bibitem{web1}
Smithsonianmag.com, ``What really sparked the hindenburg disaster.''
  {https://www.smithsonianmag.com/science-nature/what-really-sparked-the-hindenburg-disaster-85867521/}.

\bibitem{web2}
History.com, ``The hindenburg disaster.''
  {https://www.history.com/this-day-in-history/the-hindenburg-disaster}.

\bibitem{reiff2009acoustic}
C.~G. Reiff, ``Acoustic source localization and cueing from an aerostat during
  the nato set-093 field experiment,'' in {\em Unattended Ground, Sea, and Air
  Sensor Technologies and Applications XI}, vol.~7333, p.~73330M, International
  Society for Optics and Photonics, 2009.

\bibitem{Gore}
Gore, ``Gore cables \& cable assemblies.'' {https://www.gore.com/}.

\bibitem{Deregtcables}
Deregt, ``Deregt aerostats and uavs cables.'' {www.deregtcables.com}.

\bibitem{Allsoppwebsite1}
Allsopp, ``Allsopp helikites ltd..'' {https://helikites.com/}.

\bibitem{AllsoppH}
Allsopp, ``Allsopp helikites ltd..'' {https://www.4gremote.com/}.

\bibitem{Craftsmen}
C.~I, ``Craftsmen industries, inc..'' {https://craftsmenindustrial.com/}.

\bibitem{gawande2007design}
V.~Gawande, P.~Bilaye, A.~Gawale, R.~Pant, and U.~Desai, ``Design and
  fabrication of an aerostat for wireless communication in remote areas,'' in
  {\em 7th AIAA ATIO Conf, 2nd CEIAT Int'l Conf on Innov and Integr in Aero
  Sciences, 17th LTA Systems Tech Conf; followed by 2nd TEOS Forum}, p.~7832,
  2007.

\bibitem{chopra2011new}
P.~Chopra, R.~Manchanda, R.~Mehrotra, and S.~Jain, ``A new topology for telecom
  and broad band services in spars, remote and hilly areas,'' {\em WSEAS
  TRANSACTIONS on COMMUNICATIONS}, vol.~10, no.~9, pp.~273--286, 2011.

\bibitem{FortressUAV}
F.~U.~A. V, ``Fortress uav, uav repair \& maintenance experts.''
  {https://www.fortressuav.com/}.

\bibitem{Comprehensivecom}
Comprehensivecom, ``Comprehensive communication services.''
  {https://www.comprehensivecom.net/}.

\bibitem{flickr}
Flickr. {https://www.flickr.com/photos/keyslibraries/15004011277/}.

\bibitem{bushnaq2020cellular}
O.~M. Bushnaq, M.~A. Kishk, A.~{\c{C}}elik, M.-S. Alouini, and T.~Y.
  Al-Naffouri, ``Cellular traffic offloading through tethered-uav deployment
  and user association,'' {\em arXiv preprint arXiv:2003.00713}, 2020.

\bibitem{kishk2019capacity}
M.~A. Kishk, A.~Bader, and M.-S. Alouini, ``Capacity and coverage enhancement
  using long-endurance tethered airborne base stations,'' {\em arXiv preprint
  arXiv:1906.11559}, 2019.

\bibitem{teixeira2017enabling}
F.~B. Teixeira, T.~Oliveira, M.~Lopes, C.~Leoc{\'a}dio, P.~Salazar, J.~Ruela,
  R.~Campos, and M.~Ricardo, ``Enabling broadband internet access offshore
  using tethered balloons: The bluecom+ experience,'' in {\em OCEANS
  2017-Aberdeen}, pp.~1--7, IEEE, 2017.

\bibitem{gajra2014soptas}
K.~M. Gajra and R.~S. Pant, ``Soptas: solar powered tethered aerostat system,''
  in {\em 2014 1st International Conference on Non Conventional Energy (ICONCE
  2014)}, pp.~65--68, IEEE, 2014.

\bibitem{kishk20203}
M.~A. Kishk, A.~Bader, and M.-S. Alouini, ``On the 3-d placement of airborne
  base stations using tethered uavs,'' {\em IEEE Transactions on
  Communications}, 2020.

\bibitem{carlson2018surface}
D.~F. Carlson, T.~{\"O}zg{\"o}kmen, G.~Novelli, C.~Guigand, H.~Chang,
  B.~Fox-Kemper, J.~Mensa, S.~Mehta, E.~Fredj, H.~Huntley, {\em et~al.},
  ``Surface ocean dispersion observations from the ship-tethered aerostat
  remote sensing system,'' {\em Frontiers in Marine Science}, vol.~5, p.~479,
  2018.

\bibitem{teixeira2016tethered}
F.~B. Teixeira, T.~Oliveira, M.~Lopes, J.~Ruela, R.~Campos, and M.~Ricardo,
  ``Tethered balloons and tv white spaces: A solution for real-time marine data
  transfer at remote ocean areas,'' in {\em 2016 IEEE Third Underwater
  Communications and Networking Conference (UComms)}, pp.~1--5, IEEE, 2016.

\bibitem{campos2016bluecom+}
R.~Campos, T.~Oliveira, N.~Cruz, A.~Matos, and J.~M. Almeida, ``Bluecom+:
  Cost-effective broadband communications at remote ocean areas,'' in {\em
  OCEANS 2016-Shanghai}, pp.~1--6, IEEE, 2016.

\bibitem{akita2012feasibility}
D.~Akita, ``Feasibility study of a sea-anchored stratospheric balloon for
  long-duration flights,'' {\em Advances in space research}, vol.~50, no.~4,
  pp.~508--515, 2012.

\bibitem{fesen2015method}
R.~Fesen and Y.~Brown, ``A method for establishing a long duration,
  stratospheric platform for astronomical research,'' {\em Experimental
  Astronomy}, vol.~39, no.~3, pp.~475--493, 2015.

\bibitem{aglietti2008aerostat}
G.~Aglietti, T.~Markvart, A.~Tatnall, and S.~Walker, ``Aerostat for electrical
  power generation—concept feasibility,'' {\em Proceedings of the Institution
  of Mechanical Engineers, Part G: Journal of Aerospace Engineering}, vol.~222,
  no.~1, pp.~29--39, 2008.

\bibitem{chiba2017aerodynamic}
K.~Chiba, R.~Nishikawa, M.~Onda, S.~Satori, and R.~Akiba, ``Aerodynamic
  influences on a tethered high-altitude lighter-than-air platform system to
  its behavior,'' {\em Aerospace Science and Technology}, vol.~70,
  pp.~405--411, 2017.

\bibitem{bely1995high}
P.~Y. Bely, R.~Ashford, and C.~D. Cox, ``High-altitude aerostats as
  astronomical platforms,'' in {\em Space Telescopes and Instruments},
  vol.~2478, pp.~101--116, International Society for Optics and Photonics,
  1995.

\bibitem{badesha2002dynamic}
S.~Badesha and J.~Bunn, ``Dynamic simulation of high altitude tethered balloon
  system subject to thunderstorm windfield,'' in {\em AIAA Atmospheric Flight
  Mechanics Conference and Exhibit}, p.~4614, 2002.

\bibitem{badesha2002sparcl}
S.~S. Badesha, ``Sparcl: a high-altitude tethered balloon-based optical
  space-to-ground communication system,'' in {\em Free-Space Laser
  Communication and Laser Imaging II}, vol.~4821, pp.~181--193, International
  Society for Optics and Photonics, 2002.

\bibitem{badesha1996very}
S.~Badesha, A.~Euler, and L.~Schroder, ``Very high altitude tethered balloon
  parametric sensitivity study,'' in {\em 34th Aerospace Sciences Meeting and
  Exhibit}, p.~579, 1996.

\bibitem{badesha1996very1}
S.~Badesha, A.~Euler, and L.~Schroder, ``Very high altitude tethered balloon
  trajectory simulation,'' in {\em 21st Atmospheric Flight Mechanics
  Conference}, p.~3440, 1996.

\bibitem{aglietti2009dynamic}
G.~Aglietti, ``Dynamic response of a high-altitude tethered balloon system,''
  {\em Journal of Aircraft}, vol.~46, no.~6, pp.~2032--2040, 2009.

\bibitem{izet2011low}
K.~Izet-{\"U}nsalan and D.~{\"U}nsalan, ``A low cost alternative for
  satellites-tethered ultra-high altitude balloons,'' in {\em Proceedings of
  5th International Conference on Recent Advances in Space
  Technologies-RAST2011}, pp.~13--16, IEEE, 2011.

\bibitem{chim2006employing}
Y.-C. Chim, K.-W. Seah, S.-L. Sim, K.-L. Khoo, Y.-S. Lee, Y.-C. Foo, and Y.-C.
  Lai, ``Employing ad-hoc networking with aerial communications nodes for
  wireless tactical experimentation,'' in {\em MILCOM 2006-2006 IEEE Military
  Communications conference}, pp.~1--7, IEEE, 2006.

\bibitem{weiwei2015integrated}
Z.~Weiwei, ``The integrated panoramic surveillance system based on tethered
  balloon,'' in {\em 2015 IEEE Aerospace Conference}, pp.~1--7, IEEE, 2015.

\bibitem{ACCwebsite}
A.~C. C, ``Air combat command, tethered aerostat radar system, 2006.''
  {https://www.acc.af.mil/About-Us/Fact-Sheets/Display/Article/199135/tethered-aerostat-radar-system/}.

\bibitem{dusane2017elevated}
C.~R. Dusane, A.~V. Wani, R.~S. Pant, D.~Chakraborty, and B.~Chakravarthy, ``An
  elevated balloon-kite hybrid platform for surveillance,'' in {\em 23rd AIAA
  Lighter-Than-Air Systems Technology Conference}, p.~3995, 2017.

\bibitem{sharma2014design}
N.~Sharma, R.~Sehgal, R.~Sehgal, and R.~Pant, ``Design fabrication and
  deployment of a tethered aerostat system for aerial surveillance,'' in {\em
  National Level Conference on Advances in Aerial/Road Vehicle and its
  Application, MIT, Manipal}, pp.~18--19, 2014.

\bibitem{prior2015tethered}
S.~D. Prior, ``Tethered drones for persistent aerial surveillance
  applications,'' {\em Defence Global}, pp.~78--79, 2015.

\bibitem{krisztian2016military}
K.~Kriszti{\'a}n, ``Military ballooning in point of hungarian defense force’s
  communication support,'' {\em Rep{\"u}l{\'e}studom{\'a}nyi
  K{\"o}zlem{\'e}nyek}, vol.~28, no.~1, pp.~27--40, 2016.

\bibitem{hariyanto2009emergency}
H.~Hariyanto, H.~Santoso, and A.~K. Widiawan, ``Emergency broadband access
  network using low altitude platform,'' in {\em International Conference on
  Instrumentation, Communication, Information Technology, and Biomedical
  Engineering 2009}, pp.~1--6, IEEE, 2009.

\bibitem{qiantori2012emergency}
A.~Qiantori, A.~B. Sutiono, H.~Hariyanto, H.~Suwa, and T.~Ohta, ``An emergency
  medical communications system by low altitude platform at the early stages of
  a natural disaster in indonesia,'' {\em Journal of medical systems}, vol.~36,
  no.~1, pp.~41--52, 2012.

\bibitem{alsamhi2018tethered}
S.~H. Alsamhi, M.~S. Ansari, O.~Ma, F.~Almalki, and S.~K. Gupta, ``Tethered
  balloon technology in design solutions for rescue and relief team emergency
  communication services,'' {\em Disaster Med. Public Health Prep}, pp.~1--8,
  2018.

\bibitem{valcarce2013airborne}
A.~Valcarce, T.~Rasheed, K.~Gomez, S.~Kandeepan, L.~Reynaud, R.~Hermenier,
  A.~Munari, M.~Mohorcic, M.~Smolnikar, and I.~Bucaille, ``Airborne base
  stations for emergency and temporary events,'' in {\em International
  conference on personal satellite services}, pp.~13--25, Springer, 2013.

\bibitem{barrado2010wildfire}
C.~Barrado, R.~Messeguer, J.~L{\'o}pez, E.~Pastor, E.~Santamaria, and P.~Royo,
  ``Wildfire monitoring using a mixed air-ground mobile network,'' {\em IEEE
  Pervasive Computing}, vol.~9, no.~4, pp.~24--32, 2010.

\bibitem{alsamhi2019performance}
S.~Alsamhi, F.~Almalki, O.~Ma, M.~Ansari, and M.~Angelides, ``Performance
  optimization of tethered balloon technology for public safety and emergency
  communications,'' {\em Telecommunication Systems}, pp.~1--10, 2019.

\bibitem{alsamhi2018disaster}
S.~H. Alsamhi, M.~S. Ansari, and N.~S. Rajput, ``Disaster coverage predication
  for the emerging tethered balloon technology: capability for preparedness,
  detection, mitigation, and response,'' {\em Disaster medicine and public
  health preparedness}, vol.~12, no.~2, pp.~222--231, 2018.

\bibitem{kanoria2012winged}
A.~A. Kanoria and R.~S. Pant, ``Winged aerostat systems for better station
  keeping for aerial surveillance,'' in {\em Advanced Materials Research},
  vol.~433, pp.~6871--6879, Trans Tech Publ, 2012.

\bibitem{badesha2006optical}
S.~S. Badesha, A.~D. Goldfinger, and T.~W. Jerardi, ``Optical communication
  system using a high altitude tethered balloon,'' May~16 2006.
\newblock US Patent 7,046,934.

\bibitem{alsamhi2019tethered}
S.~Alsamhi, M.~S. Ansari, L.~Zhao, S.~N. Van, S.~Gupta, A.~A. Alammari, A.~H.
  Saber, M.~Y. Hebah, M.~A.~A. Alasali, H.~M. Aljabali, {\em et~al.},
  ``Tethered balloon technology for green communication in smart cities and
  healthy environment,'' in {\em 2019 First International Conference of
  Intelligent Computing and Engineering (ICOICE)}, pp.~1--7, IEEE, 2019.

\bibitem{alsamhi2016network}
S.~Alsamhi, S.~K. Gapta, N.~Rajput, and R.~Saket, ``Network architectures
  exploiting multiple tethered balloon constellations for coverage extension,''
  in {\em 6th international conference on advances in engineering sciences and
  applied mathematics Kuala Lumpur (Malaysia)}, 2016.

\bibitem{bilaye2008low}
P.~Bilaye, V.~Gawande, U.~Desai, A.~Raina, and R.~Pant, ``Low cost wireless
  internet access for rural areas using tethered aerostats,'' in {\em 2008 IEEE
  Region 10 and the Third international Conference on Industrial and
  Information Systems}, pp.~1--5, IEEE, 2008.

\bibitem{chen2006spectral}
X.~Chen and L.~Vierling, ``Spectral mixture analyses of hyperspectral data
  acquired using a tethered balloon,'' {\em Remote Sensing of Environment},
  vol.~103, no.~3, pp.~338--350, 2006.

\bibitem{wu2018experimental}
J.~Wu, E.~Li, X.~Shen, S.~Yao, Z.~Tong, C.~Hu, Z.~Liu, S.~Liu, S.~Tan, and
  S.~Han, ``Experimental results of the balloon-borne spectral camera based on
  ghost imaging via sparsity constraints,'' {\em IEEE Access}, vol.~6,
  pp.~68740--68748, 2018.

\bibitem{inoue2000blimp}
Y.~Inoue, S.~Morinaga, and A.~Tomita, ``A blimp-based remote sensing system for
  low-altitude monitoring of plant variables: a preliminary experiment for
  agricultural and ecological applications,'' {\em International Journal of
  Remote Sensing}, vol.~21, no.~2, pp.~379--385, 2000.

\bibitem{vierling2006short}
L.~A. Vierling, M.~Fersdahl, X.~Chen, Z.~Li, and P.~Zimmerman, ``The short wave
  aerostat-mounted imager (swami): A novel platform for acquiring remotely
  sensed data from a tethered balloon,'' {\em Remote sensing of environment},
  vol.~103, no.~3, pp.~255--264, 2006.

\bibitem{jo2015mapping}
Y.-H. Jo, J.~Sha, J.-I. Kwon, K.~Jun, and J.~Park, ``Mapping bathymetry based
  on waterlines observed from low altitude helikite remote sensing platform,''
  {\em Acta Oceanologica Sinica}, vol.~34, no.~9, pp.~110--116, 2015.

\bibitem{lee2019estimate}
J.-S. Lee, J.-Y. Baek, D.~Jung, J.-S. Shim, H.-S. Lim, and Y.-H. Jo, ``Estimate
  of coastal water depth based on aerial photographs using a low-altitude
  remote sensing system,'' {\em Ocean Science Journal}, vol.~54, no.~3,
  pp.~349--362, 2019.

\bibitem{garstang1971fluctuations}
M.~Garstang, M.~Murday, W.~R. Seguin, J.~D. Brown, and N.~E. Laseur,
  ``Fluctuations in humidity, temperature, and horizontal wind as measured by a
  subcloud tethered-balloon system,'' {\em IEEE Transactions on Geoscience
  Electronics}, vol.~9, no.~4, pp.~199--208, 1971.

\bibitem{balasuriya2017development}
A.~Balasuriya, T.~Pennington, T.~Scudere, M.~McCann, R.~Thayer, and R.~Wronski,
  ``Development of an autonomous mobile marine meteorological
  station—swims,'' in {\em OCEANS 2017-Anchorage}, pp.~1--4, IEEE, 2017.

\bibitem{hu2019atmospheric}
D.~Hu, B.~Qi, R.~Du, H.~Yang, J.~Wang, and J.~Zhuge, ``An atmospheric vertical
  detection system using the multi-rotor uav,'' in {\em 2019 International
  Conference on Meteorology Observations (ICMO)}, pp.~1--4, IEEE, 2019.

\bibitem{krishnamoorthy}
S.~Krishnamoorthy, A.~Komjathy, M.~T. Pauken, J.~A. Cutts, R.~F. Garcia,
  D.~Mimoun, A.~Cadu, A.~Sournac, J.~M. Jackson, V.~H. Lai, {\em et~al.},
  ``Detection of artificially generated seismic signals using balloon-borne
  infrasound sensors,'' {\em Geophysical Research Letters}, vol.~45, no.~8,
  pp.~3393--3403, 2018.

\bibitem{krishnamoorthy20}
S.~Krishnamoorthy, V.~H. Lai, A.~Komjathy, M.~T. Pauken, J.~A. Cutts, R.~F.
  Garcia, D.~Mimoun, J.~M. Jackson, D.~C. Bowman, E.~Kassarian, {\em et~al.},
  ``Aerial seismology using balloon-based barometers,'' {\em IEEE Transactions
  on Geoscience and Remote Sensing}, vol.~57, no.~12, pp.~10191--10201, 2019.

\bibitem{deschesnes2005design}
F.~Deschesnes and M.~Nahon, ``Design improvements for a multi-tethered aerostat
  system,'' in {\em AIAA Atmospheric Flight Mechanics Conference and Exhibit},
  p.~6126, 2005.

\bibitem{lambert2003design}
C.~Lambert, A.~Saunders, C.~Crawford, and M.~Nahon, ``Design of a one-third
  scale multi-tethered aerostat system for precise positioning of a radio
  telescope receiver,'' in {\em CASI Flight Mechanics and Operations
  Symposium}, pp.~1--12, 2003.

\bibitem{fitzsimmons2000steady}
J.~T. Fitzsimmons, B.~Veidt, and P.~E. Dewdney, ``Steady-state analysis of the
  multi-tethered aerostat platform for the large adaptive reflector
  telescope,'' in {\em Radio Telescopes}, vol.~4015, pp.~476--487,
  International Society for Optics and Photonics, 2000.

\bibitem{nahon2002dynamics}
M.~Nahon, G.~Gilardi, and C.~Lambert, ``Dynamics/control of a radio telescope
  receiver supported by a tethered aerostat,'' {\em Journal of Guidance,
  Control, and Dynamics}, vol.~25, no.~6, pp.~1107--1115, 2002.

\bibitem{bajoria2017design}
A.~Bajoria, N.~K. Mahto, C.~K. Boppana, and R.~S. Pant, ``Design of a tethered
  aerostat system for animal and bird hazard mitigation,'' in {\em 2017 First
  International Conference on Recent Advances in Aerospace Engineering
  (ICRAAE)}, pp.~1--6, IEEE, 2017.

\bibitem{pant2016tethered}
R.~S. Pant, ``Tethered aerostat systems for agricultural applications in
  india,'' {\em The Institution of Engineers (India)}, p.~82, 2016.

\bibitem{silva2014mapping}
B.~Silva, L.~Lehnert, K.~Roos, A.~Fries, R.~Rollenbeck, E.~Beck, and J.~Bendix,
  ``Mapping two competing grassland species from a low-altitude helium
  balloon,'' {\em IEEE Journal of Selected Topics in Applied Earth Observations
  and Remote Sensing}, vol.~7, no.~7, pp.~3038--3049, 2014.

\bibitem{fiorentin2018minlu}
P.~Fiorentin, C.~Bettanini, E.~Lorenzini, A.~Aboudan, G.~Colombatti,
  S.~Ortolani, and A.~Bertolo, ``Minlu: An instrumental suite for monitoring
  light pollution from drones or airballoons,'' in {\em 2018 5th IEEE
  International Workshop on Metrology for AeroSpace (MetroAeroSpace)},
  pp.~274--278, IEEE, 2018.

\bibitem{aglietti2009harnessing}
G.~S. Aglietti, S.~Redi, A.~R. Tatnall, and T.~Markvart, ``Harnessing
  high-altitude solar power,'' {\em IEEE Transactions on Energy Conversion},
  vol.~24, no.~2, pp.~442--451, 2009.

\bibitem{aglietti2008solar1}
G.~Aglietti, T.~Markvart, A.~Tatnall, and S.~Walker, ``Solar power generation
  using high altitude platforms feasibility and viability,'' {\em Progress in
  Photovoltaics: Research and Applications}, vol.~16, no.~4, pp.~349--359,
  2008.

\bibitem{ghosh2017power}
K.~Ghosh, A.~Guha, and S.~P. Duttagupta, ``Power generation on a solar
  photovoltaic array integrated with lighter-than-air platform at low
  altitudes,'' {\em Energy Conversion and Management}, vol.~154, pp.~286--298,
  2017.

\bibitem{fotouhi2019survey}
A.~Fotouhi, H.~Qiang, M.~Ding, M.~Hassan, L.~G. Giordano, A.~Garcia-Rodriguez,
  and J.~Yuan, ``Survey on uav cellular communications: Practical aspects,
  standardization advancements, regulation, and security challenges,'' {\em
  IEEE Communications Surveys \& Tutorials}, vol.~21, no.~4, pp.~3417--3442,
  2019.

\bibitem{finn2014study}
R.~L. Finn, D.~Wright, L.~Jacques, and P.~De~Hert, ``Study on privacy, data
  protection and ethical risks in civil remotely piloted aircraft systems
  operations,'' {\em Final Report, Luxembourg: Publications Office of the
  European Union}, vol.~39, 2014.

\bibitem{FFAwebsite}
F.~A. A, ``The federal aviation administration, title 14: Aeronautics and
  space, part 101 — moored balloons, kites, amateur rockets, unmanned free
  balloons, and certain model aircraft, and part 107 — small unmanned
  aircraft systems.'' {https://www.ecfr.gov/}.

\bibitem{LOCwebsite}
L.~of~C, ``Library of congress, regulation of drones.''
  {https://www.loc.gov/law/help/regulation-of-drones/}.

\bibitem{stocker2017review}
C.~St{\"o}cker, R.~Bennett, F.~Nex, M.~Gerke, and J.~Zevenbergen, ``Review of
  the current state of uav regulations,'' {\em Remote sensing}, vol.~9, no.~5,
  p.~459, 2017.

\bibitem{FCCwebsite}
F.~C. C, ``The federal communications commission (fcc).''
  {https://www.fcc.gov/}.

\bibitem{elistair_SC_1}
Elisair, ``Paris airport maintenance use case.''
  {https://elistair.com/airport-maintenance-use-case/}.

\bibitem{datafromsky}
DataFromSky, ``Deep traffic video analysis - datafromsky.''
  {https://datafromsky.com/}.

\bibitem{elistair_SC_2}
Elisair, ``Roundabout traffic monitoring use case in lyon.''
  {https://elistair.com/roundabout-traffic-monitoring-use-case/}.

\bibitem{CBP}
C.B.P, ``U.s. customs and border protection: U.s. border patrol southwest
  border apprehensions.''
  {https://www.cbp.gov/newsroom/stats/usbp-sw-border-apprehensions}.

\bibitem{TCOM_CS1}
TCOM, ``Tcom case study: Border patrol in southern texas.''
  {https://tcomlp.com/aerostat-system-border-patrol/}.

\bibitem{NOFO}
NOFO, ``The norwegian clean seas association for operating companies.''
  {https://www.nofo.no/}.

\bibitem{AAA}
AAA, ``Aerostats all australia (aaa) mobile coverage.''
  {https://www.bal.com.au/AAA.pdf}.

\bibitem{LT}
L.~T, ``Lindstrand technologies.'' {https://www.lindstrandtech.com/}.

\bibitem{CNIM}
CNIM, ``Cnim air space.'' {https://cnim-air-space.com/fr/}.

\bibitem{ADASI}
ADASI, ``Adasi: Get the edge with the autonomous advantage.''
  {https://adasi.ae/}.

\bibitem{Vigilance}
vigilance, ``Vigilance.'' {http://www.vigilance.nl/}.

\bibitem{DroneAC}
Drone, ``Drone aviation corp.'' {https://droneaviationcorp.com/}.

\bibitem{SkyDoc}
SkyDoc, ``Skydoc balloon.'' {http://www.skydocballoon.com/}.

\bibitem{Equinox}
E.~I. S, ``Equinox innovative systems.''
  {https://www.equinoxinnovativesystems.com/}.

\bibitem{TDS}
T.~D. S, ``Tethered drone systems.'' {https://tethereddronesystems.co.uk/}.

\bibitem{Hoverfly}
Hoverfly, ``Hoverfly technology.'' {https://hoverflytech.com/}.

\bibitem{Fotokite}
Fotokite, ``Fotokite.'' {https://fotokite.com/}.

\bibitem{AWE}
AWE, ``Airborne wind europe.'' {https://airbornewindeurope.org/}.

\bibitem{spooky}
SA, ``Spooky action robotics.'' https://spookyactionrobotics.com/.

\bibitem{aeromana}
Aeromana, ``Aeromana.'' https://www.aeromana.com/.

\bibitem{6GSummit}
S.~A. I. t. T. A. C.~T. Ben~Glass, ``6g summit, connecting the unconnected.''
  http://6gsummit.org/.

\bibitem{cakaj2011range}
S.~Cakaj, B.~Kamo, V.~Koli{\c{c}}i, and O.~Shurdi, ``The range and horizon
  plane simulation for ground stations of low earth orbiting (leo)
  satellites.,'' {\em IJCNS}, vol.~4, no.~9, pp.~585--589, 2011.

\bibitem{geyer2016earth}
M.~Geyer {\em et~al.}, ``Earth-referenced aircraft navigation and surveillance
  analysis,'' tech. rep., John A. Volpe National Transportation Systems Center
  (US), 2016.

\bibitem{li2018placement}
P.~Li and J.~Xu, ``Placement optimization for uav-enabled wireless networks
  with multi-hop backhauls,'' 2018.

\bibitem{alzidaneen2019resource}
A.~Alzidaneen, A.~Alsharoa, and M.-S. Alouini, ``Resource and placement
  optimization for multiple uavs using backhaul tethered balloons,'' {\em IEEE
  Wireless Communications Letters}, vol.~9, no.~4, pp.~543--547, 2019.

\bibitem{selim2018post}
M.~Y. Selim and A.~E. Kamal, ``Post-disaster 4g/5g network rehabilitation using
  drones: Solving battery and backhaul issues,'' in {\em 2018 IEEE Globecom
  Workshops (GC Wkshps)}, pp.~1--6, IEEE, 2018.

\bibitem{zedini2016performance}
E.~Zedini, H.~Soury, and M.-S. Alouini, ``On the performance analysis of
  dual-hop mixed fso/rf systems,'' {\em IEEE Transactions on Wireless
  Communications}, vol.~15, no.~5, pp.~3679--3689, 2016.

\bibitem{alzenad2018fso}
M.~Alzenad, M.~Z. Shakir, H.~Yanikomeroglu, and M.-S. Alouini, ``Fso-based
  vertical backhaul/fronthaul framework for 5g+ wireless networks,'' {\em IEEE
  Communications Magazine}, vol.~56, no.~1, pp.~218--224, 2018.

\bibitem{kishk2020aerial}
M.~A. Kishk, A.~Bader, and M.-S. Alouini, ``Aerial base station deployment in
  6g cellular networks using tethered drones: The mobility and endurance
  tradeoff,'' 2020.

\bibitem{bushnaq2020optimal}
O.~Bushnaq, M.~Kishk, A.~{\c{C}}elik, M.~Alouini, and T.~Al-Naffouri, ``Optimal
  deployment of tethered drones for maximum cellular coverage in user
  clusters,'' {\em arXiv: 2003.00713}, 2020.

\bibitem{belmekki2019cooperative}
B.~E.~Y. Belmekki, A.~Hamza, and B.~Escrig, ``Cooperative vehicular
  communications at intersections over nakagami-m fading channels,'' {\em
  Vehicular Communications}, vol.~19, p.~100165, 2019.

\bibitem{2020outagecopMW}
B.~E.~Y. Belmekki, A.~Hamza, and B.~Escrig, ``Outage analysis of cooperative
  noma in millimeter wave vehicular network at intersections,'' in {\em 2020
  IEEE 91st Vehicular Technology Conference (VTC2020-Spring)}, pp.~1--6, IEEE,
  2020.

\bibitem{sudheesh2017sum}
P.~Sudheesh, M.~Mozaffari, M.~Magarini, W.~Saad, and P.~Muthuchidambaranathan,
  ``Sum-rate analysis for high altitude platform (hap) drones with tethered
  balloon relay,'' {\em IEEE Communications Letters}, vol.~22, no.~6,
  pp.~1240--1243, 2017.

\bibitem{belmekki2020performance1}
B.~E.~Y. Belmekki, A.~Hamza, and B.~Escrig, ``On the performance of 5g
  non-orthogonal multiple access for vehicular communications at road
  intersections,'' {\em Vehicular Communications}, vol.~22, p.~100202, 2020.

\bibitem{belmekki2020performance2}
B.~E.~Y. Belmekki, A.~Hamza, and B.~Escrig, ``Performance analysis of
  cooperative noma at intersections for vehicular communications in the
  presence of interference,'' {\em Ad hoc Networks}, vol.~98, p.~102036, 2020.

\bibitem{pan2021flying}
G.~Pan and M.-S. Alouini, ``Flying car transportation system: Advances,
  techniques, and challenges,'' {\em IEEE Access}, 2021.

\bibitem{saeed2020point}
N.~Saeed, H.~Almorad, H.~Dahrouj, T.~Y. Al-Naffouri, J.~S. Shamma, and M.-S.
  Alouini, ``Point-to-point communication in integrated satellite-aerial
  networks: State-of-the-art and future challenges,'' {\em arXiv preprint
  arXiv:2012.06182}, 2020.

\bibitem{AltaerosST}
Altaeros, ``Altaeros supertower.'' {https://smallcells.3g4g.co.uk/2019/04/}.

\bibitem{SNH}
S.~M. H, ``Nbn's first satellite, sky muster, launches successfully into
  orbit.''
  https://www.smh.com.au/technology/nbns-first-satellite-sky-muster-launches-successfully-into-orbit-20151001-gjymxv.html.

\bibitem{mynaric}
Mynaric, ``Mynaric, connectivity for the skies \& beyond.''
  https://mynaric.com/.

\bibitem{data2003prediction}
P.~Data, ``Prediction methods required for the design of terrestrial broadband
  millimetric radio access systems operating in a frequency range of about
  20-50 ghz,'' {\em Draft New Reco. ITU-R P.[DOC. 3/47], Working Party K},
  vol.~3, 2003.

\bibitem{al2014optimal}
A.~Al-Hourani, S.~Kandeepan, and S.~Lardner, ``Optimal lap altitude for maximum
  coverage,'' {\em IEEE Wireless Communications Letters}, vol.~3, no.~6,
  pp.~569--572, 2014.

\bibitem{azari2017ultra}
M.~M. Azari, F.~Rosas, K.-C. Chen, and S.~Pollin, ``Ultra reliable uav
  communication using altitude and cooperation diversity,'' {\em IEEE
  Transactions on Communications}, vol.~66, no.~1, pp.~330--344, 2017.

\bibitem{khuwaja2018survey}
A.~A. Khuwaja, Y.~Chen, N.~Zhao, M.-S. Alouini, and P.~Dobbins, ``A survey of
  channel modeling for uav communications,'' {\em IEEE Communications Surveys
  \& Tutorials}, vol.~20, no.~4, pp.~2804--2821, 2018.

\bibitem{abualhaol2010performance}
I.~Y. Abualhaol and M.~M. Matalgah, ``Performance analysis of multi-carrier
  relay-based uav network over fading channels,'' in {\em 2010 IEEE Globecom
  Workshops}, pp.~1811--1815, IEEE, 2010.

\bibitem{newhall2003wideband}
W.~G. Newhall, R.~Mostafa, C.~Dietrich, C.~R. Anderson, K.~Dietze, G.~Joshi,
  and J.~H. Reed, ``Wideband air-to-ground radio channel measurements using an
  antenna array at 2 ghz for low-altitude operations,'' in {\em IEEE Military
  Communications Conference, 2003. MILCOM 2003.}, vol.~2, pp.~1422--1427, IEEE,
  2003.

\bibitem{cai2017low}
X.~Cai, A.~Gonzalez-Plaza, D.~Alonso, L.~Zhang, C.~B. Rodr{\'\i}guez, A.~P.
  Yuste, and X.~Yin, ``Low altitude uav propagation channel modelling,'' in
  {\em 2017 11th European Conference on Antennas and Propagation (EUCAP)},
  pp.~1443--1447, IEEE, 2017.

\bibitem{ye2017air}
X.~Ye, X.~Cai, X.~Yin, J.~Rodr{\'\i}guez-Pi{\~n}eiro, L.~Tian, and J.~Dou,
  ``Air-to-ground big-data-assisted channel modeling based on passive sounding
  in lte networks,'' in {\em 2017 IEEE Globecom Workshops (GC Wkshps)},
  pp.~1--6, IEEE, 2017.

\bibitem{goddemeier2015investigation}
N.~Goddemeier and C.~Wietfeld, ``Investigation of air-to-air channel
  characteristics and a uav specific extension to the rice model,'' in {\em
  2015 IEEE Globecom Workshops (GC Wkshps)}, pp.~1--5, IEEE, 2015.

\bibitem{frew2008airborne}
E.~W. Frew and T.~X. Brown, ``Airborne communication networks for small
  unmanned aircraft systems,'' {\em Proceedings of the IEEE}, vol.~96, no.~12,
  2008.

\bibitem{yanmaz2013achieving}
E.~Yanmaz, R.~Kuschnig, and C.~Bettstetter, ``Achieving air-ground
  communications in 802.11 networks with three-dimensional aerial mobility,''
  in {\em 2013 Proceedings IEEE INFOCOM}, pp.~120--124, IEEE, 2013.

\bibitem{simunek2013uav}
M.~Simunek, F.~P. Font{\'a}n, and P.~Pechac, ``The uav low elevation
  propagation channel in urban areas: Statistical analysis and time-series
  generator,'' {\em IEEE Transactions on Antennas and Propagation}, vol.~61,
  no.~7, pp.~3850--3858, 2013.

\bibitem{karapantazis2005broadband}
S.~Karapantazis and F.~Pavlidou, ``Broadband communications via high-altitude
  platforms: A survey,'' {\em IEEE Communications Surveys \& Tutorials},
  vol.~7, no.~1, pp.~2--31, 2005.

\bibitem{cuevas2004channel}
J.~L. Cuevas-Ru{\'\i}z and J.~A. Delgado-Pen{\'\i}n, ``Channel modeling and
  simulation in haps systems,'' {\em European Wireless 2004}, pp.~24--27, 2004.

\bibitem{zhao2020ka}
J.~Zhao, Q.~Wang, Y.~Li, J.~Zhou, and W.~Zhou, ``Ka-band based channel modeling
  and analysis in high altitude platform (hap) system,'' in {\em 2020 IEEE 91st
  Vehicular Technology Conference (VTC2020-Spring)}, pp.~1--5, IEEE, 2020.

\bibitem{trichili2020roadmap}
A.~Trichili, M.~A. Cox, B.~S. Ooi, and M.-S. Alouini, ``Roadmap to free space
  optics,'' {\em JOSA B}, vol.~37, no.~11, pp.~A184--A201, 2020.

\bibitem{nistazakis2009average}
H.~E. Nistazakis, E.~A. Karagianni, A.~D. Tsigopoulos, M.~E. Fafalios, and
  G.~S. Tombras, ``Average capacity of optical wireless communication systems
  over atmospheric turbulence channels,'' {\em Journal of Lightwave
  Technology}, vol.~27, no.~8, pp.~974--979, 2009.

\bibitem{al2001mathematical}
A.~Al-Habash, L.~C. Andrews, and R.~L. Phillips, ``Mathematical model for the
  irradiance probability density function of a laser beam propagating through
  turbulent media,'' {\em Optical engineering}, vol.~40, no.~8, pp.~1554--1562,
  2001.

\bibitem{tsiftsis2009optical}
T.~A. Tsiftsis, H.~G. Sandalidis, G.~K. Karagiannidis, and M.~Uysal, ``Optical
  wireless links with spatial diversity over strong atmospheric turbulence
  channels,'' {\em IEEE Transactions on Wireless Communications}, vol.~8,
  no.~2, pp.~951--957, 2009.

\bibitem{nistazakis2011performance}
H.~Nistazakis, V.~Assimakopoulos, and G.~Tombras, ``Performance estimation of
  free space optical links over negative exponential atmospheric turbulence
  channels,'' {\em Optik}, vol.~122, no.~24, pp.~2191--2194, 2011.

\bibitem{jurado2011unifying}
A.~Jurado-Navas, J.~M. Garrido-Balsells, J.~F. Paris, A.~Puerta-Notario, and
  J.~Awrejcewicz, ``A unifying statistical model for atmospheric optical
  scintillation,'' {\em Numerical simulations of physical and engineering
  processes}, vol.~181, no.~8, pp.~181--205, 2011.

\bibitem{alquwaiee2016asymptotic}
H.~AlQuwaiee, H.-C. Yang, and M.-S. Alouini, ``On the asymptotic capacity of
  dual-aperture fso systems with generalized pointing error model,'' {\em IEEE
  Transactions on Wireless Communications}, vol.~15, no.~9, pp.~6502--6512,
  2016.

\bibitem{farid2007outage}
A.~A. Farid and S.~Hranilovic, ``Outage capacity optimization for free-space
  optical links with pointing errors,'' {\em Journal of Lightwave technology},
  vol.~25, no.~7, pp.~1702--1710, 2007.

\bibitem{yang2014free}
F.~Yang, J.~Cheng, and T.~A. Tsiftsis, ``Free-space optical communication with
  nonzero boresight pointing errors,'' {\em IEEE Transactions on
  Communications}, vol.~62, no.~2, pp.~713--725, 2014.

\bibitem{gappmair2011ook}
W.~Gappmair, S.~Hranilovic, and E.~Leitgeb, ``Ook performance for terrestrial
  fso links in turbulent atmosphere with pointing errors modeled by hoyt
  distributions,'' {\em IEEE communications letters}, vol.~15, no.~8,
  pp.~875--877, 2011.

\bibitem{vavoulas2012weather}
A.~Vavoulas, H.~G. Sandalidis, and D.~Varoutas, ``Weather effects on fso
  network connectivity,'' {\em IEEE/OSA Journal of Optical Communications and
  Networking}, vol.~4, no.~10, pp.~734--740, 2012.

\bibitem{nadeem2009weather}
F.~Nadeem, V.~Kvicera, M.~S. Awan, E.~Leitgeb, S.~S. Muhammad, and G.~Kandus,
  ``Weather effects on hybrid fso/rf communication link,'' {\em IEEE journal on
  selected areas in communications}, vol.~27, no.~9, pp.~1687--1697, 2009.

\bibitem{data2007prediction}
P.~Data, ``Prediction methods required for the design of terrestrial
  line-of-sight systems,'' {\em Recommendation ITU-R}, p.~530, 2007.

\bibitem{esmail2016experimental}
M.~A. Esmail, H.~Fathallah, and M.-S. Alouini, ``An experimental study of fso
  link performance in desert environment,'' {\em IEEE Communications Letters},
  vol.~20, no.~9, pp.~1888--1891, 2016.

\end{thebibliography}

\begin{IEEEbiography}
    [{\includegraphics[width=2.6cm,height=2.8cm,clip,keepaspectratio]{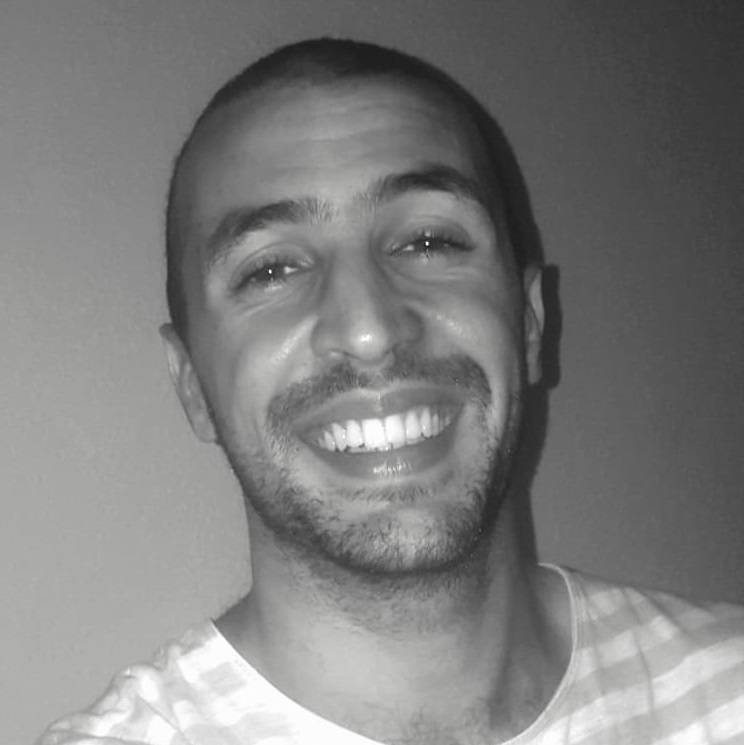}}]{Baha Eddine Youcef Belmekki}
received the B.S. degree in Electronics, and the M.Sc. degree in Wireless Communications and Networking from University of Science and Technology Houari Boumediene, Algiers, Algeria, in 2011 and 2013. From 2013 to 2014, he was with the Department of Radio Access Network, Huawei Technologies, Algiers, Algeria. From 2014 to 2016, he was an Assistant Professor with University of Science and Technology Houari Boumediene, Algiers, Algeria. He obtained his Ph.D. degrees in Signal Processing from the National Polytechnic Institute of Toulouse in 2020. He is currently working as teaching and research assistant at the National Polytechnic Institute of Toulouse, France.
His research interests include cooperative communications, millimeter wave communications, non-orthogonal multiple access systems, stochastic geometry analysis of vehicular networks, and networked flying platforms.
\end{IEEEbiography}
\vspace{-15cm}

\begin{IEEEbiography}
    [{\includegraphics[width=2.6cm,height=2.8cm,clip,keepaspectratio]{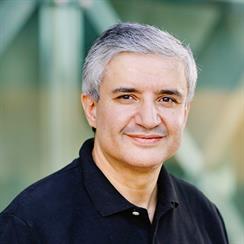}}]
    {Mohamed-Slim Alouini}
(S’94-M’98-SM’03-
F’09) was born in Tunis, Tunisia. He received
the Ph.D. degree in Electrical Engineering from
the California Institute of Technology (Caltech),
Pasadena, CA, USA, in 1998. He served as a
faculty member in the University of Minnesota,
Minneapolis, MN, USA, then in the Texas A\&M
University at Qatar, Education City, Doha, Qatar
before joining King Abdullah University of
Science and Technology (KAUST), Thuwal,
Makkah Province, Saudi Arabia as a Professor
of Electrical and Computer Engineering in 2009. His current research
interests include the modeling, design, and performance analysis of wireless
communication systems.
\end{IEEEbiography}

\end{document}